%% file: DDF_decay.v0.tex
\documentclass[a4paper,11pt]{article}

\usepackage[
backend=biber,
style=numeric-comp, 
sorting=none 
]{biblatex}

\usepackage[utf8]{inputenc}
\usepackage[british]{babel}
\usepackage{csquotes}

\usepackage{manfnt}
\usepackage{braket}

\usepackage{graphicx}
\usepackage{subcaption} 
\usepackage{color} 
\usepackage{float} 

\renewcommand{\arraystretch}{1.7}

\usepackage{tabularx}

\usepackage{makecell}

\usepackage{multirow}
\usepackage{ltablex}
\keepXColumns

\newlength{\mylen}
\newsavebox{\mybox}
\newcommand{\conditionalrow}[6]{%
  #1 & $#2$ & [$#3$] & $#4$ & $#5$ & $#6$%
\typeout{=== BOX DEBUG ===}%
\typeout{Content1: #1}%
\typeout{Content6: #6}%
}

\usepackage{pdflscape}
\usepackage{lscape} 

\usepackage{graphicx}
\usepackage{subcaption} 
\usepackage{color} 

\usepackage{authblk}

\usepackage{hyperref}
\usepackage{amsmath}
\usepackage{amssymb}
\usepackage{amsfonts}
\usepackage{enumitem}

\usepackage{cleveref} 

\numberwithin{equation}{section}
\usepackage{ytableau}

\newcommand{\COMMENTOOK}[1]{}

\newcommand{\wrt}{{w.r.t.} }

\newcommand{\lc}{{lightcone} }

\newcommand{\sN}[1]{ { [#1] } }  

\newcommand{\oh}{ \frac{1}{2} }
\newcommand{\ap}{ {\alpha'} }

\newcommand{\sdap}{ \sqrt{2\alpha'} }

\newcommand{\one}{ \mathbb{I} }

\newcommand{\R}{ \mathbb{R} }

\newcommand{\cA}{{\cal A} }
\newcommand{\cC}{{\cal C} }

\newcommand{\cT}{{\cal T} }

\newcommand{\dirst}[3]{
  \begin{array}{c} \lbrack #1 \rbrack
    \\ \lbrack #2 \rbrack \\ \lbrack #3 \rbrack \end{array}
  \,
}

\begin{document}

\begin{center}
  {\Large \bf
All DDF/\lc two-particle decay widths up to level $8$ in open bosonic
string in critical dimension
  }

\end{center}

\vskip .6cm
\medskip

\vspace*{4.0ex}

\baselineskip=18pt

\begin{center}

{\large 
\rm  Samuele Critelli$^a$ and Igor Pesando$^{ab}$ }

\end{center}

\vspace*{4.0ex}
\centerline{ \it \small $^a$ Dipartimento di Fisica, Universit\`{a} di Torino}
\centerline{ \it \small $^b$ I.N.F.N., Sezione di Torino}
  
\centerline{ \it \small 
  Via P.\ Giuria 1, I-10125 Torino}
  

\vspace*{1.0ex}
\centerline{\small E-mail:
samuele.critelli@edu.unito.it,
igor.pesando@to.infn.it}

\vspace*{5.0ex}

\renewcommand{\check}{\bar }

\centerline{\bf Abstract}
\bigskip

We carry out an ``experimental analysis'' in which we explicitly
compute all possible Abelian two-particle decay channels (excluding
the tachyon) of open bosonic massive states up to level (8), amounting
to approximately 2,000,000 cases. The aim is to develop intuition
about which states are the most stable and to identify the dominant
decay channels.

Our results show that, for all levels considered, the ratio of decay
widths between the slowest- and fastest-decaying states is of order
one.

In most cases, the dominant polarized decay channel accounts for at least $10\%$
of the total decay width, while the first five channels together
contribute roughly $60\%$.
Even when we sum about $10000$ polarized channels.

Dominant polarized channels always involve at least one photon 
and all strings ultimately decay into a photon carrying energy
at the string scale.

The most stable states are made using the \lc/DDF oscillators
$A_{-1}$ and some $A_{-2}$ and are close relatives of the states on
the leading Regge trajectory.

Finally, we discuss how the computation on the \lc differs
from the equal-time computation due to ``surges'', i.e., decays into a tachyon and a higher-mass state,
and we hypothesize that, in bosonic string theory, decays into tachyons
constitute the dominant decay channels.

%

%

\vfill

\vfill \eject
\baselineskip18pt

\tableofcontents











\section{Introduction and conclusions}
\label{sec:intro_conclu}

String theory is a strong candidate for a theory of quantum gravity
and, as such, it should be able to account for both spacelike and
timelike singularities.

Regarding timelike singularities, it is expected to describe black
holes—or black-hole–like objects observed astrophysically. In
particular, it should reproduce Hawking radiation if black holes, or
more generally any objects with an event horizon, exist, since this
effect is a robust prediction for all such systems. Alternatively, if
black holes do not exist as fundamental objects, the theory should be
capable of computing their microscopic entropy.

On the other end of the spectrum lie spacelike (or null)
singularities, which have proven difficult to handle within string
theory (see, e.g., the reviews cited
in \cite{Cornalba:2003kd,Craps:2006yb}). However, as pointed out
in \cite{Arduino:2020axy} and further discussed
in \cite{Arduino:2022mir,Pesando:2022amk}, it has been overlooked in
the literature that the issues arising in temporal orbifold
backgrounds already appear at tree level in open string theory and are
therefore not necessarily, or directly, of gravitational origin. In
particular, divergences arise as early as the three-point level when
at least one external state is massive. Nevertheless, it can be shown
that on-shell open string tree-level amplitudes remain finite once a
natural noncommutative structure is introduced \cite{IPRS2026}.

Independently of the problem of singularities, it is also natural to
ask whether observable string signatures could be detected in
gravitational waves, either in the coalescence spectra of “black
holes” or from primordial cosmic strings \cite{Frey:2024jqy} (see also
the review in \cite{Villa:2025wgl}).

In all these contexts, massive string states play an important, if not
central, role. Indeed, Hawking radiation may be interpreted as arising
from the decay of highly excited open strings; massive string states
are responsible for the structure of open string three-point
amplitudes in temporal orbifolds; and cosmic strings themselves can be
viewed as highly excited string configurations.


It was realized early on that essentially all massive string
decays \cite{DiGiacomo:1970xh, Neveu:1970iq, Green:1971zk}, as well as
their associated decay widths, must be taken into account in order to
construct a consistent S-matrix.

The problem of massive string decays has been addressed in the
literature using two main approaches.

The first focuses on the decay
of special classes of states, primarily those lying on the leading
Regge trajectory.
The second approach studies averaged properties,
under the assumption that these capture the behavior of a “typical”
string state. However, it is not clear that this assumption is
justified \cite{Firrotta:2024fvi}.

For the latter approach to be meaningful, it is crucial to understand
whether most states have comparable decay widths. Indeed, the inferred
properties rely on averaging over states in which each degree of
freedom is weighted equally.
If most states decay with similar widths,
this statistical mixture is preserved over time. Conversely, if the
decay widths are broadly distributed and not sharply peaked, the
composition of the mixture evolves, and the derived properties are
generically not reliable.

Other approaches have also been
explored \cite{Dai:1989cp,Manes:2001cs,Firrotta:2024fvi}.
In particular, the latter work introduces an elegant method to extract
tachyon and photon emission rates from the imaginary part of the
forward four-point amplitude.

In this work, we adopt a different perspective on the decay of massive
strings. We follow a deliberately simple and direct approach,
computing “experimentally” all decay channels that do not involve
tachyons, together with their partial widths, and then summing them to
obtain the total decay width.

We perform this analysis in the critical dimension using the
DDF/\lc formalism\footnote{It has recently been shown that, for
the bosonic open string, on-shell amplitudes in the two formalisms
coincide
exactly \cite{Biswas:2024epj,Biswas:2024mdu,Biswas:2024unn}}.
While
this method is not elegant and does not scale efficiently to very high
levels, it has the advantage of providing a detailed, fine-grained
view of the decay processes. The computation necessarily relies
heavily on symbolic manipulation, for which we employ the open-source
system maxima \cite{maxima}.

We restrict our analysis to $U(1)$ open strings and not to the color
ordered amplitudes.
This choice allows us to highlight a distinction between even and odd
excitation levels. After summing over the two orderings, the total
U(1) amplitude is nonvanishing only when $N_1+N_2+N_3$ 
 is even (where $N_r$  denote the levels of the $r$th states).
 As a consequence, even-level states can decay into two photons,
 whereas odd-level states cannot; instead, they decay into a photon
 and a spin-2 state.

There are, however, some limitations to our approach:
\begin{itemize}
\item
The use of DDF/\lc states allows one to work directly with physical states,
but the price to pay is that they correspond to mixed-spin configurations.
Nevertheless, DDF/\lc operators provide a complete and orthonormal basis of
physical states, and are therefore a convenient choice.
A caveat is in order: one may wonder whether the features we observe are an
artifact of this mixing or are instead basis-independent.
Indeed, the mixing involves coefficients spanning an enormous range, up to
$10^{2000}$ already at level $20$, when the states are normalized to have
relatively prime integer coefficients
\cite{Pesando:2024lqa,Bucciotti:2025dnh}.
There is however one selection rule discussed in
section \ref{sec:selection rules} which is quite restrictive at low levels. 
In order to extract decay properties of pure spin states, one should also
compute the corresponding interference terms, which we have not included.
In principle, interference effects could render some states significantly
more stable.
This raises the question of identifying the orthogonal basis associated with
the longest-lived states. Such a basis could be constructed, for instance, via
a Gram--Schmidt procedure combined with a maximization of the decay
time or directly using the Lorentz algebra \cite{Pesando:2024lqa,Bucciotti:2025dnh}.

\item
Some of the most stable states we identify do not generalize to arbitrary
levels.
For example, the only massive states on the leading Regge trajectory of the
form
$A^2_{-1}\,A^3_{-1}|k\rangle$ $\dots,$
$A^2_{-1}\,A^3_{-1}\dots\,A^{25}_{-1}|k\rangle$
exist only up to level $N=24$, yet they are among the most stable ones.

\item
We work in the critical dimension, and our results may depend sensitively on
the number of non-compact dimensions.
In particular, phase space factors tend to suppress (``kill'') decay widths as
the number of non-compact dimensions increases (e.g.\ $4$ vs.\ $26$)
\cite{Mitchell:1988qe}.

\end{itemize}

We compute all decays of massive bosonic open string states up to
level $8$ into non tachyonic states,  
amounting to approximately $2\cdot 10^6$ processes. We then address the following questions:
\begin{enumerate}
\item
  Is there any systematic trend in the total decay width (excluding final states  
  containing tachyons)?

\item
  Do all states at a given level have approximately the same total width?

%

%
\item
  Which are the most stable states?

\item
  Which are the dominant polarized decay channels?

\item
  What fraction of the total width (excluding tachyonic final states) is  
  contributed by the dominant polarized decay channels?
  
\end{enumerate}

The motivation for addressing these questions is to gain insight into whether  
the notion of a ``typical'' string state used in the literature is
meaningful,
and whether the resulting  massless spectrum is predominantly soft or hard.


The answers to the previous questions are:
\begin{enumerate}
\item
It appears that the total decay widths (excluding tachyonic final states)  
tend to become approximately constant at sufficiently high levels; see  
Figure \ref{fig:slowest_Fastest}.

\item
All DDF/\lc states seem to exhibit increasingly similar total decay widths  
(excluding tachyons), in the sense that the ratio between the fastest- and  
slowest-decaying states approaches $1$; see  
Figure \ref{fig:slowest_Fastest_ratio}.

\item
The most stable states appear to lie on the leading Regge trajectory
(or with some $A_{-2}$ ) and are essentially of the form
$\prod_{k=1}^N A^k_{-1} |k\rangle$
or
$\prod_{k=1}^N ( A^k_{-2} )^{m_k}  ( A^k_{-1} )^{n_k}  |k\rangle$
with few $A_{-2}$;
moreover because of the selection rule \ref{sec:selection rules}
most of them do not mix with other states and therefore the results are
completely stable see  
tables \ref{summaryL2}-\ref{summaryL8}.

\item
All $U(1)$ strings ultimately decay into hard photons. More precisely, for $N\le 8$,  
the fastest decay channels always involve at least one photon carrying energy  
at the string mass scale. In particular, for $N\le 7$, the dominant decay  
channels are into two photons for even $N$  
(see tables \ref{N=2_state_1}, \ref{N=2_state_2}, \ref{N=2_state_3},  
\ref{app:N=4}, \ref{app:N=6} and supplementary material), and into one photon plus one spin-$2$ state for  
odd $N$ (see tables \ref{app:N=3}, \ref{app:N=5}, \ref{app:N=7} and supplementary material), whereas for  
$N=8$ the fastest decays typically involve at least one photon, though not  
exclusively (see table \ref{app:N=8} and supplementary material).

\item
The fastest decay channels account for at least about $10\%$ of the total decay  
width (see Figure \ref{fig:ratio_fastest_partial_total}), while the five fastest  
channels together account for at least about $60\%$  
(see Figure \ref{fig:ratio_5_fastest_partials_total}).

\end{enumerate}

We can now compare with the results obtained in the literature.

On a general ground we can ask how much we can trust the generalization
of the findings beyond $N=8$.
The true honest answer is that we do not know since there are arguments in
favor and against.
If we consider the results in in the graph in \cite{Okada:1989sd} we
see that the results from low $N$ extend in a smooth and linear way to
higher $N$.
There are however three catches: they consider the leading Regge
trajectory in the superstring and so $D=10$ and they found a decay
rate rising linearly in $\sqrt{N}$.
On the other side if we consider figure 6 in \cite{Mitchell:1988qe}
we see a dependence on $N$ and a decrease of the total decay width as
$1/\sqrt{N-1}$.
Again we cannot really compare since they deal with the open bosonic string
leading Regge trajectory in a space where the non compact directions
are $4$.

%
Let us consider the total width which is the main subject of this research.
Our result roughly  matches the ones in \cite{Green:1971zk,Mitchell:1988qe}
even if the results obtained in these papers include the tachyon as a
final state.

There is however a mismatch of the actual values
\wrt \cite{Mitchell:1988qe}.
Most of this paper deals with the leading Regge trajectory and
an effective dimension $4$ since they
assume the internal dimensions to be very small so that it is not
possible to excited KK momentum.
However at the end of paragraph 6 they quote the total decay width
in critical dimension for level $20$ leading Regge state
to be $\Gamma_{20}=3.78\cdot 10^{-10}$
and
for level $40$ to be $\Gamma_{20}=7.79\cdot 10^{-10}$
(it is nor clear which $g^2$ is used or whether it is factored out).
These results do not seem to be consistent with their general
statement according to which $\Gamma$ should slowly decrease.
Even so the order of magnitude is clear.
We get $\Gamma\sim 10^{-20} g^2$ without tachyons.
This order of magnitude for the total width is also the value for DDF/\lc states like
$A^2_{-1}\,A^3_{-1}|k\rangle$,
$A^2_{-1}\,A^3_{-1}\,A^4_{-1}|k\rangle$,
\dots
$A^2_{-1}\,\dots\,A^{9}_{-1}|k\rangle$
which are specific polarizations of the leading Regge trajectory and
do not mix.

It is then reasonable to ask which checks we have on our results.
First of all we have checked
that we (obviously) have the same theoretical expression
for the decay width
(compare our eq. \eqref{app:eq:adimensional_decay_width} with their eq. 10). 
Then we have a rigorous check and an intuitive one
besides having checked that the results are independent on the order
of the decay particles.

Let us start with he intuitive one.
Partial decay widths involve
$log()$s as for example in table  \ref{N=3_state_1}
which cancel in the total decay width .
And this happens also for level $N=8$ when summing sometimes
$O(10000)$ partial channels (see table \ref{summaryL8}).

As far the rigorous one we notice that the total decay width (without
tachyon in the final state) of the three $N=2$ states is equal.
To get this result we have to sum over $4$, $4$ and $5$ partial
channels and the partial widths are very different
(see tables \ref{N=2_state_1},\ref{N=2_state_2},\ref{N=2_state_3}).

%

One could object that tachyonic final states are treated differently.
However, space $SO(25)$ transformations act only on the polarization
structure while preserving the physical state and the null velocity of
the decaying particle.  
As a result, they do not mix decay channels involving tachyons with
those that do not involve tachyons.

Another objection concerns the possible mixing but as we show in
section \ref{sec:selection rules} there is no mixing among these
states because of a selection rule.

Actually we can prove a slightly stronger result: the partial
widths into states of fixed levels are independent on the polarization
of the decaying state.
In fact consider the decay of a state of level $N_1$ in its center of
mass.
It has polarization $S_1$ and decays into two
states with with levels $N_2$, $N_3$,
polarizations $S_2$ and $S_3$ and momenta $\pm \vec p$.
The partial width is
\begin{equation}
\Gamma_{N_2, N_3}(S_1)
=
\int_\Omega \mu(p) 
\sum_{S_2, S_3}
| \langle S_2, S_3, \vec p | S_1 \rangle|^2
,
\end{equation}
with $\mu(p) \propto d^{D-1} \vec p\, \delta(\vec p^2 -p_*^2)$ as
follows from eq. \eqref{app:eq:adimensional_decay_width}.
Then the partial width of the rotated polarization $R S_1$ is
\begin{align}
\Gamma_{N_2, N_3}(R S_1)
=&
\int_\Omega \mu(p) 
\sum_{S_2, S_3}
| \langle S_2, S_3, \vec p | R S_1 \rangle|^2
=
\int_{R \Omega} \mu( R p') 
\sum_{R S_2', R S_3'}
| \langle R S_2', R S_3', R \vec p' | R S_1 \rangle|^2
\nonumber\\
=&
\int_{R\Omega} \mu(p' ) 
\sum_{S_2', S_3'}
| \langle S_2', S_3', \vec p' |S_1 \rangle|^2
,
\end{align}
where we have first renamed the dummy variables and then used
rotational invariance.
Invariance for the partial width follows from invariance of the
integration volume $\Omega$.
This is always true for equal time quantization where $\Omega$ is a
sphere.
However decay partial widths computed in equal time formalism and \lc
one match only for decays without tachyons since only in this case the
two formalisms give the same integration region $\Omega$.
The \lc one differs because of the integration region
$\Omega$ which is dictated by having positive \lc energies $k_{+} > 0$.

In any case we have verified that
the equal-time quantization decay widths of the three level $2$ states
into two tachyons for the three states are equal
($\frac{1}{4}\Gamma_ {N=2 \rightarrow T T}=$
$\Gamma_ {N=2 \rightarrow T T}^{\mbox{color ordered}}
=
{{5^{{{19}\over{2}}}}\over{341532874683\,2^{{{53}\over{2}}}\,\pi^{12}}}
$
$=1.4577\cdot 10^{-19}$).

This width is $10^6$ bigger that the total width into non tachyonic
states (see table \ref{summaryL2}) and
strongly suggests that decay widths into tachyons are responsible
for the discrepancy.


%
Let us now address the issue of tachyons in \lc and equal-time quantization.
In the presence of tachyons, \lc\ quantization allows for the occurrence
of ``surges,'' in which the mass of the final non-tachyonic particle can
be larger than or {\sl equal to} the mass of the initial non-tachyonic
decaying particle (see appendix~\ref{app:kinematics} for a discussion).

This phenomenon arises because one can have a tachyon with negative
equal-time energy, $E < 0$, while still having positive light-cone
energy, $\sqrt{2}\,k_+ = E + p_1 > 0$.

However, this comes at a cost: the system is no longer invariant under
$SO(25)$, but only under $SO(24)$. More precisely, the integration region
$\Omega$ is not invariant under rotations, i.e.\ $R\Omega \neq \Omega$.
As a consequence, in \lc\ quantization the decay of massive particles
into tachyons can depend on their polarization.

It follows that the true \lc\ Hamiltonian vacuum should be viewed as a
stationary superposition (``mixture'') of states in which decays and surges are in
equilibrium.

We choose to set aside this subtle issue, in line with the standard
treatment in the literature.

The paper is organized as follows.
In section \ref{sec:N=2} we give an overview of our approach using
the $N=2$ states as an example.
We discuss our notation and how the results are presented in the
following sections.
We derive the selection rule too.
Finally in section \ref{sec:results} we present a discussion of the
results of the actual decays, mostly as plots and tables.
For each of them we give the total width\footnote{We actually give the
adimensional total width divided by the squared string constant as
given in eq. \eqref{eq:adim_Gamma}.} 
and number of decay channels in
tables \ref{summaryL2}-\ref{summaryL8}.
For any level we give in the appendixes \ref{app:N=3}-\ref{app:N=8}
the 3 fastest decaying states of the 5 fastest decaying states
while in the supplementary material we give the 5 fastest and 5 slowest
decay amplitudes for all states.

These are the data necessary to give reasonably founded answers to the
questions we posed.

In the appendix \ref{app:kinematics} we discuss the kinematics which
is used in \ref{app:decay_width} to rederive and fix the notations of
the decay width $\Gamma$.
Finally in appendix \ref{app:Further_details} we give further details
on the implementation used in the program.

\section{The simplest case: level $N=2$ decay in details}
\label{sec:N=2}
In this section we make explicit all the steps we used in computing
all the decay amplitudes and the decay widths.
We consider only decays without tachyons.

\subsection{Level decays}
Th first step is to determine all the decays kinematically allowed and
non zero for a $U(1)$ open string.

Since the total 3 point amplitude is given by the sum of the color
ordered amplitudes ad
$ \langle V_1\, V_2\, V_3 \rangle  + \langle V_1\, V_3\, V_2 \rangle$
and we consider open string with $U(1)$ symmetry the previous
amplitude is equal to
$( 1+ (-1)^{N_1+N_2+N_3}) \langle V_1\, V_2\, V_3 \rangle $
where $N_r$ is the level of the $r$-th state ($r=1, 2, 3$)
hence it is not zero only when $N_1+N_2+N_3\equiv 0~~~\mod 2$.

We get therefore that the only allowed level decay is
\begin{equation}
[2,\, 1,\, 1]
,
\label{eq:N2_level_decay}
\end{equation}  
which means a state of level 2 can decay into two states of level $1$.

This happens because we have excluded the tachyon.
If we had included the tachyon we would have got the decays
\begin{equation}
[2,\, 1,\, 0]
,\,
[2,\, 0,\, 0]
,
\end{equation}  
and the infinite surges
\begin{equation}
[2,\, 0,\, 3]
,\,
[2,\, 0,\, 4]
\dots
\end{equation}  
These surges are not allowed in the usual equal time quantization but
are present in \lc since we require that the \lc energy be positive $\sqrt{2} k_{+} = E+ k_1>0$ and
not that the usual energy be positive $E>0$.

\subsection{Direction level decays}


The previous level decay can be embedded into the transverse space (we
are working in the lightcone formalism) in several different ways.

We must classify all such embeddings up to an appropriate equivalence relation.

The simplest choice is to require that the direction levels of state 1
form a non-decreasing sequence as the transverse direction index $i$ increases.
In doing so we define a canonical ordering of the whole state.

Explicitly from the level decay
$[2,\, 1,\, 1]$
in \eqref{eq:N2_level_decay}
we get the following direction level decays
\begin{align}
\,[ [ [2, 0], [0, 1], [0, 1]],
\nonumber\\
\,[ [2, 0, 0], [0, 1, 0], [0, 0, 1]],
\nonumber\\
\,[[2], [1], [1]],
\nonumber\\
\,[[2, 0], [1, 0], [0, 1]],
\nonumber\\
\,[[1, 1, 0], [0, 0, 1], [0, 0, 1]],
\nonumber\\
\,[[1, 1, 0, 0], [0, 0, 1, 0], [0, 0, 0, 1]],
\nonumber\\
\,[[1, 1], [0, 1], [0, 1]],
\nonumber\\
\,[[1, 1], [0, 1], [1, 0]],
\nonumber\\
\,[[1, 1, 0], [0, 1, 0], [0, 0, 1]]]
.
\end{align}

%
Let us explain the meaning of the previous notation.
For example $[ [2, 0, 0], [0, 1, 0], [0, 0, 1]]$ means
that
\begin{itemize}
\item
the direction state 1 $[2, 0, 0]$
has a level 2 excitation in direction $i=2$ and level 0 excitations
in directions $i=3,4$,
\item
the direction state 2 $[0, 1, 0]$ has a level 1 excitation in
direction $i=3$ and level 0 excitations in directions $i=2,4$,
\item
the excitation state 3 $[0, 0, 1]$ has a level 1 excitation in
direction $i=4$
and level 0 excitations in directions $i=2,3$.
\end{itemize}

The previous direction level decays are the really independent ones,
i.e. the direction level decays up to an equivalence relation.
All the other direction level decays and the derived decays have the
same decay width and their contribution to the total width can be
easily computed by computing the degeneracy.

Let us make an example.
The following states are equivalent 
\begin{align}
[ [2, 0, 0], [0, 1, 0], [0, 0, 1]]
\,
\equiv&
\,
[ [2, 0, 0, 0], [0, 0, 1, 0], [0, 0, 0, 1]]
\nonumber\\
\equiv&
\,
[ [2, 0, 0, 0, 0], [0, 0, 0, 1, 0], [0, 0, 0, 0, 1]]
\dots
,
\end{align}
but in this specific case since the 2nd and 3rd particles are equal we
have also the equivalences
\begin{equation}
[ [2, 0, 0], [0, 1, 0], [0, 0, 1]]
\,
\equiv
\,
[ [2, 0, 0], [0, 0, 1], [0, 1, 0]]
\,
\equiv
\,
[ [2, 0, 0, 0], [0, 0, 0, 1], [0, 0, 1, 0]]
\dots
.
\end{equation}

\subsection{Decays}
Finally each direction level may be realize in different way.
For example the level $2$ excitation in direction $i=2$ of the
direction state $[2, 0, 0]$ may be realized as the
{\sl normalized } states\footnote{
We use the usual DDF algebra $[A^i_m,\, A^j_n]= m \delta^{i j}\, \delta_{m+n,0}$.}
\begin{equation}
\frac{1}{\sqrt{1^2\, 2!}} \left( A^{i=2}_{-1}\right)^2\, |0\rangle,
~~~~
\frac{1}{\sqrt{2^1\, 1!}} A^{i=2}_{-2}\, |0\rangle
,
\end{equation}  
which we denote
\begin{equation}
[1,\, 2],
~~~~
[2,\,1]
,
\end{equation}
where the first number refers to the level of the oscillator and the
second to the power.

Therefore direction state $[2, 0, 0]$ becomes actually two different 
physical states
\begin{equation}
\{ [[2,1], [], []],~~ [1,2], [], []] \},
\end{equation}  
where $[]$ denotes the state without excitation.

At higher levels we have more complex cases.
For example at direction level 5 for direction $i=3$ we could have
the excitation $[3,1,1,2]$ which means
$\frac{1}{\sqrt{3^1\, 1!\, 1^2\, 2!}} A^{i=3}_{-3}\, \left( A^{i=3}_{-1}\right)^2 |0\rangle$.

\subsection{Decay amplitude for a decay}
Given a decay we use the standard expression ($i=2,\dots 25$)
\begin{align}
\langle V_3|
=&
\prod_{r=1}^3 \langle 0_{a [r]},\, x^i_{[r]}=0|\,
e^{
\sum_{r=1}^3\, \sum_{m=1}^\infty
A^i_{[r] m}\, {\cal P}^i\, N_{[r] m}
+
\oh \sum_{r, s=1}^3\, \sum_{m, n=1}^\infty
A^i_{[r] m}\, A^i_{[s] n}\, N_{[r] m, [s] n}
}
,
\nonumber\\
{\cal P}^i\
&= k^i_{[r+1]} \alpha_{[r]} - k^i_{[r]} \alpha_{[r+1]}
,
\nonumber\\
N_{[r] m}
&=
\frac{1}{ \alpha_{[r]}}
\frac{1}{m!} 
\prod_{k=1}^{m-1} \left( m \frac{-\alpha_{[r+1]}}{\alpha_{[r]}} - k\right)
,
\nonumber\\
N_{[r] m, [s] n}
&=
N_{[r] m}\, N_{[s] n}\,
\frac{-m n \alpha_{[1]}\, \alpha_{[2]}\, \alpha_{[3]} }
{ n \alpha_{[r]} + m \alpha_{[s]} }
,
\label{eq:LC_V3}
\end{align}
with $\alpha_{[r]} = 2 k_{[r] +}$ ($r=1,2,3$ and $r=1$ is the state
which decays)
to compute the amplitude.

We perform the computation the state 1 center of mass.
Because of this the modulus of state 2 and 3 momenta are fixed and
equal to $p_*$ (see app. \ref{app:kinematics} for details) and 
${\cal P}^i=\sqrt{2} m_i\, p_*\,  n^i$ ($i=2, \dots 25$)
where $(n^1, n^i)$ is a versor in $\R^{25}$.
We have introduced this versor because we have to consider all the
inequivalent momenta (see appendix \ref{app:decay_width} for details)
since the states have spin.
The spin implies a non trivial angular distribution which turns out to be
erratic \cite{Gross:2021gsj,Firrotta:2022cku,Bianchi:2022mhs,Hashimoto:2022bll,Firrotta:2023wem,Bianchi:2023uby,Das:2023xge,Savic:2024ock,Firrotta:2024qel,Bianchi:2024fsi,Bhattacharya:2024szw}.
This erraticity is however due to the spin
mixing \cite{Bianchi:2023uby,Bianchi:2024fsi,Pesando:2025ztr}.

\subsection{Decay width}
After we have computed the decay amplitude with its dependence on all
angles, we square it, normalize it and perform the exact angular
integration in order to get the final partial decay width $\Gamma$
(see appendix \ref{app:decay_width} for details).
It is noteworthy that sometimes up to about $100$ digits must be
retained in the numerical evaluation of the exact result in order to
get a positive partial decay width $\Gamma$.

The explicit results for the decay of $N=2$ states are contained in
the 3 tables \ref{N=2_state_1}, \ref{N=2_state_2}
and \ref{N=2_state_3}.

Each table has an header showing the state whose decays we consider,
the total width in floating point and the exact result and the number
of inequivalent decay channels.
It then follows the list of the actual inequivalent decays.
For each inequivalent decay it is given the partial width in floating point
and the exact result. 
In order to compute the contribution of the decays equivalent to the
given inequivalent decay the classical degeneracy, the quantum
degeneracy (due to Bose symmetry) are given.
To precisely compute the contribution from this class it is also
necessary to give the space time QFT normalization of the
state\footnote{
For example
the final state
$ A^{i=2}_{[2]-1} |k_{[2]}\rangle\, \oplus\,
  A^{i=2}_{[3]-1} |k_{[3]}\rangle $
corresponds to the QFT state
$\frac{1}{2!}
\left( a^\dagger_{\mbox{photon}}(\epsilon^{i=2}) \right)^2
|0_{\mbox{QFT}}\rangle $
with $a^\dagger_{\mbox{photon}}(\epsilon^i)$ the QFT creation operator for a
photon with polarization $\epsilon^i$
This QFT state requires a normalization factor $\frac{1}{2!}$.
}
.  
Finally the ratio between the sum of
partial widths of all decays  equivalent to the given inequivalent decay
and the total decay with is given between square brackets $[\dots]$.

\subsection{Selection rule}
\label{sec:selection rules}

As discussed in the introduction, we restrict our analysis to the
decay widths of mixed-spin states arising in the DDF/\lc
formalism.
Spin mixing allows for nontrivial decay widths between different states.

A simple selection rule determines which states can mix under decay.
It originates from the ``anomalous'' part of the light-cone boosts \cite{Pesando:2024lqa}, specifically from terms that replace a product
$A_{-m}^j \, A_{-n}^j$ with a single operator $A^i_{-m-n}$.

%
As discussed in appendix \ref{app:decay_width}
the mixing arises from
\begin{equation}
\Gamma_{(S_1, S_1')}
=
\int_\Omega \mu(p)\, \sum_{S_2, S_3}
\langle S_2, S_3, \vec p \mid S_1 \rangle\,
\left( \langle S_2, S_3, \vec p \mid S_1' \rangle \right)^*
.
\end{equation}

Since we integrate over all directions of the transverse momentum,
only terms containing an even number of momenta in each transverse
direction contribute.

Given the structure of the vertex~\eqref{eq:LC_V3}, the number of
momenta modulo $2$ in each direction is equal to the number of
excitations in that direction modulo $2$, independently on their levels.

In computing the mixing, the two final particles with polarizations
$S_2$ and $S_3$ are identical. Each contributes terms with the same
number of momenta modulo $2$; therefore, their combined contribution
to the mixing amplitude is always even and can be disregarded for the
purpose of the selection rule.

We can thus restrict our attention to the momenta associated with the
decaying state.

It then follows that only those initial states with the same number of
momenta modulo $2$ in each transverse direction can mix.


For example for $N=2$ we have the following cases of momenta modulo
$2$ per direction
\begin{align}
\begin{array} {c c}
\mbox{Momenta} & \mbox{states}
\\
(0) & [[1,2]]
\\
(1) & [[2,1]]
\\
(1,1) & [[1,1], [1,1]]
\end{array}
,
\end{align}
this means that in $N=2$ there is no mixing.
Differently for $N=3$ we have
\begin{align}
\begin{array} {c c}
\mbox{Momenta} & \mbox{states}
\\
(1) & [[3,1]];\, [[1,3]];\, [[1,2], [1,1]];\, [[2,1, 1,1]]
\\
(1,1) & [[2,1], [1,1]]
\\
(1,1,1) & [[1,1], [1,1], [1,1]]
\end{array}
.
\end{align}


%
%
\begin{longtable}[l]{m{0.4\textwidth} m{0.16\textwidth}
m{0.18\textwidth} m{0.10\textwidth} m{0.10\textwidth}
m{0.04\textwidth} }
\caption{$N=2$ state 1  }
\label{N=2_state_1}
\\\hline\hline
State & & Total width & Channels & &
\endfirsthead
\\
\multirow{2}{0.3\textwidth}{$\, [[1,2]] $ } 
& & $5.3406e-25$ 
& $4$  
& &
\\
& \multicolumn{5}{c}{
\makecell{${{1}\over{341532874683\,2^{{{45}\over{2}}}\,\pi^{12}}}$} 
  } 
\\ 
 \hline
decay & width & ratio  & cl. deg. & q. deg. & s.t. norm.
\\
$\, \begin{pmatrix} \left[ 1 , 2 \right] \cr \left[ 1 , 1 \right] \cr \left[ 1 , 1 \right] \cr \ifx\endpmatrix\undefined}\else\end{pmatrix}\fi  $ 
& & & & & \\ 
& $ 8.7230e-25 $ 
& [$ 8.1666e-1 $]  
& $ 1 $  
& $ 1 $ 
& $ {{1}\over{2}} $  
\\
\multicolumn{6}{ c }{ \makecell{${{9437}\over{61665657928875\,2^{{{55}\over{2}}}\,\pi^{12}}}$} 
} 
\\
\hline
$\, \begin{pmatrix} \left[ 1 , 2 \right] &\left[  \right] \cr \left[ 1 , 1 \right] &\left[  \right] \cr \left[  \right] &\left[ 1 , 1 \right] \cr \ifx\endpmatrix\undefined}\else\end{pmatrix}\fi  $ 
& & & & & \\ 
& $ 2.1876e-27 $ 
& [$ 9.4212e-2 $]  
& $ 46 $  
& $ 23 $ 
& $ 1 $  
\\
\multicolumn{6}{ c }{ \makecell{${{71}\over{184996973786625\,2^{{{55}\over{2}}}\,\pi^{12}}}$} 
} 
\\
\hline
$\, \begin{pmatrix} \left[ 1 , 2 \right] &\left[  \right] \cr \left[  \right] &\left[ 1 , 1 \right] \cr \left[  \right] &\left[ 1 , 1 \right] \cr \ifx\endpmatrix\undefined}\else\end{pmatrix}\fi  $ 
& & & & & \\ 
& $ 3.6858e-27 $ 
& [$ 7.9366e-2 $]  
& $ 23 $  
& $ 23 $ 
& $ {{1}\over{2}} $  
\\
\multicolumn{6}{ c }{ \makecell{${{4831}\over{7471031633690625\,2^{{{55}\over{2}}}\,\pi^{12}}}$} 
} 
\\
\hline
$\, \begin{pmatrix} \left[ 1 , 2 \right] &\left[  \right] &\left[  \right] \cr \left[  \right] &\left[ 1 , 1 \right] &\left[  \right] \cr \left[  \right] &\left[  \right] &\left[ 1 , 1 \right] \cr \ifx\endpmatrix\undefined}\else\end{pmatrix}\fi  $ 
& & & & & \\ 
& $ 2.0600e-29 $ 
& [$ 9.7586e-3 $]  
& $ 506 $  
& $ 253 $ 
& $ 1 $  
\\
\multicolumn{6}{ c }{ \makecell{${{1}\over{276704875321875\,2^{{{55}\over{2}}}\,\pi^{12}}}$} 
} 
\\
\hline
\end{longtable}
%
%
%

%
%
\begin{longtable}[l]{m{0.4\textwidth} m{0.16\textwidth} m{0.18\textwidth} m{0.10\textwidth} m{0.10\textwidth} m{0.04\textwidth} }
\caption{$N=2$ state 2  }
\label{N=2_state_2}
\\\hline\hline
State & & Total width & Channels & &
\endfirsthead
\\
\multirow{2}{0.3\textwidth}{$\, [[2,1]] $ } 
& & $5.3406e-25$ 
& $4$  
& &
\\
& \multicolumn{5}{c}{
\makecell{${{1}\over{341532874683\,2^{{{45}\over{2}}}\,\pi^{12}}}$} 
  } 
\\ 
 \hline
decay & width & ratio  & cl. deg. & q. deg. & s.t. norm.
\\
$\, \begin{pmatrix} \left[ 2 , 1 \right] &\left[  \right] \cr \left[ 1 , 1 \right] &\left[  \right] \cr \left[  \right] &\left[ 1 , 1 \right] \cr \ifx\endpmatrix\undefined}\else\end{pmatrix}\fi  $ 
& & & & & \\ 
& $ 2.3075e-26 $ 
& [$ 9.9375e-1 $]  
& $ 46 $  
& $ 23 $ 
& $ 1 $  
\\
\multicolumn{6}{ c }{ \makecell{${{6553}\over{6474894082531875\,2^{{{51}\over{2}}}\,\pi^{12}}}$} 
} 
\\
\hline
$\, \begin{pmatrix} \left[ 2 , 1 \right] &\left[  \right] \cr \left[  \right] &\left[ 1 , 1 \right] \cr \left[  \right] &\left[ 1 , 1 \right] \cr \ifx\endpmatrix\undefined}\else\end{pmatrix}\fi  $ 
& & & & & \\ 
& $ 2.5001e-28 $ 
& [$ 5.3835e-3 $]  
& $ 23 $  
& $ 23 $ 
& $ {{1}\over{2}} $  
\\
\multicolumn{6}{ c }{ \makecell{${{71}\over{6474894082531875\,2^{{{51}\over{2}}}\,\pi^{12}}}$} 
} 
\\
\hline
$\, \begin{pmatrix} \left[ 2 , 1 \right] &\left[  \right] &\left[  \right] \cr \left[  \right] &\left[ 1 , 1 \right] &\left[  \right] \cr \left[  \right] &\left[  \right] &\left[ 1 , 1 \right] \cr \ifx\endpmatrix\undefined}\else\end{pmatrix}\fi  $ 
& & & & & \\ 
& $ 1.1738e-30 $ 
& [$ 5.5604e-4 $]  
& $ 506 $  
& $ 253 $ 
& $ 1 $  
\\
\multicolumn{6}{ c }{ \makecell{${{1}\over{19424682247595625\,2^{{{51}\over{2}}}\,\pi^{12}}}$} 
} 
\\
\hline
$\, \begin{pmatrix} \left[ 2 , 1 \right] \cr \left[ 1 , 1 \right] \cr \left[ 1 , 1 \right] \cr \ifx\endpmatrix\undefined}\else\end{pmatrix}\fi  $ 
& & & & & \\ 
& $ 3.2983e-28 $ 
& [$ 3.0879e-4 $]  
& $ 1 $  
& $ 1 $ 
& $ {{1}\over{2}} $  
\\
\multicolumn{6}{ c }{ \makecell{${{281}\over{19424682247595625\,2^{{{51}\over{2}}}\,\pi^{12}}}$} 
} 
\\
\hline
\end{longtable}
%
%
%

%
%
\begin{longtable}[l]{m{0.4\textwidth} m{0.16\textwidth} m{0.18\textwidth} m{0.10\textwidth} m{0.10\textwidth} m{0.04\textwidth} }
\caption{$N=2$ state 3  }
\label{N=2_state_3}
\\\hline\hline
State & & Total width & Channels & &
\endfirsthead
\\
\multirow{2}{0.3\textwidth}{$\, [[1,1],[1,1]] $ } 
& & $5.3406e-25$ 
& $5$  
& &
\\
& \multicolumn{5}{c}{
\makecell{${{1}\over{341532874683\,2^{{{45}\over{2}}}\,\pi^{12}}}$} 
  } 
\\ 
 \hline
decay & width & ratio  & cl. deg. & q. deg. & s.t. norm.
\\
$\, \begin{pmatrix} \left[ 1 , 1 \right] &\left[ 1 , 1 \right] \cr \left[ 1 , 1 \right] &\left[  \right] \cr \left[  \right] &\left[ 1 , 1 \right] \cr \ifx\endpmatrix\undefined}\else\end{pmatrix}\fi  $ 
& & & & & \\ 
& $ 4.8960e-25 $ 
& [$ 9.1675e-1 $]  
& $ 2 $  
& $ 1 $ 
& $ 1 $  
\\
\multicolumn{6}{ c }{ \makecell{${{695203}\over{32374470412659375\,2^{{{51}\over{2}}}\,\pi^{12}}}$} 
} 
\\
\hline
$\, \begin{pmatrix} \left[ 1 , 1 \right] &\left[ 1 , 1 \right] &\left[  \right] \cr \left[ 1 , 1 \right] &\left[  \right] &\left[  \right] \cr \left[  \right] &\left[  \right] &\left[ 1 , 1 \right] \cr \ifx\endpmatrix\undefined}\else\end{pmatrix}\fi  $ 
& & & & & \\ 
& $ 9.0099e-28 $ 
& [$ 7.4229e-2 $]  
& $ 88 $  
& $ 44 $ 
& $ 1 $  
\\
\multicolumn{6}{ c }{ \makecell{${{101}\over{5111758486209375\,2^{{{49}\over{2}}}\,\pi^{12}}}$} 
} 
\\
\hline
$\, \begin{pmatrix} \left[ 1 , 1 \right] &\left[ 1 , 1 \right] &\left[  \right] \cr \left[  \right] &\left[  \right] &\left[ 1 , 1 \right] \cr \left[  \right] &\left[  \right] &\left[ 1 , 1 \right] \cr \ifx\endpmatrix\undefined}\else\end{pmatrix}\fi  $ 
& & & & & \\ 
& $ 2.4720e-28 $ 
& [$ 5.0914e-3 $]  
& $ 22 $  
& $ 22 $ 
& $ {{1}\over{2}} $  
\\
\multicolumn{6}{ c }{ \makecell{${{1}\over{92234958440625\,2^{{{51}\over{2}}}\,\pi^{12}}}$} 
} 
\\
\hline
$\, \begin{pmatrix} \left[ 1 , 1 \right] &\left[ 1 , 1 \right] \cr \left[ 1 , 1 \right] &\left[  \right] \cr \left[ 1 , 1 \right] &\left[  \right] \cr \ifx\endpmatrix\undefined}\else\end{pmatrix}\fi  $ 
& & & & & \\ 
& $ 1.8628e-27 $ 
& [$ 3.4879e-3 $]  
& $ 2 $  
& $ 2 $ 
& $ {{1}\over{2}} $  
\\
\multicolumn{6}{ c }{ \makecell{${{23}\over{281517134023125\,2^{{{51}\over{2}}}\,\pi^{12}}}$} 
} 
\\
\hline
$\, \begin{pmatrix} \left[ 1 , 1 \right] &\left[ 1 , 1 \right] &\left[  \right] &\left[  \right] \cr \left[  \right] &\left[  \right] &\left[ 1 , 1 \right] &\left[  \right] \cr \left[  \right] &\left[  \right] &\left[  \right] &\left[ 1 , 1 \right] \cr \ifx\endpmatrix\undefined}\else\end{pmatrix}\fi  $ 
& & & & & \\ 
& $ 1.0173e-30 $ 
& [$ 4.4000e-4 $]  
& $ 462 $  
& $ 231 $ 
& $ 1 $  
\\
\multicolumn{6}{ c }{ \makecell{${{1}\over{22413094901071875\,2^{{{51}\over{2}}}\,\pi^{12}}}$} 
} 
\\
\hline
\end{longtable}
%
%
%
%
%


\section{General results up to level $8$}
\label{sec:results}

After we compute all decays we can pass to the ``experimental
analysis''
whose results are reported earlier in section \ref{sec:intro_conclu}.

\subsection{How different are the total decay widths?}
\label{sec:Ratio}
\begin{figure}[H]
  \centering
    \begin{subfigure}[b]{1\textwidth}
    \centering    
    \includegraphics[width=0.8\textwidth]{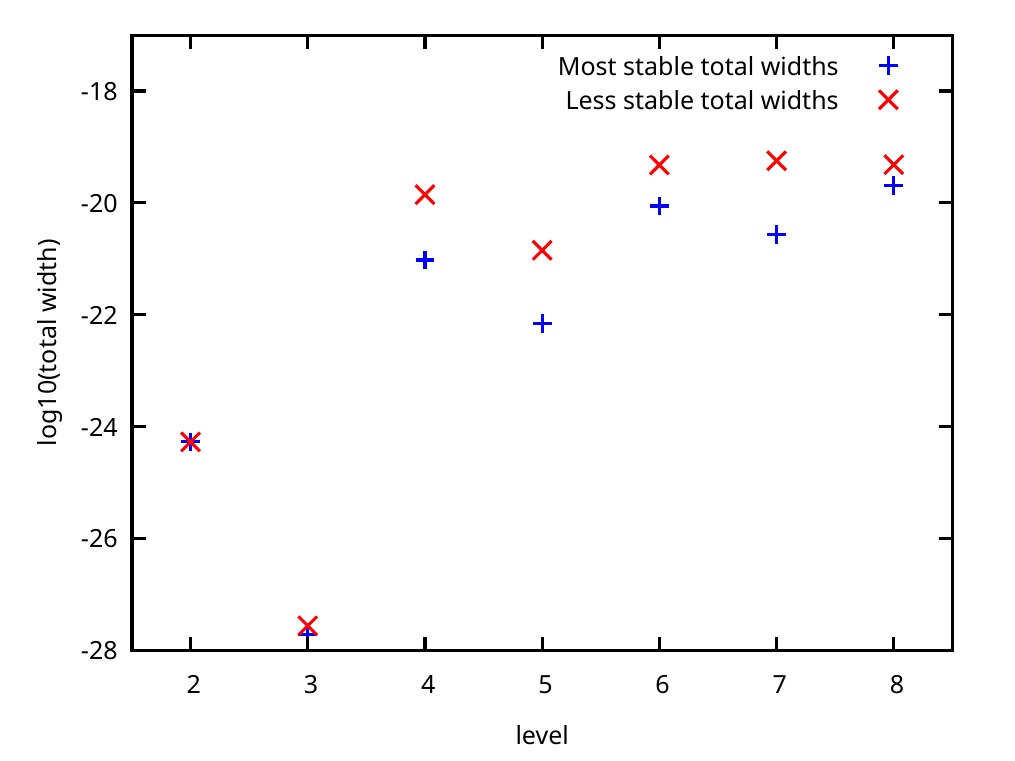}
    \end{subfigure}
\caption{%
We plot for any level $N\le8$ the $log_{10}$
of the total decay width divided by
$4 e^{-2 \phi_0}= g_o^2...$
for the most stable and less stable states.
}    
\label{fig:slowest_Fastest}
\end{figure}

The question we would like to explore is
whether the mixture of possible string states where each dof is
weighed in the same way evolves with time.
To answer this question we plot in figure \ref{fig:slowest_Fastest}
the most stable state total decay
width and the less stable state total decay width.
We see that the two values converge when $N$ is increasing.
We notice also that even and odd $N$ have a slightly different
behavior and range.
This is not unexpected since we discuss
in \ref{sec:main_decay_channels} and we have already noticed before
the fastest polarized decays are into 2 photons for even levels
or into 1 photon and 1 spin 2 for odd levels.

\begin{figure}[H]
  \centering
    \begin{subfigure}[b]{1\textwidth}
    \centering    
    \includegraphics[width=0.8\textwidth]{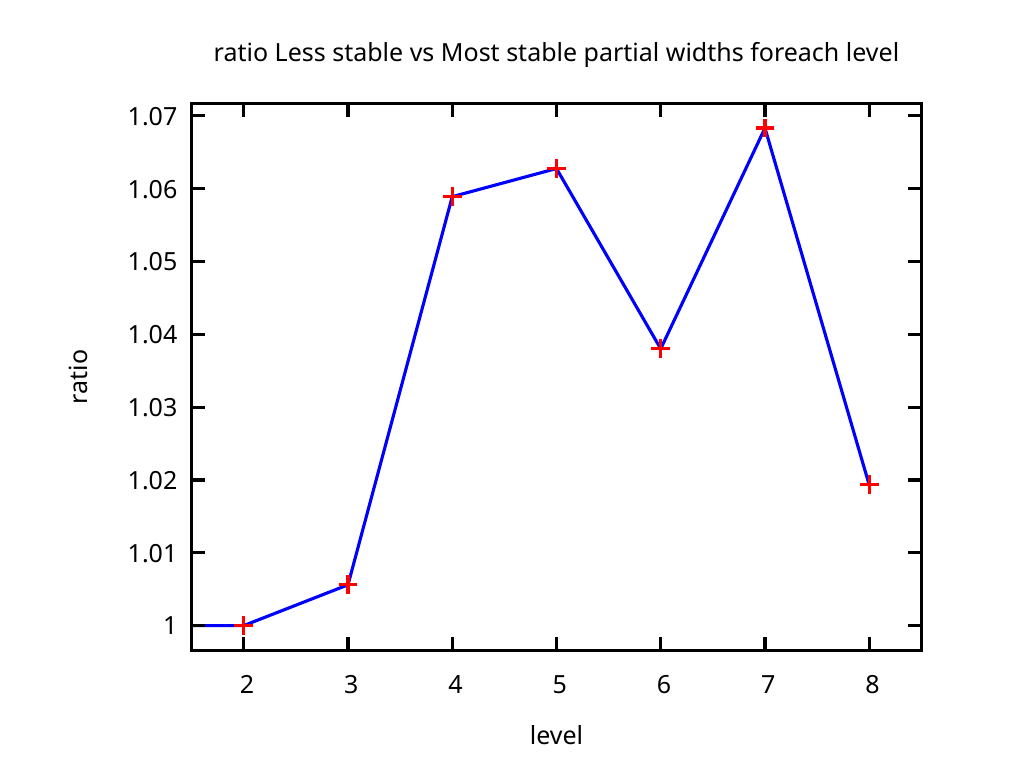}
    \end{subfigure}
\caption{%
We plot for any level the ratio between the total decay width for the
less stable state and
most stable one in order to try to identify any possible trend.
}    
\label{fig:slowest_Fastest_ratio}
\end{figure}

The same result can be obtained by considering their ratios in figure
\ref{fig:slowest_Fastest_ratio}.

\newpage

\subsection{ Which percentage of the total width is explained by the
fastest decay (biggest partial width)?}

\begin{figure}[H]
\centering
\begin{subfigure}[b]{0.48\textwidth}
\centering    
\includegraphics[width=\textwidth]{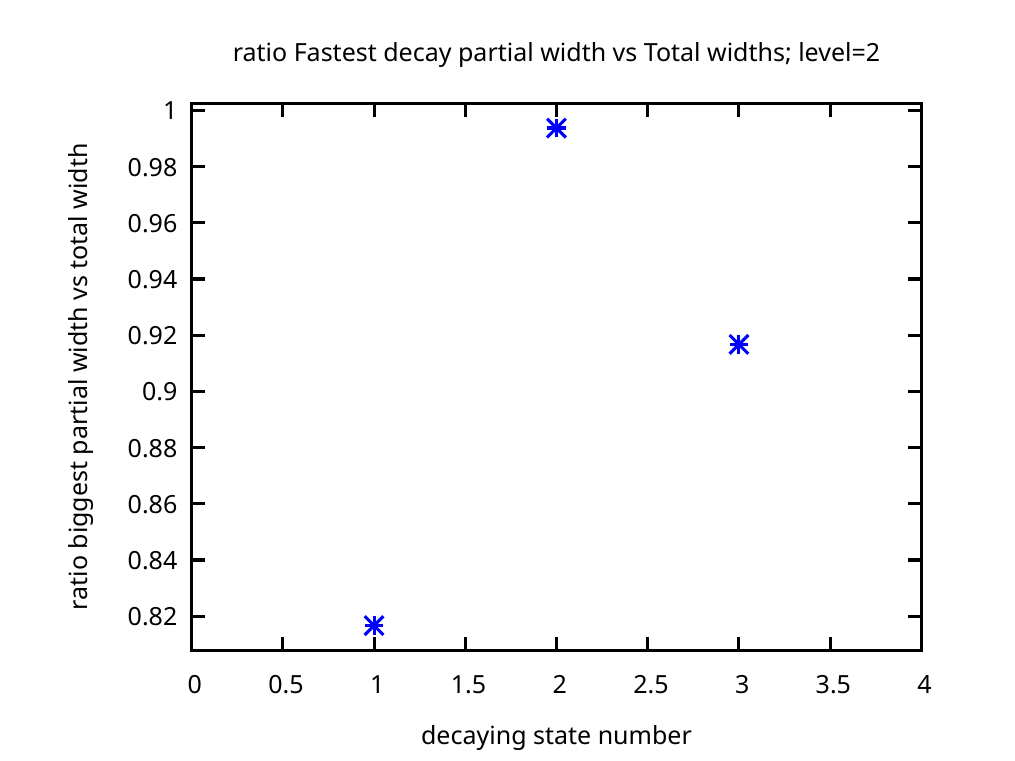}
\end{subfigure}
\begin{subfigure}[b]{0.48\textwidth}
\centering    
\includegraphics[width=\textwidth]{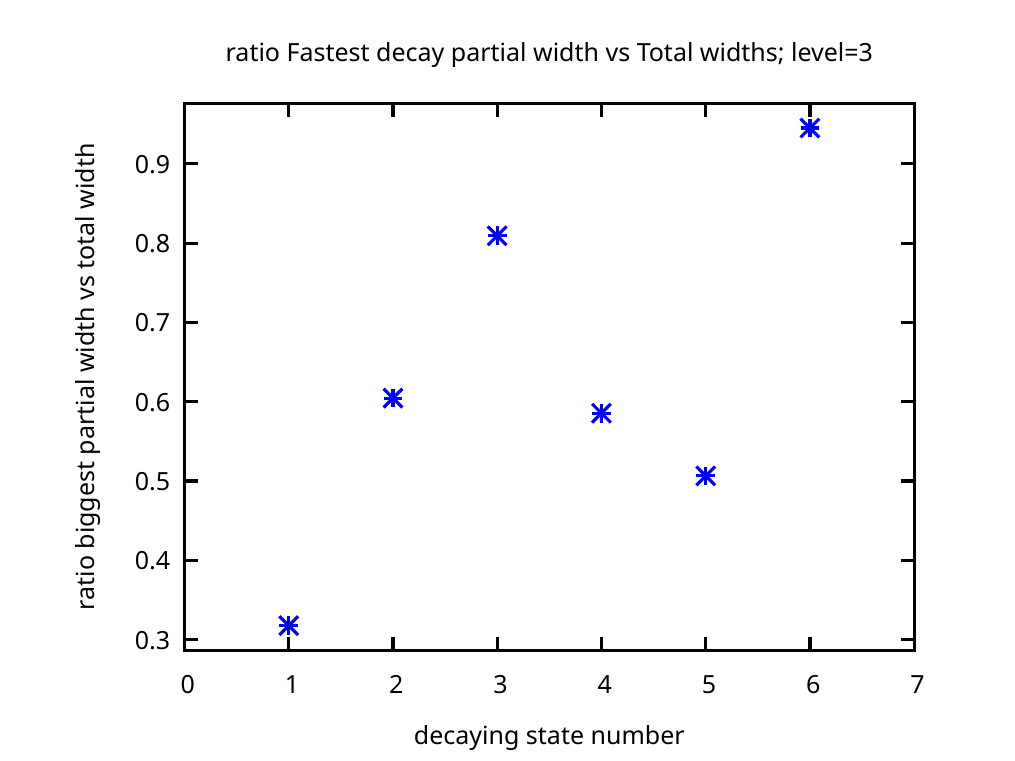}
\end{subfigure}
\begin{subfigure}[b]{0.48\textwidth}
\centering    
\includegraphics[width=\textwidth]{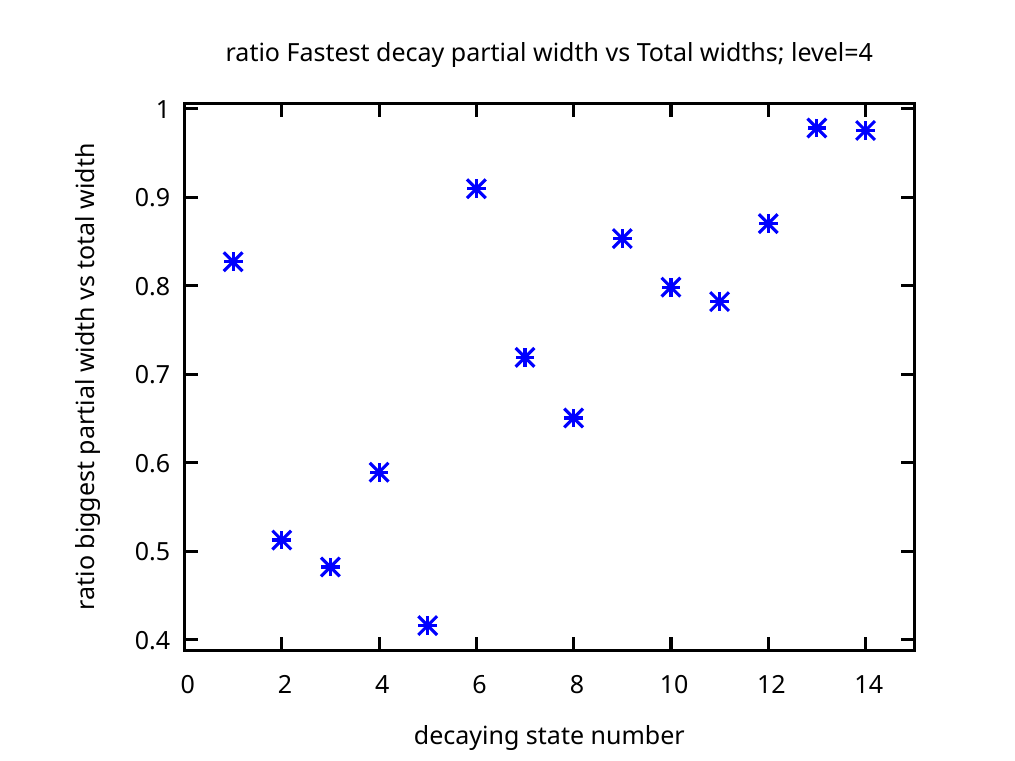}
\end{subfigure}
\begin{subfigure}[b]{0.48\textwidth}
\centering    
\includegraphics[width=\textwidth]{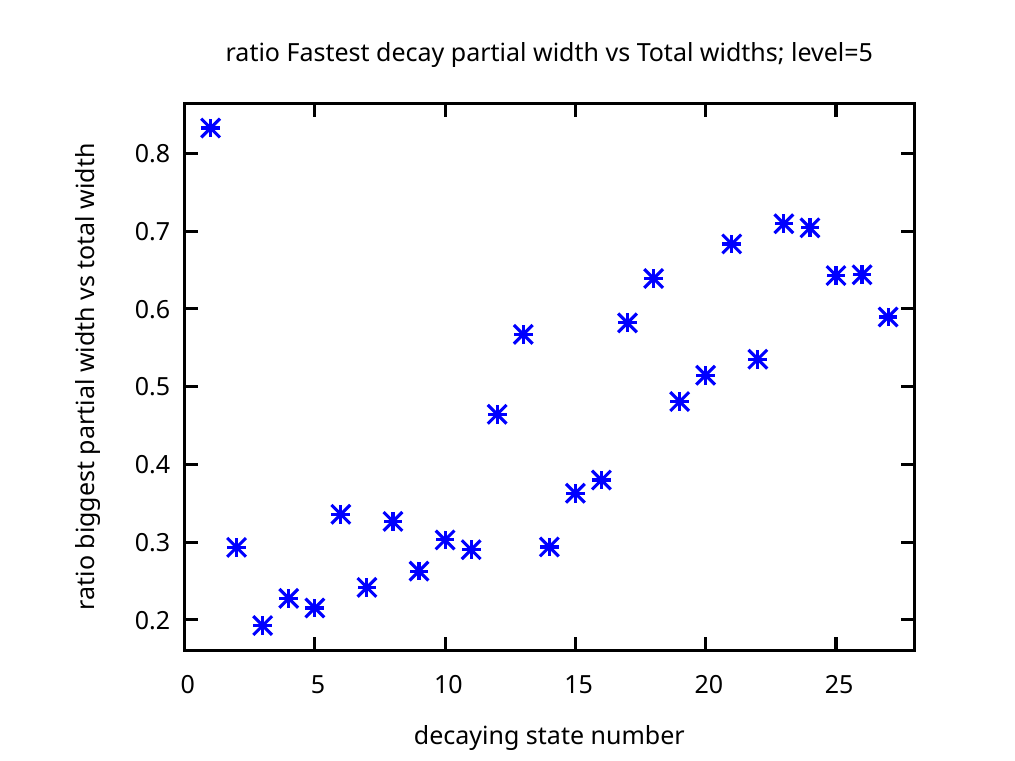}
\end{subfigure}
\begin{subfigure}[b]{0.48\textwidth}
\centering    
\includegraphics[width=\textwidth]{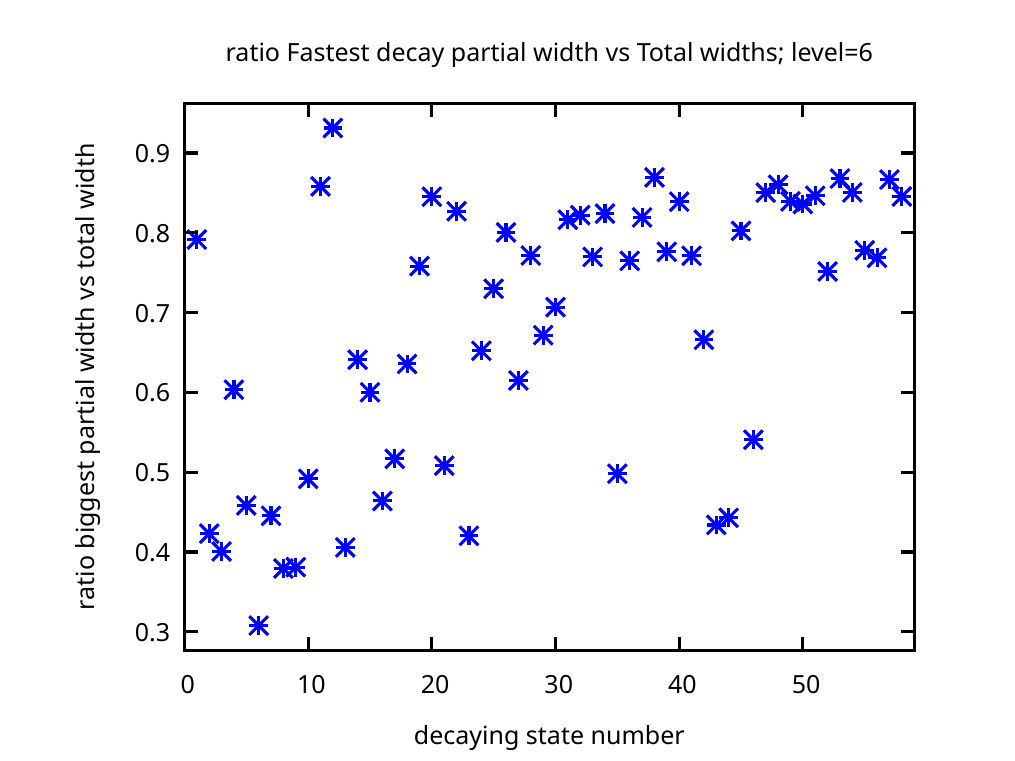}
\end{subfigure}
\begin{subfigure}[b]{0.48\textwidth}
\centering    
\includegraphics[width=\textwidth]{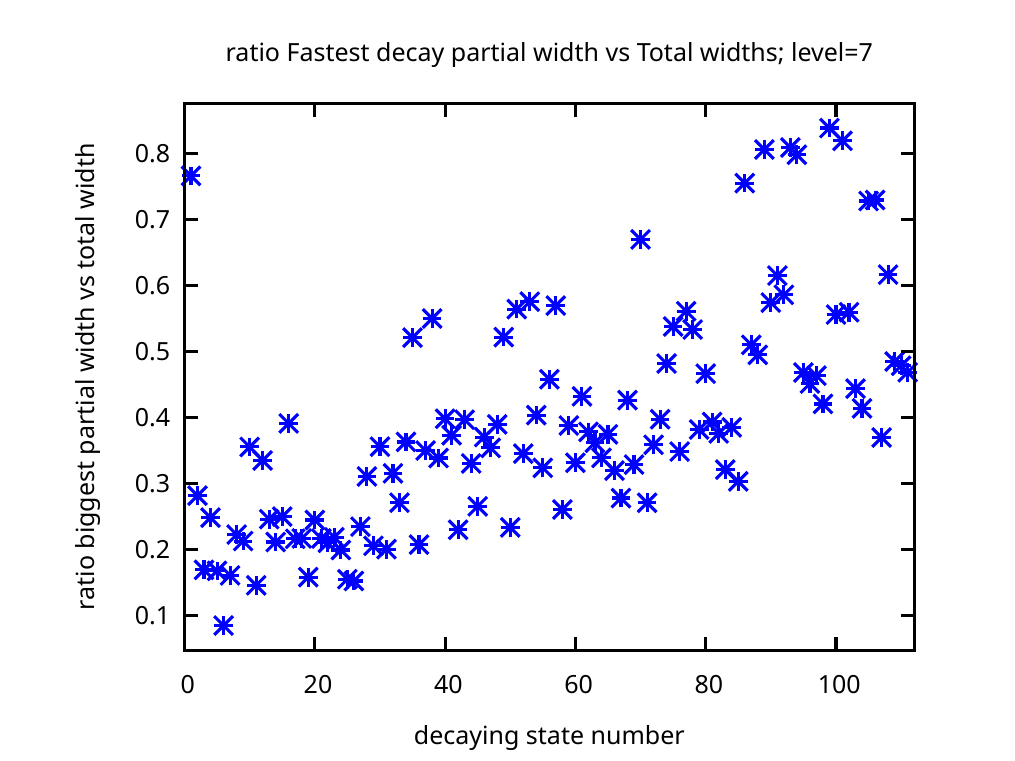}
\end{subfigure}
\begin{subfigure}[b]{0.6\textwidth}
\centering    
\includegraphics[width=\textwidth]{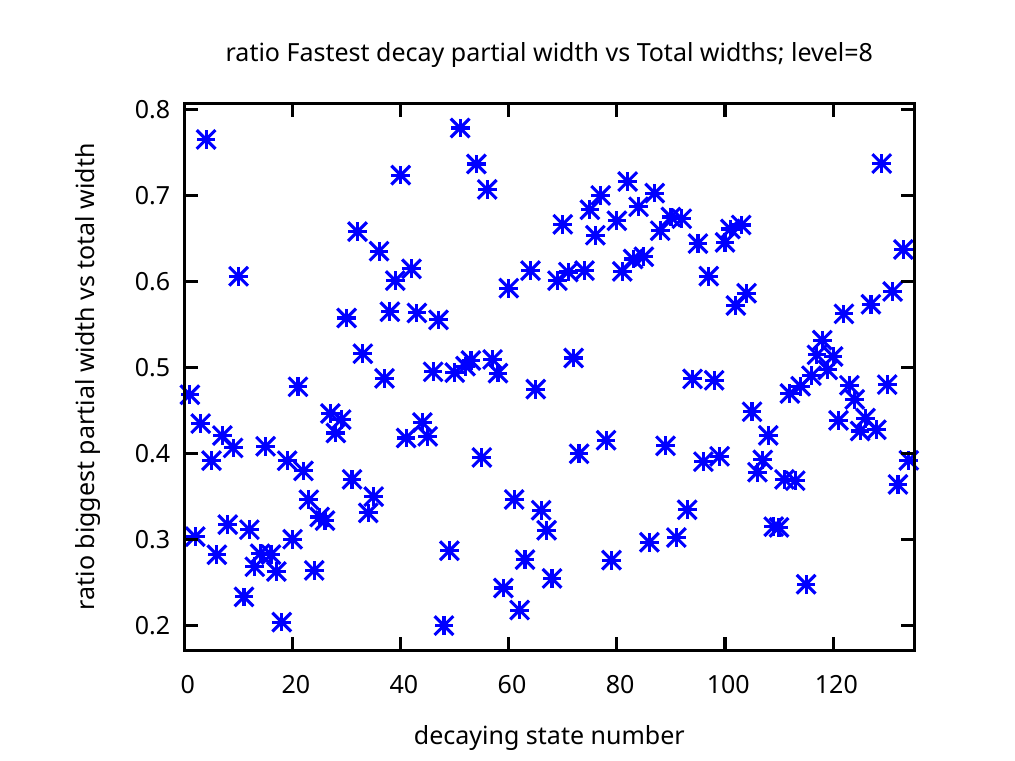}
\end{subfigure}
\caption{%
We plot for any level and any state
the ratio between 
fastest decay partial width (biggest partial width) and
the state total width  in order to try to identify any possible trend.
Notice that the fastest decay is not the sum of all the decays having
the same final particles as the fastest decay but is the sum of all
decays which are equivalent up to permutations of transverse space
directions which leave the decaying state invariant
}    
\label{fig:ratio_fastest_partial_total}
\end{figure}

\begin{figure}[H]
\centering
\begin{subfigure}[b]{0.48\textwidth}
\centering    
\includegraphics[width=\textwidth]{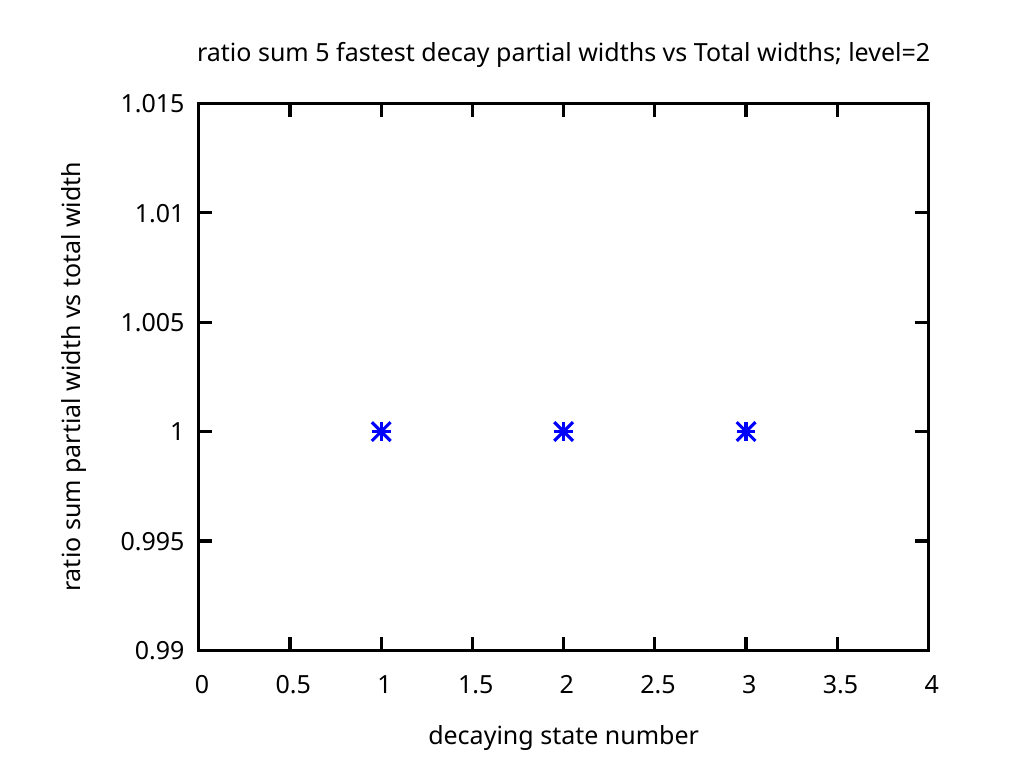}
\end{subfigure}
\begin{subfigure}[b]{0.48\textwidth}
\centering    
\includegraphics[width=\textwidth]{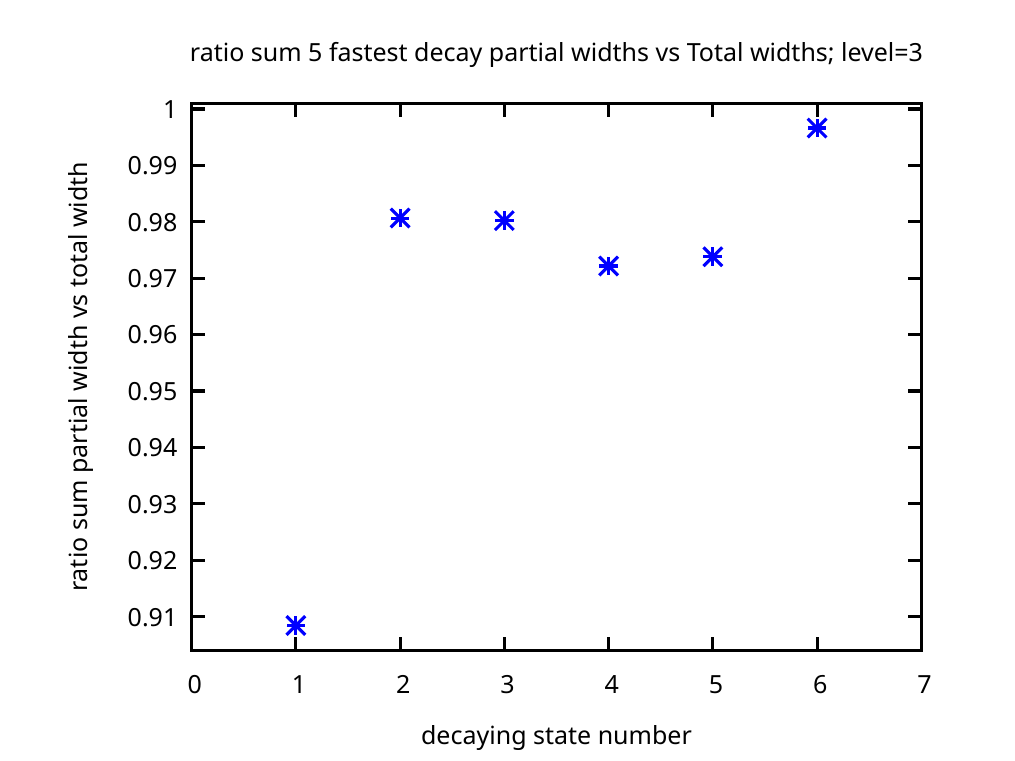}
\end{subfigure}
\begin{subfigure}[b]{0.48\textwidth}
\centering    
\includegraphics[width=\textwidth]{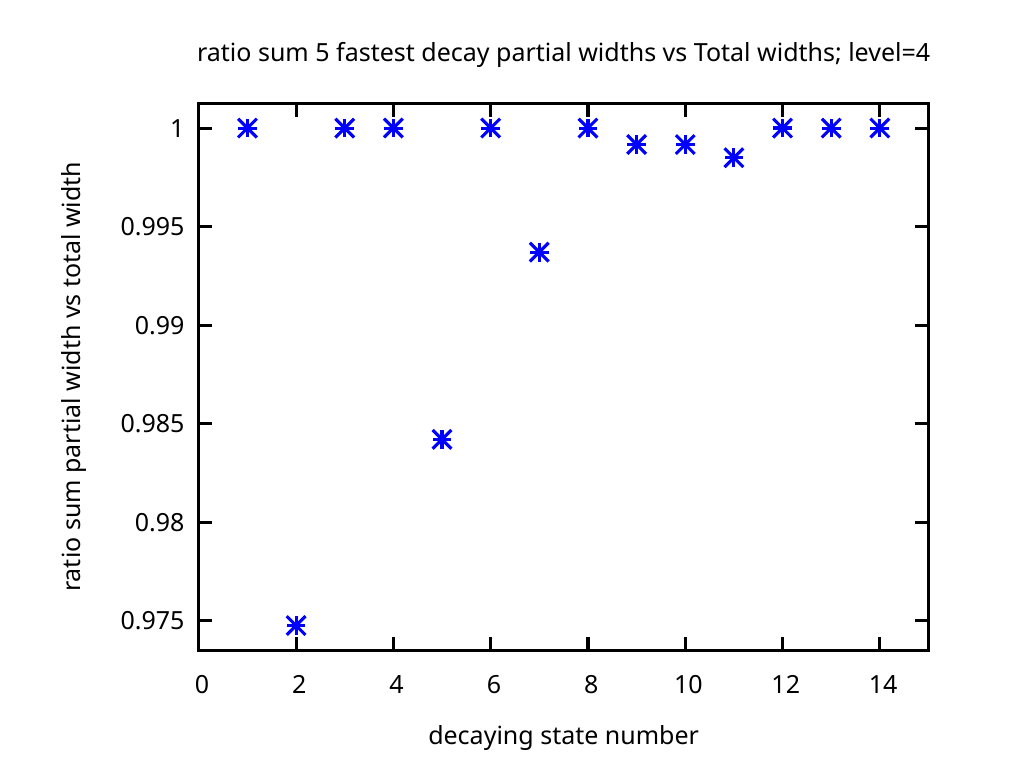}
\end{subfigure}
\begin{subfigure}[b]{0.48\textwidth}
\centering    
\includegraphics[width=\textwidth]{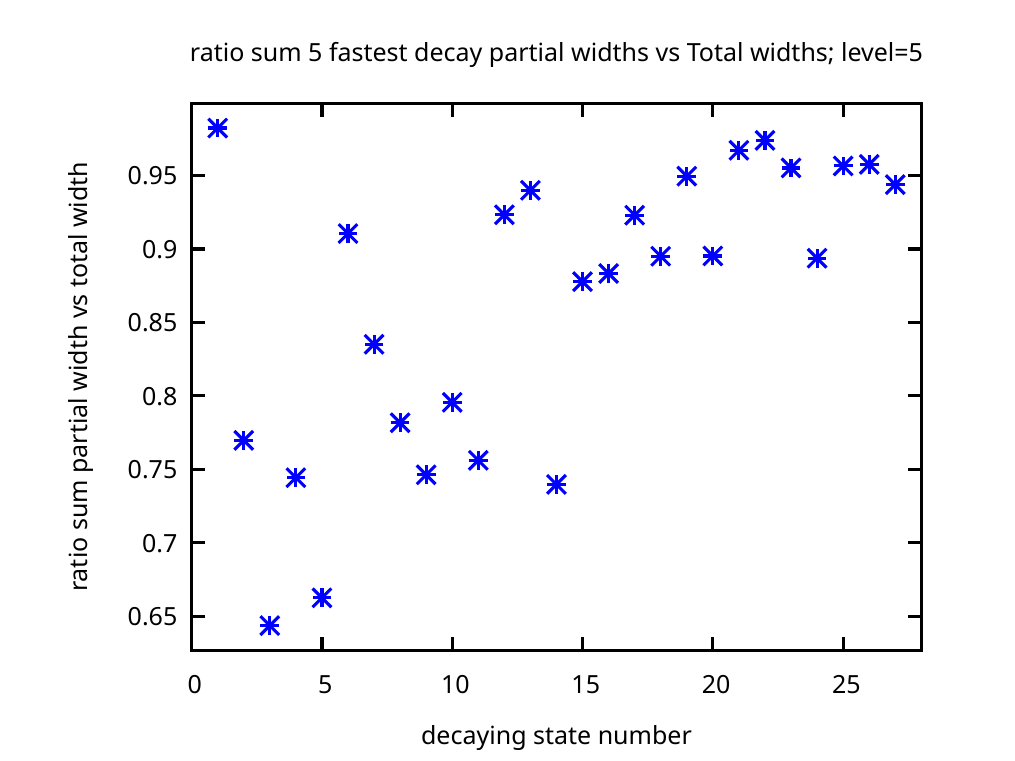}
\end{subfigure}
\begin{subfigure}[b]{0.48\textwidth}
\centering    
\includegraphics[width=\textwidth]{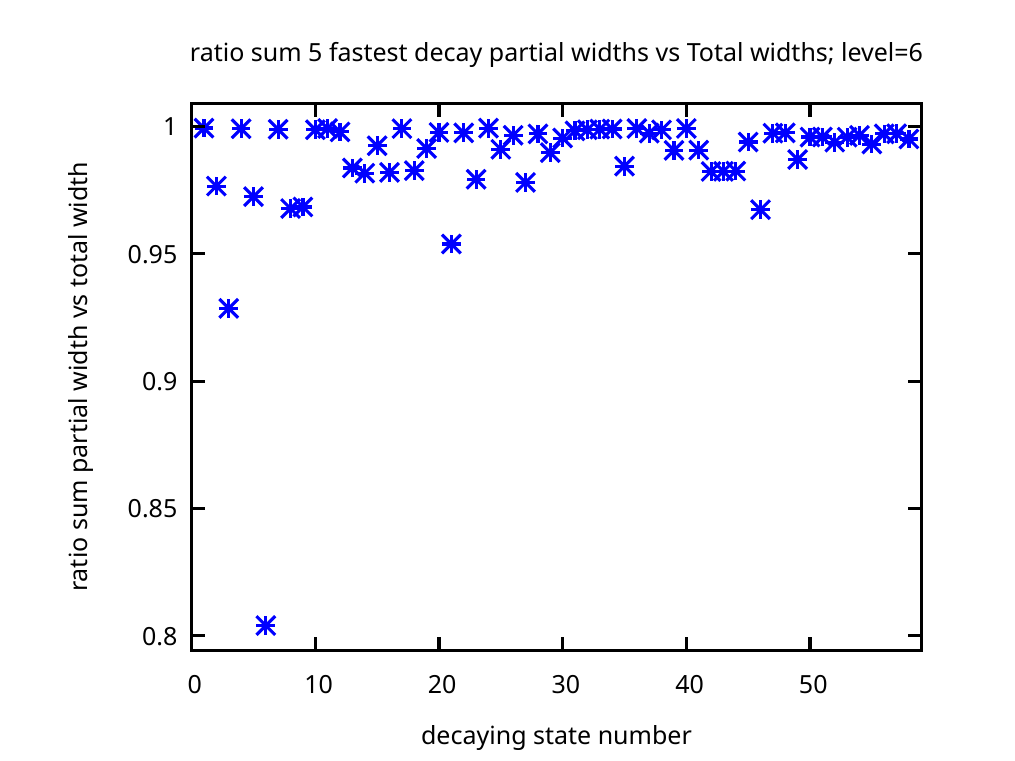}
\end{subfigure}
\begin{subfigure}[b]{0.48\textwidth}
\centering    
\includegraphics[width=\textwidth]{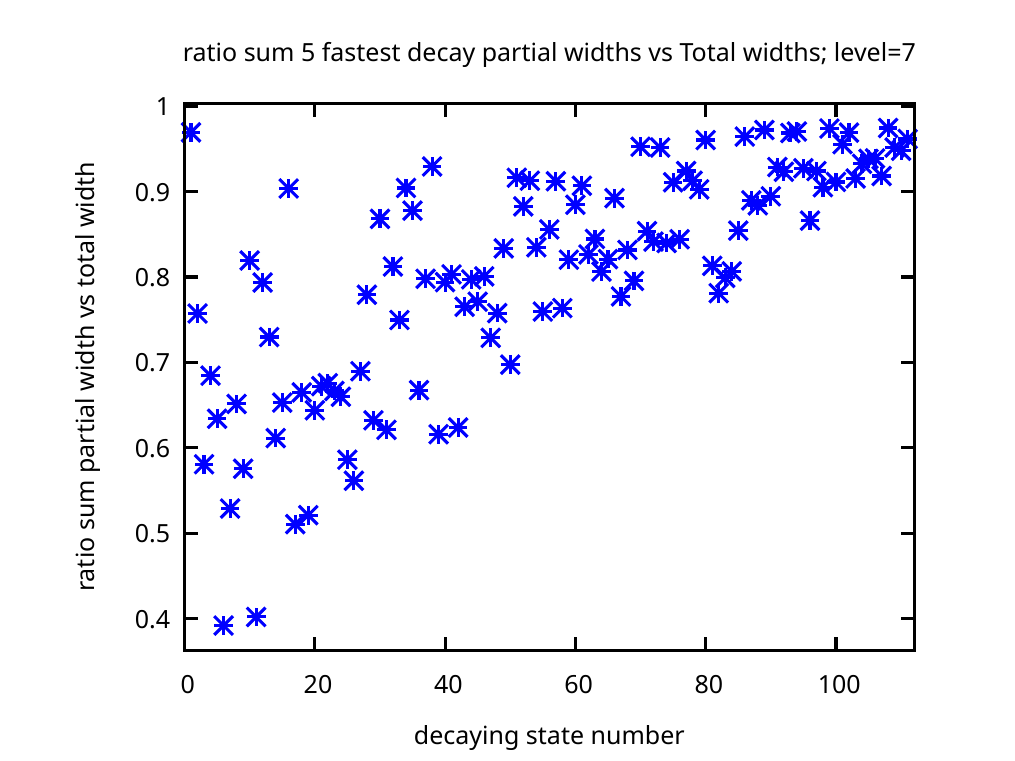}
\end{subfigure}
\begin{subfigure}[b]{0.6\textwidth}
\centering    
\includegraphics[width=\textwidth]{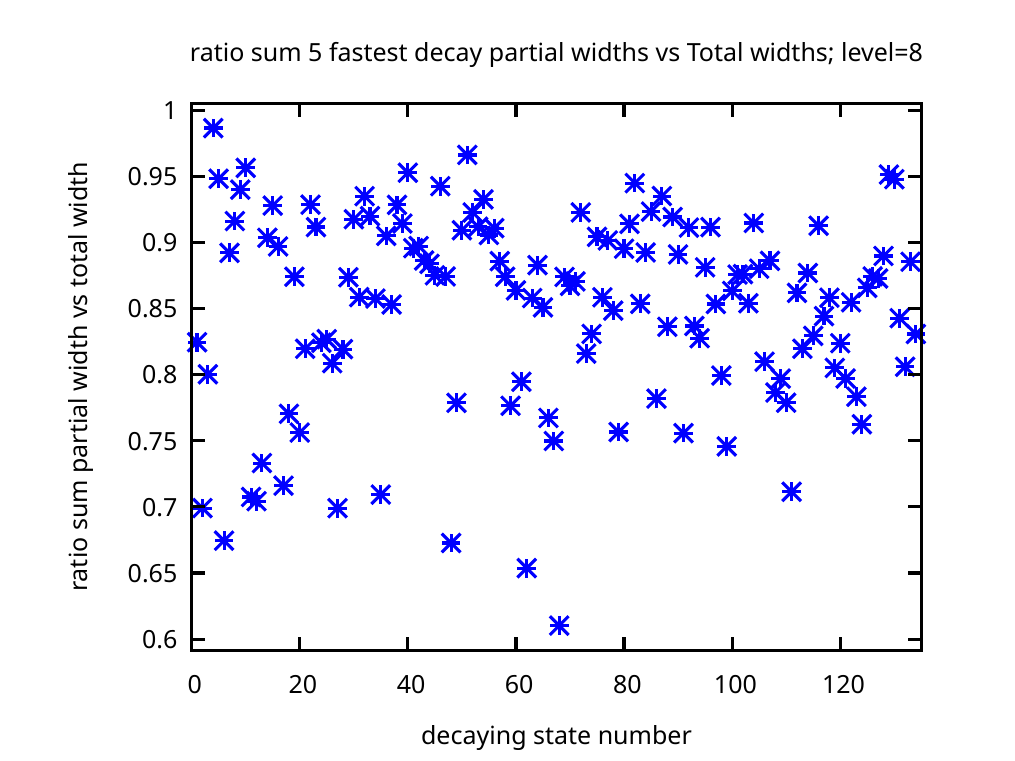}
\end{subfigure}
\caption{%
We plot for any level and any state
the ratio between the sum of the 5
fastest decay partial widths (biggest partial widths) and
the state total width  in order to try to identify any possible trend.
Notice that the fastest decay is not the sum of all the decays having
the same final particles as the fastest decay but is the sum of all
decays which are equivalent up to permutations of transverse space
directions which leave the decaying state invariant
}    
\label{fig:ratio_5_fastest_partials_total}
\end{figure}

Another important question is how important are the fastest partial
decays since it there is a pattern we can easily estimate the total widths.
To this purpose we plot in
figure \ref{fig:ratio_fastest_partial_total}
for each state the ratio of the fastest
partial decay width to the total decay width.
We observe that for any state and any level the fastest decay channel
accounts for at least 

\subsection{Most stable and less stable states}
\label{sec:most_less_stable}

Using the same experimental approach as before we look at the most
stable states in the tables \ref{summaryL2}-\ref{summaryL8}.
It is somewhat clear that the most stable states are often of the form
$\prod_{i=2} (A_{-1}^i)^{n_i} | k \rangle$
or 
$\prod_{i=2} (A_{-1}^i)^{n_i} (A_{-2}^i)^{m_i} | k \rangle$.
States with oscillators of higher level are generically less stable
but the difference is not enormous as noticed before in section \ref{sec:Ratio}.
These  states are expected to be mainly  leading Regge trajectory
even if only the states
$\prod_{i=2}^N A_{-1}^i | k \rangle$
with $N\le 24$ are really pure leading Regge trajectory since
they have no anomalous Lorentz transformations \cite{Pesando:2024lqa}
and these states do no exist beyond level $N=24$.

Less stable states are no so easy to characterize but they usually
involve higher excitation operators.

\subsection{Main partial decay channels}
\label{sec:main_decay_channels}
Since this is an ``experimental'' paper we consider the data in tables
\ref{N=2_state_1},\ref{N=2_state_2},\ref{N=2_state_3}
and in appendixes \ref{app:N=3}-\ref{app:N=8}.
These data are partial since they do not show all possible states.
All data are however in the supplementary material.
The conclusion is clear.
For $N\le7$ the fastest decay channels are in two photons
for $N$ even
(see
tables \ref{N=2_state_1},\ref{N=2_state_2},\ref{N=2_state_3},\ref{app:N=4},\ref{app:N=6})
and one photon and one spin $2$
for $N$ odd (see tables \ref{app:N=3},\ref{app:N=5},\ref{app:N=7})
while
for $N=8$ the fastest decays involve at least but not always one
photon (see tables \ref{app:N=8}).

The probable explanation in $D=26$ is that the phase space volume is
bigger in these configurations.
It could be the right explanation if all ``coupling constants'' were
of the same order.



\printbibliography


\appendix

\section{Notations}
Index are
\begin{equation}
  \mu,\nu=0,1,\dots d-1=25,~~~~
  I=1,2,\dots d-1=25,~~~~
  i=2,3,\dots d-1=25
\end{equation}

We denote a state
\begin{align}
  &\prod_{i=2}^{25}\, \prod_{n=1}^{L_i}\, \left( \alpha^{i}_{-n} \right)^{N_{i\,n}}
  =
  \left( \alpha^{2}_{-L_2} \right)^{N_{2\,L_2}}\,  \left( \alpha^{2}_{-L_2+1} \right)^{N_{2\,L_2-1}}\,
  \dots \left( \alpha^{2}_{-1} \right)^{N_{2\,1}}\,
  \left( \alpha^{3}_{-L_3} \right)^{N_{3\,L_1}}\,  \dots  \left( \alpha^{2}_{-1} \right)^{N_{2\,1}}\,
  \dots
    \left( \alpha^{25}_{-L_{25}} \right)^{N_{25\,L_{25}}}\,  \dots  \left( \alpha^{25}_{-1} \right)^{N_{25\,1}}
    ,
\end{align}
as a list
\begin{align}
\Rightarrow&
  [
  [L_2, N_{2\, L_2},\, L_2-1, N_{2\, L_2-1},\, \dots 2, N_{2\, 2},\,
  1, N_{2\, 1}],\,
  [L_3, N_{3\, L_3},\,\dots 1, N_{3\, 1}],\,
  \dots
  [L_{25}, N_{25\, L_{25}},\,\dots 1, N_{25\, 1}],\,
]
.
\end{align}

The \lc coordinates are
\begin{equation}
  x^\pm = \frac{1}{\sqrt{2}}( x^0 \pm x^1),~~~~
  k_\pm = \frac{1}{\sqrt{2}}( k_0 \pm k_1)
  .
\end{equation}

\section{Kinematics}
\label{app:kinematics}
In the center of mass we consider a decay described as level decay as
$N_1 \rightarrow N_2 + N_3$
then the simplest kinematics is when the decay happens in directions
$x^0$ and $x^1$.
In this case there is only one non vanishing momentum component which
we denote $  p_{1\, \sN 2 *} = p^1_{\sN 2 *} = p_*$
in which case we have to solve
\begin{equation}
m_1 = \sqrt{p^2 + m_2^2} + \sqrt{p^2 +m_3^2}
,
\end{equation}
with $\ap m_r^2= N_r -1$.
The explicit solution reads
\begin{equation}
  p^1_{\sN 2 *}
  =
  p_*
  =
  \sqrt{
    \frac{N_1 -1 }{4 \ap}
    \left[
      1 + \left( \frac{N_2-N_3}{N_1-1} \right)^2
      -2 \left( \frac{N_2+N_3-2}{N_1-1} \right)
    \right]
  }
  ,
  \label{app:eq:p*}
\end{equation}
since we consider massive states $N_1\ge 2$ and the expression is well
defined.

We have obviously to require that $p^1_{\sN 2 *} $ is real.
If we do not consider the possibility of a decay in one or more
tachyons this simply amounts to $N_1 \ge N_2 +N_3 -1$.

If we consider a tachyon in equal time quantization we have in the
center of mass of the decaying particle
\begin{equation}
m_1 = \sqrt{p^2 + m_2^2} + \sqrt{p^2 - |m_T|^2}
,
\end{equation}
which implies $m_1\ge \sqrt{|m_T|^2 + m_2^2}$ since $p\ge |m_T|$
and $\vec p$ may have all possible directions.

However on \lc what matters is $\sqrt{2} k_+= E+ p_1$ which must be
positive hence for the tachyon not all $SO(25)$ rotations are allowed
since we must have
$p \cos\theta + \sqrt{p^2 - |m_T|^2}>0$,
i.e.
$\cos\theta > -\sqrt{1 - \frac{|m_T|^2}{p^2}}>0$.
This is also true for the decay into two tachyons!

On the other side since on \lc what matters is $\sqrt{2} k_+= E+ p_1$ which
must be positive we can write
\begin{equation}
m_1 = \sqrt{p^2 + m_2^2} - \sqrt{p^2 - |m_T|^2}
,
\end{equation}
since for $p>|m_T|$ we get a positive $\sqrt{2} k_+= p - \sqrt{p^2 -
|m_T|^2}$.

This equation can be solved when $m_1^2 < m_2^2 + |m_T|^2$ as
\begin{equation}
p^2 = |m_T|^2 + \left( \frac{ m_2^2 + |m_T|^2 -m_1^1}{2 m_1} \right)^2
.
\end{equation}

In particular it can be solved for $m_2=m_1$, i.e. a particle can
decay in itself plus a tachyon as
\begin{equation}
p= |m_T| \sqrt{ 1 + \frac{|m_T|^2}{4 m_1^2} },
~~~~~
E_2= m_1 + \frac{|m_T|^2}{2 m_1},
~~~~
E_T=- \frac{|m_T|^2}{2 m_1}
.
\end{equation}  
\section{Decay width}
\label{app:decay_width}

We would now derive the decay width starting from the color ordered
decay amplitude $A_{1 2 3}(k_{\sN 1}, k_{\sN 2}, k_{\sN 3})$
in the massive particle center of mass and arbitrary decay particles
directions, i.e. we assume the kinematics
\begin{align}
  k_{\sN 1 \mu} =& \left( M_{\sN 1}, 0, \vec 0 \right),
  ~~~~
  k_{\sN 2 \mu} =& \left( -E_{\sN 2}, -p_{\sN 2 2}, -\vec p_{\sN 2} \right),
  ~~~~
  k_{\sN 3 \mu} =& \left( -E_{\sN 3}, +p_{\sN 2 2}, +\vec p_{\sN 2}
  \right)
  ,
  \label{app:setup_cm_momenta}
\end{align}
with $E_{\sN r}>0$  and $\mu=0,1, i$ and $i=2, \dots d-1=25$.

This means that we can only consider the usual decays since surges
have one negative $E$ and the decay width must be computed using \lc
formalism.
In the usual case \lc and equal time formalisms give the same answer
as it should.

We start from the S matrix definition as
$S = \one + i T$ which in our case reads
\begin{equation}
  S_{1\rightarrow 2 + 3} = i   T_{1\rightarrow 2 + 3}
  = i\, (2\pi)^d\, \delta^d(\sum_r k_{\sN r} )\,   \cT_{1\rightarrow 2+3}
  .
\end{equation}

In the Abelian open string we can connect the $T_{1\rightarrow 2+3}$ matrix element
with the string amplitude as
\begin{equation}
  T_{1\rightarrow 2 + 3}
  =
  2\,
   g_o^3\, \cC_{D_2}\,
  A_{1 2 3}(k_{\sN 1}, k_{\sN 2}, k_{\sN 3})
  =
  (2\pi)^d\, \delta^d(\sum_{r=1}^3 k_{\sN r} )\,
  2\,
   g_o^3\, \cC_{D_2}\,
   \cA_{1 2 3}(k_{\sN 1}, k_{\sN 2}, k_{\sN 3})
   ,
\end{equation}
when $N_1+N_2+N_3\equiv 0 ~\mod 2$
and
where the factor $2$ in the last equation is due to
the sum over the non cyclically equivalent
color ordered amplitudes
and
$A_{1 3 2} = (-)^{N_1+N_2+N_3} A_{1 2 3}$.

In QFT we normalize the on shell states as
\begin{equation}
  \langle p_I | q_I \rangle =
  (2\pi)^{d-1}\, 2 E(p_I)\, \delta^{d-1}(p_I - q_I)
 , 
\end{equation}
and $S$ and $T$ are adimensional.
It then follows that
the dimension of the matrix element is
\begin{equation}
[T_{1\dots M \rightarrow 1'\dots N'} ]
=L^{-d} ( L^{\oh(d-2)} )^{M+N}
.
\end{equation}
At the same time all string states have adimensional polarizations and the
momentum stripped amplitude
$\cA_{1\dots M+N}(k_{\sN r})$ is adimensional too.
This means that the dimension of
$A_{1\dots M+N}(k_{\sN r})$ ($r,s, \dots =1,\dots M+N$)
is the same of the momentum conservation delta, i.e. $L^d$.
Since in general
\begin{equation}
  A_{1\dots M+N}(k_{\sN r})
  =
  (2\pi)^d\, \delta^d(\sum_{r=1}^{M+N} k_{\sN r} )\,
  g_o^{M+N}\, \cC_{D_2}\,
  \sum_{\sigma\in S_{M+N-1}}
   \cA_{1 \sigma(2 \dots N+M)}(k_{\sN r})
  ,
\end{equation}
with $S_{M+N-1}$ the permutation group of $M+N-1$ elements.
It then follows that
\begin{equation}
  [g_o] = L^{\oh(d-2)},~~~~
  [C_{D_2}] =L^{-d}.
\end{equation}
From unitarity we get
\begin{equation}
  C_{D_2}
  =
  \frac{1}{\ap g_o^2}
  =
  \frac{2}{(\sdap)^d }
  e^{-2 \phi_0}
  ,
\end{equation}
where $\phi_0$ is the vacuum expectation value of the dilaton.

The decay width we are interested in is
\begin{align}
  V T\, \hat \Gamma_{1\rightarrow 2+3}
  =&
  \int
  \prod_{r=2}^3\,\frac{d^d k_{\sN r}}{(2 \pi)^{d-1}}
  \delta(k_{\sN  r}^2+m_{\sN r}^2)
  \theta(k_{\sN r\, 0})\,
  | T_{1\rightarrow 2 + 3} |^2
  \nonumber\\
  =&
  V T\,
  4 \frac{ (g_o^3\, \cC_{D_2})^2 }{(2 \pi)^{d-2}}
  \int   \prod_{r=2}^3\,\frac{d^{d-1} k_{\sN r}^I}{2 E_{\sN r}}
  |\cA_{1 2 3}(k_{\sN 1}, k_{\sN 2}, k_{\sN 3})|^2
,
\label{app:eq:decay_start_point}
\end{align}
in the center of mass after solving for the momenta delta function we
can write
\begin{align}
  \hat \Gamma_{1\rightarrow 2+3}
  =&
  4 \frac{(g_o^3\, \cC_{D_2})^2}{ (2 \pi)^{d-2}}
  \frac{1}{ 4 E_{\sN 2}(p_*)\, E_{\sN 3}(p_*)}
  \frac{E_{\sN 2}(p_*)\, E_{\sN 3}(p_*)}{m_1\, p_*}
  p_*^{d-2}
  \int   d^{d-2}\Omega
  |\cA_{1 2 3}|^2\Big|_{k_{\sN 2 I}=-k_{\sN 3 I}=p_* n_{I}}
  \nonumber\\
    =&
  \frac{(g_o^3\, \cC_{D_2})^2}{ (2 \pi)^{d-2}}
  \frac{  p_*^{d-3}}{m_1}
  \int   d^{d-2}\Omega
  |\cA_{1 2 3}|^2\Big|_{k_{\sN 2 I}=-k_{\sN 3 I}=p_* n_{I}}
,
\label{app:eq:decay_gamma_hat}
 \end{align}
where $p_*$ is the modulus of the outgoing particles given in
eq. (\eqref{app:eq:p*})
and
$n_I$ is the $d-1=25$ versor whose components are as usual
\begin{align}
  n_1 &= \cos \theta_1,
  \nonumber\\
  n_2 &= \sin \theta_1 \cos \theta_2,
  \nonumber\\
  n_{d-2} &= \sin \theta_1 \dots \sin \theta_{d-3}\, \cos \theta_{d-1},
  \nonumber\\
  n_{d-1} &= \sin \theta_1 \dots \sin \theta_{d-3}\, \sin\theta_{d-1}
  .
\end{align}
The dimension of $\hat \Gamma_{1\rightarrow 2+3}$ is $L^{-2}$.
Finally we can compute the decay width $\Gamma_{1\rightarrow 2+3}$ ad
\begin{equation}
  \hat \Gamma_{1\rightarrow 2+3}
  =
  2 m_1 \Gamma_{1\rightarrow 2+3}
  .
\end{equation}

Finally we can express the adimensional decay width as
\begin{align}
\sdap\,\Gamma_{1\rightarrow 2+3}
=&
4 e^{-2 \phi_0}
  \frac{1}{ (2 \pi)^{d-2}}
  \frac{  ( \sdap p_*)^{d-3}}{(\sdap m_1)^2}
  \int   d^{d-2}\Omega
  |\cA_{1 2 3}|^2\Big|_{\sdap k_{\sN 2 I}=- \sdap k_{\sN 3 I}=\sdap p_* n_{I}}
.
\label{app:eq:adimensional_decay_width}
\end{align}     
In the main text the quantities are 
\begin{equation}
\label{eq:adim_Gamma}
\sdap\,\Gamma_{1\rightarrow 2+3}
/ \left( 4 e^{-2 \phi_0} \right).
\end{equation}

Let us consider the mixing.
We can start from the analogous expression
of \eqref{app:eq:decay_start_point} 
\begin{align}
  \hat \Gamma_{1\rightarrow 2+3 \rightarrow 1'}
  =&
  4 \frac{(g_o^3\, \cC_{D_2})^2}{ (2 \pi)^{d-2}}
  \frac{1}{ 4 E_{\sN 2}(p_*)\, E_{\sN 3}(p_*)}
  \frac{E_{\sN 2}(p_*)\, E_{\sN 3}(p_*)}{m_1\, p_*}
  p_*^{d-2}
  \nonumber\\
  &
  \int   d^{d-2}\Omega
  \cA_{1 2 3}\Big|_{k_{\sN 2 I}=-k_{\sN 3 I}=p_* n_{I}}
  \left( \cA_{1' 2 3}\Big|_{k_{\sN 2 I}=-k_{\sN 3 I}=p_* n_{I}} \right)^*
\end{align}
and obtain the essentially the same final result (for the $U(1)$ case
which generates the factor $4$)
\begin{align}
\sdap\,\Gamma_{1\rightarrow 2+3 \rightarrow 1'}
=&
  \frac{4 e^{-2 \phi_0}}{ (2 \pi)^{d-2}}
  \frac{  ( \sdap p_*)^{d-3}}{(\sdap m_1)^2}
  \int   d^{d-2}\Omega
  \cA_{1 2 3}\Big|_{k_{\sN 2 I}=-k_{\sN 3 I}=p_* n_{I}}
  \left( \cA_{1' 2 3}\Big|_{k_{\sN 2 I}=-k_{\sN 3 I}=p_* n_{I}} \right)^*
.
\label{app:eq:adimensional_mix_width}
\end{align}     

\section{Details on the logic used in the program}
\label{app:Further_details}

\subsection{Minimal set of decays}
To determine the minimal set of decays we have
given a canonical and unique representation of the decay.
This is obtained by ordering the directional form of the decay.
The directional form describes the decay from
the space directions point of view.
Let us start with a few examples on how to map the decay to its
directional form and the compute their canonical form.

\begin{align}
[ [22, 1, 1, 3], [3,1], [], [] ]
\rightarrow
&
[ [1, 2], [], [], [1,2] ]
   \oplus
[ [], [], [2,1], [] ]
\nonumber\\
\Rightarrow&
  \dirst{22, 1, 1, 3}{1,2}{}
  \dirst{3,1}{}{}
  \dirst{}{}{1,2}
  \dirst{}{1,2}{}
,
\end{align}

The canonical form  of the decay is then
\begin{align}
  \dirst{22, 1, 1, 3}{1,2}{}
  \dirst{3,1}{}{}
  \dirst{}{1,2}{}
  \dirst{}{}{1,2}
.
\end{align}

In the case of equal final particles  the canonical directional form
takes in account the swapping of the two final particles,
for example
\begin{equation}
  \dirst{[2,1]}{[1,1]}{}
  \dirst{[1,1]}{}{1,1}
  ,~~~~
  \dirst{[2,1]}{}{[1,1]}
  \dirst{[1,1]}{1,1}{}
  ,
\end{equation}
are not equivalent if in the canonical ordering the swap of the two
final particles is not considered.

\subsection{Degeneracy}
Since we have computed only the independent decays we need to
compute the degeneracy of the independent decays.

As discussed above to determine the minimal set of decays we have
given a canonical and unique  representation of the decay.

There are essentially two different cases:
\begin{itemize}
\item
  the final decay states are different particles;
\item
  the final decay states are the same particle.
\end{itemize}

In the case of different final particles we count the degeneracy
by considering all permutations of the ordering of each of the two final
states.
For each canonical decay we consider the permutations which generate a decay
which is equivalent to the canonical decay.
These permutations are obtained by compounding the action of two
classes of permutations: the permutations in the
directions of the initial states and the permutations in the
directions transverse to the directions of the initial state.
In order to respect the constraint that the decay generated is
equivalent to the canonical decay we must restrict the the permutations in the
directions of the initial states to the ones which do not change the
initial state.

In the case of identical final particles (i.e. same polaritazion and
eventually different momenta) we have first to take in account
the symmetry factor $\oh$.
At the same time we have to count properly the degeneration by
considering all inequivalent decays.
The set of decays on which we impose an equivalence relation in order
to determine the inequivalent decays are
the decays generated by a permutation on the canonical decay
and
the decays generated by a permutation on the ``swap canonical decay''
which is generated from the canonical decay by swapping the two final
particles.
The equivalence relation comes from the 
ordering the directional form of the decay.
The ordering is applied {\sl independently} to the directions in which
the initial state is not the vacuum and the other directions.
Let us consider some examples.
\begin{itemize}
\item
We consider the decay
\begin{equation}
  [2,1] [] \rightarrow [1,1] [] \oplus [] [1,1]
  ,
\end{equation}
it has directional canonical form
\begin{equation}
  \dirst{2,1}{1,1}{}
  \dirst{}{}{1,1}
  ,
\end{equation}
and ``swap directional canonical decay''
\begin{equation}
  \dirst{2,1}{}{1,1}
  \dirst{}{1,1}{}
  ,
\end{equation}
which are equivalent and considered as one decay in the minimal set
because of the swap action.
They are however non equivalent when computing the degeneration since
we have an overall symmetry factor $\oh$.

We do not get any further inequivalent decays because there are no permutations
to use.

\item
Consider the more complex case of the decay
\begin{equation}
  [2,1] [1,1] [] [] \rightarrow [1,1] [] [] [1,1] \oplus [] [] [1,1] []
  ,
\end{equation}  
which has directional canonical form
\begin{equation}
  \dirst{2,1}{1,1}{}
  \dirst{1,1}{}{}
  \dirst{}{}{1,1}
  \dirst{}{1,1}{}
  ,
\end{equation}
and ``swap directional canonical decay''
\begin{equation}
  \dirst{2,1}{}{1,1}
  \dirst{1,1}{}{}
  \dirst{}{1,1}{}
  \dirst{}{}{1,1}
  ,
\end{equation}
which are equivalent and considered as one decay in the minimal set
but different decays in counting the degeneracy.
Moreover from the directional canonical form we get the inequivalent
decay
\begin{equation}
  \dirst{2,1}{1,1}{}
  \dirst{1,1}{}{}
  \dirst{}{1,1}{}
  \dirst{}{}{1,1}
  ,
\end{equation}
by considering the permutations in directions where the initial state
is the vacuum.
Similarly for the directional swap canonical form.

We do not get any further inequivalent decays because there are no permutations
to use in the directions where the initial state extends.

\end{itemize}

\subsection{Averaging over the transverse directions}
We use the following expression for the average over the transverse
directions
\begin{align}
\frac{1}{\int d^{d-2} \Omega\, 1}
&
\int d^{d-2} \Omega\,
(n_2)^{2 b_2}\,
(n_3)^{2 b_3}\,
\dots
(n_{d-1})^{2 b_{d-1}}\,
\nonumber\\
=&
(\sin\theta_1)^{2 \sum_{k=2}^{d-1} b_k}
\frac{
\prod_{k=2}^{d-1} (2 b_k -1)||
}{
d (d+2)\dots
(d-2 \sum_{k=2}^{d-1} b_k)
}
,
\end{align}     
with $0!! = (-1)!! =1$.

\subsection{Integrating over the space direction in the \lc}
Instead of using the angle $\theta_1$ it turns out
it is better to use polynomial expressions
using the variable $x=-\frac{k_{[2] +}}{k_{[1] +}}$, explicitly
\begin{equation}
x
=
\frac{E_{[2]} + p_* \cos \theta_1}{m_1}
,
\end{equation}
so that
\begin{equation}
\sin \theta_1
=
\frac{1}{p_*}
\sqrt{ -m_1^2 x^2 - 2 m_1  E_{[2]} x - m_2^2}
.
\end{equation}  
However the square root $\sqrt{...}$ disappears because $\sin \theta_1$
enters with even powers when $d-2$ is even,
in facts $d ^{d-2} \Omega$ is proportional to 
$
d \theta_1\, (\sin \theta_1)^{d-2}
$.

\newpage
\section{Summary of total decay widths and number of channels}
\label{app:summary}

\input{Tables_from_level2_to_8state1_num_width_channels.V1.0.tex}
\newpage
\section{Level $3$}
\label{app:N=3}

%
%
{\hskip-10em}
\begin{longtable}[l]{m{0.4\textwidth} m{0.16\textwidth} m{0.18\textwidth} m{0.10\textwidth} m{0.10\textwidth} m{0.04\textwidth} }
\caption{$N= 3$ state 1  }
\label{N=3_state_1}
\\\hline\hline
State & & Total width & Channels & &
\endfirsthead
\\
\multirow{2}{0.3\textwidth}{$\, [[2,1],[1,1]] $ } 
& & $1.9357e-28$ 
& $35$  
& &
\\
& \multicolumn{5}{c}{
\makecell{${{41}\over{229167963856690583961600\,\pi^{12}}}$} 
  } 
\\ 
 \hline
decay & width & ratio  & cl. deg. & q. deg. & s.t. norm.
\\
$\, \begin{pmatrix} \left[ 2 , 1 \right] &\left[ 1 , 1 \right] \cr \left[ 2 , 1 \right] &\left[  \right] \cr \left[  \right] &\left[ 1 , 1 \right] \cr \ifx\endpmatrix\undefined}\else\end{pmatrix}\fi  $ 
& $ 6.1508e-29 $ 
& [$ 3.1776e-1 $]  
& $ 1 $  
& $ 1 $ 
& $ 1 $  
\\
\multicolumn{6}{ c }{ \makecell{ $+{{167291\,\log 2}\over{183119118336000\,\pi^{12}}}$  $-\left({{21571422587290891861421}\over{34065474075351270270015897600000\,\pi^{12}}}\right)$ \\} 
} 
\\
\hline
$\, \begin{pmatrix} \left[ 2 , 1 \right] &\left[ 1 , 1 \right] &\left[  \right] \cr \left[ 1 , 1 \right] &\left[ 1 , 1 \right] &\left[  \right] \cr \left[  \right] &\left[  \right] &\left[ 1 , 1 \right] \cr \ifx\endpmatrix\undefined}\else\end{pmatrix}\fi  $ 
& $ 1.9947e-30 $ 
& [$ 2.2671e-1 $]  
& $ 22 $  
& $ 22 $ 
& $ 1 $  
\\
\multicolumn{6}{ c }{ \makecell{ $+{{9052635756367038307}\over{34065474075351270270015897600000\,\pi^{12}}}$  $-\left({{1369\,\log 2}\over{3570822807552000\,\pi^{12}}}\right)$ \\} 
} 
\\
\hline
$\, \begin{pmatrix} \left[ 2 , 1 \right] &\left[ 1 , 1 \right] &\left[  \right] \cr \left[  \right] &\left[ 1 , 1 \right] &\left[ 1 , 1 \right] \cr \left[ 1 , 1 \right] &\left[  \right] &\left[  \right] \cr \ifx\endpmatrix\undefined}\else\end{pmatrix}\fi  $ 
& $ 1.8932e-30 $ 
& [$ 2.1517e-1 $]  
& $ 22 $  
& $ 22 $ 
& $ 1 $  
\\
\multicolumn{6}{ c }{ \makecell{ $+{{14638070138226933121}\over{11355158025117090090005299200000\,\pi^{12}}}$  $-\left({{6641\,\log 2}\over{3570822807552000\,\pi^{12}}}\right)$ \\} 
} 
\\
\hline
$\, \begin{pmatrix} \left[ 2 , 1 \right] &\left[ 1 , 1 \right] \cr \left[  \right] &\left[ 2 , 1 \right] \cr \left[ 1 , 1 \right] &\left[  \right] \cr \ifx\endpmatrix\undefined}\else\end{pmatrix}\fi  $ 
& $ 2.3672e-29 $ 
& [$ 1.2230e-1 $]  
& $ 1 $  
& $ 1 $ 
& $ 1 $  
\\
\multicolumn{6}{ c }{ \makecell{ $+{{1503\,\log 2}\over{29389488128000\,\pi^{12}}}$  $-\left({{1207554822730897689011}\over{34065474075351270270015897600000\,\pi^{12}}}\right)$ \\} 
} 
\\
\hline
$\, \begin{pmatrix} \left[ 2 , 1 \right] &\left[ 1 , 1 \right] &\left[  \right] \cr \left[ 1 , 1 \right] &\left[  \right] &\left[ 1 , 1 \right] \cr \left[  \right] &\left[ 1 , 1 \right] &\left[  \right] \cr \ifx\endpmatrix\undefined}\else\end{pmatrix}\fi  $ 
& $ 2.3275e-31 $ 
& [$ 2.6453e-2 $]  
& $ 22 $  
& $ 22 $ 
& $ 1 $  
\\
\multicolumn{6}{ c }{ \makecell{ $+{{61032001815512516717}\over{2271031605023418018001059840000\,\pi^{12}}}$  $-\left({{27689\,\log 2}\over{714164561510400\,\pi^{12}}}\right)$ \\} 
} 
\\
\hline
\end{longtable}
\newpage
{\hskip-10em}
\begin{longtable}[l]{m{0.4\textwidth} m{0.16\textwidth} m{0.18\textwidth} m{0.10\textwidth} m{0.10\textwidth} m{0.04\textwidth} }
\caption{$N= 3$ state 2  }
\label{N=3_state_2}
\\\hline\hline
State & & Total width & Channels & &
\endfirsthead
\\
\multirow{2}{0.3\textwidth}{$\, [[3,1]] $ } 
& & $2.4516e-28$ 
& $16$  
& &
\\
& \multicolumn{5}{c}{
\makecell{${{1}\over{4413248603682766848000\,\pi^{12}}}$} 
  } 
\\ 
 \hline
decay & width & ratio  & cl. deg. & q. deg. & s.t. norm.
\\
$\, \begin{pmatrix} \left[ 3 , 1 \right] &\left[  \right] \cr \left[ 2 , 1 \right] &\left[  \right] \cr \left[  \right] &\left[ 1 , 1 \right] \cr \ifx\endpmatrix\undefined}\else\end{pmatrix}\fi  $ 
& $ 6.4452e-30 $ 
& [$ 6.0467e-1 $]  
& $ 23 $  
& $ 23 $ 
& $ 1 $  
\\
\multicolumn{6}{ c }{ \makecell{ $+{{55628046293432910107}\over{1101106232738626917818695680000\,\pi^{12}}}$  $-\left({{13\,\log 2}\over{178362777600\,\pi^{12}}}\right)$ \\} 
} 
\\
\hline
$\, \begin{pmatrix} \left[ 3 , 1 \right] &\left[  \right] \cr \left[  \right] &\left[ 1 , 2 \right] \cr \left[ 1 , 1 \right] &\left[  \right] \cr \ifx\endpmatrix\undefined}\else\end{pmatrix}\fi  $ 
& $ 1.8751e-30 $ 
& [$ 1.7592e-1 $]  
& $ 23 $  
& $ 23 $ 
& $ 1 $  
\\
\multicolumn{6}{ c }{ \makecell{ $+{{698881\,\log 2}\over{5713316492083200\,\pi^{12}}}$ \\ $-\left({{146711540807251942553}\over{1730309794303556585143664640000\,\pi^{12}}}\right)$ \\} 
} 
\\
\hline
$\, \begin{pmatrix} \left[ 3 , 1 \right] &\left[  \right] \cr \left[  \right] &\left[ 2 , 1 \right] \cr \left[ 1 , 1 \right] &\left[  \right] \cr \ifx\endpmatrix\undefined}\else\end{pmatrix}\fi  $ 
& $ 1.5891e-30 $ 
& [$ 1.4909e-1 $]  
& $ 23 $  
& $ 23 $ 
& $ 1 $  
\\
\multicolumn{6}{ c }{ \makecell{ $+{{1835396367645848155123}\over{12112168560124896096005652480000\,\pi^{12}}}$  $-\left({{697\,\log 2}\over{3188234649600\,\pi^{12}}}\right)$ \\} 
} 
\\
\hline
$\, \begin{pmatrix} \left[ 3 , 1 \right] &\left[  \right] \cr \left[ 1 , 1 \right] &\left[ 1 , 1 \right] \cr \left[  \right] &\left[ 1 , 1 \right] \cr \ifx\endpmatrix\undefined}\else\end{pmatrix}\fi  $ 
& $ 3.3325e-31 $ 
& [$ 3.1265e-2 $]  
& $ 23 $  
& $ 23 $ 
& $ 1 $  
\\
\multicolumn{6}{ c }{ \makecell{ $+{{17617\,\log 2}\over{1428329123020800\,\pi^{12}}}$  $-\left({{14792888714395252891}\over{1730309794303556585143664640000\,\pi^{12}}}\right)$ \\} 
} 
\\
\hline
$\, \begin{pmatrix} \left[ 3 , 1 \right] &\left[  \right] &\left[  \right] \cr \left[  \right] &\left[ 1 , 1 \right] &\left[ 1 , 1 \right] \cr \left[ 1 , 1 \right] &\left[  \right] &\left[  \right] \cr \ifx\endpmatrix\undefined}\else\end{pmatrix}\fi  $ 
& $ 1.8968e-32 $ 
& [$ 1.9574e-2 $]  
& $ 253 $  
& $ 253 $ 
& $ 1 $  
\\
\multicolumn{6}{ c }{ \makecell{ $+{{36061\,\log 2}\over{4284987369062400\,\pi^{12}}}$  $-\left({{105980715204313271629}\over{18168252840187344144008478720000\,\pi^{12}}}\right)$ \\} 
} 
\\
\hline
\end{longtable}
\newpage
{\hskip-10em}
\begin{longtable}[l]{m{0.4\textwidth} m{0.16\textwidth} m{0.18\textwidth} m{0.10\textwidth} m{0.10\textwidth} m{0.04\textwidth} }
\caption{$N= 3$ state 3  }
\label{N=3_state_3}
\\\hline\hline
State & & Total width & Channels & &
\endfirsthead
\\
\multirow{2}{0.3\textwidth}{$\, [[1,3]] $ } 
& & $2.6063e-28$ 
& $16$  
& &
\\
& \multicolumn{5}{c}{
\makecell{${{349}\over{1448762989898618634240000\,\pi^{12}}}$} 
  } 
\\ 
 \hline
decay & width & ratio  & cl. deg. & q. deg. & s.t. norm.
\\
$\, \begin{pmatrix} \left[ 1 , 3 \right] \cr \left[ 1 , 2 \right] \cr \left[ 1 , 1 \right] \cr \ifx\endpmatrix\undefined}\else\end{pmatrix}\fi  $ 
& $ 2.1091e-28 $ 
& [$ 8.0923e-1 $]  
& $ 1 $  
& $ 1 $ 
& $ 1 $  
\\
\multicolumn{6}{ c }{ \makecell{ $+{{5751377\,\log 2}\over{43167280162406400\,\pi^{12}}}$  $-\left({{1026850452233782637893889}\over{11118970738194654616133188976640000\,\pi^{12}}}\right)$ \\} 
} 
\\
\hline
$\, \begin{pmatrix} \left[ 1 , 3 \right] &\left[  \right] \cr \left[ 1 , 1 \right] &\left[ 1 , 1 \right] \cr \left[  \right] &\left[ 1 , 1 \right] \cr \ifx\endpmatrix\undefined}\else\end{pmatrix}\fi  $ 
& $ 1.1647e-30 $ 
& [$ 1.0278e-1 $]  
& $ 23 $  
& $ 23 $ 
& $ 1 $  
\\
\multicolumn{6}{ c }{ \makecell{ $+{{241\,\log 2}\over{809386503045120\,\pi^{12}}}$ $-\left({{152988883021740188209}\over{741264715879643641075545931776000\,\pi^{12}}}\right)$ \\} 
} 
\\
\hline
$\, \begin{pmatrix} \left[ 1 , 3 \right] &\left[  \right] \cr \left[ 1 , 2 \right] &\left[  \right] \cr \left[  \right] &\left[ 1 , 1 \right] \cr \ifx\endpmatrix\undefined}\else\end{pmatrix}\fi  $ 
& $ 5.7236e-31 $ 
& [$ 5.0508e-2 $]  
& $ 23 $  
& $ 23 $ 
& $ 1 $  
\\
\multicolumn{6}{ c }{ \makecell{ $+{{12497\,\log 2}\over{43167280162406400\,\pi^{12}}}$  $-\left({{106248262696036918271}\over{529474797056888315053961379840000\,\pi^{12}}}\right)$ \\} 
} 
\\
\hline
$\, \begin{pmatrix} \left[ 1 , 3 \right] &\left[  \right] \cr \left[ 1 , 1 \right] &\left[ 1 , 1 \right] \cr \left[ 1 , 1 \right] &\left[  \right] \cr \ifx\endpmatrix\undefined}\else\end{pmatrix}\fi  $ 
& $ 1.2777e-31 $ 
& [$ 1.1275e-2 $]  
& $ 23 $  
& $ 23 $ 
& $ 1 $  
\\
\multicolumn{6}{ c }{ \makecell{ $+{{1185491\,\log 2}\over{48563190182707200\,\pi^{12}}}$ \\ $-\left({{20904442591432746526303}\over{1235441193132739401792576552960000\,\pi^{12}}}\right)$ \\} 
} 
\\
\hline
$\, \begin{pmatrix} \left[ 1 , 3 \right] \cr \left[ 2 , 1 \right] \cr \left[ 1 , 1 \right] \cr \ifx\endpmatrix\undefined}\else\end{pmatrix}\fi  $ 
& $ 1.6669e-30 $ 
& [$ 6.3958e-3 $]  
& $ 1 $  
& $ 1 $ 
& $ 1 $  
\\
\multicolumn{6}{ c }{ \makecell{ $+{{11737008216419638403701}\over{26162284089869775567372209356800\,\pi^{12}}}$  $-\left({{22186859\,\log 2}\over{34279898952499200\,\pi^{12}}}\right)$ \\} 
} 
\\
\hline
\end{longtable}
\newpage
{\hskip-10em}
\begin{longtable}[l]{m{0.4\textwidth} m{0.16\textwidth} m{0.18\textwidth} m{0.10\textwidth} m{0.10\textwidth} m{0.04\textwidth} }
\caption{$N= 3$ state 4  }
\label{N=3_state_4}
\\\hline\hline
State & & Total width & Channels & &
\endfirsthead
\\
\multirow{2}{0.3\textwidth}{$\, [[1,2],[1,1]] $ } 
& & $2.7095e-28$ 
& $35$  
& &
\\
& \multicolumn{5}{c}{
\makecell{${{6313}\over{25208476024235964235776000\,\pi^{12}}}$} 
  } 
\\ 
 \hline
decay & width & ratio  & cl. deg. & q. deg. & s.t. norm.
\\
$\, \begin{pmatrix} \left[ 1 , 2 \right] &\left[ 1 , 1 \right] \cr \left[ 1 , 1 \right] &\left[ 1 , 1 \right] \cr \left[ 1 , 1 \right] &\left[  \right] \cr \ifx\endpmatrix\undefined}\else\end{pmatrix}\fi  $ 
& $ 1.5864e-28 $ 
& [$ 5.8548e-1 $]  
& $ 1 $  
& $ 1 $ 
& $ 1 $  
\\
\multicolumn{6}{ c }{ \makecell{ $+{{70367\,\log 2}\over{3396027285504000\,\pi^{12}}}$  $-\left({{24196022391075640621897}\over{1684692536090099184262604390400000\,\pi^{12}}}\right)$ \\} 
} 
\\
\hline
$\, \begin{pmatrix} \left[ 1 , 2 \right] &\left[ 1 , 1 \right] \cr \left[ 1 , 2 \right] &\left[  \right] \cr \left[  \right] &\left[ 1 , 1 \right] \cr \ifx\endpmatrix\undefined}\else\end{pmatrix}\fi  $ 
& $ 8.5108e-29 $ 
& [$ 3.1411e-1 $]  
& $ 1 $  
& $ 1 $ 
& $ 1 $  
\\
\multicolumn{6}{ c }{ \makecell{ $+{{1600751\,\log 2}\over{35318683769241600\,\pi^{12}}}$  $-\left({{1512158587029476367029}\over{48134072459717119550360125440000\,\pi^{12}}}\right)$ \\} 
} 
\\
\hline
$\, \begin{pmatrix} \left[ 1 , 2 \right] &\left[ 1 , 1 \right] &\left[  \right] \cr \left[  \right] &\left[ 1 , 1 \right] &\left[ 1 , 1 \right] \cr \left[  \right] &\left[  \right] &\left[ 1 , 1 \right] \cr \ifx\endpmatrix\undefined}\else\end{pmatrix}\fi  $ 
& $ 3.8596e-31 $ 
& [$ 3.1338e-2 $]  
& $ 22 $  
& $ 22 $ 
& $ 1 $  
\\
\multicolumn{6}{ c }{ \makecell{ $+{{69427\,\log 2}\over{485631901827072000\,\pi^{12}}}$  $-\left({{13807286865461786503}\over{139335472909707451329989836800000\,\pi^{12}}}\right)$ \\} 
} 
\\
\hline
$\, \begin{pmatrix} \left[ 1 , 2 \right] &\left[ 1 , 1 \right] &\left[  \right] \cr \left[ 1 , 1 \right] &\left[ 1 , 1 \right] &\left[  \right] \cr \left[  \right] &\left[  \right] &\left[ 1 , 1 \right] \cr \ifx\endpmatrix\undefined}\else\end{pmatrix}\fi  $ 
& $ 3.5179e-31 $ 
& [$ 2.8564e-2 $]  
& $ 22 $  
& $ 22 $ 
& $ 1 $  
\\
\multicolumn{6}{ c }{ \makecell{ $+{{20659\,\log 2}\over{485631901827072000\,\pi^{12}}}$  $-\left({{3002406538620905839}\over{101822076357093906741146419200000\,\pi^{12}}}\right)$ \\} 
} 
\\
\hline
$\, \begin{pmatrix} \left[ 1 , 2 \right] &\left[ 1 , 1 \right] &\left[  \right] \cr \left[ 1 , 2 \right] &\left[  \right] &\left[  \right] \cr \left[  \right] &\left[  \right] &\left[ 1 , 1 \right] \cr \ifx\endpmatrix\undefined}\else\end{pmatrix}\fi  $ 
& $ 1.5590e-31 $ 
& [$ 1.2658e-2 $]  
& $ 22 $  
& $ 22 $ 
& $ 1 $  
\\
\multicolumn{6}{ c }{ \makecell{ $+{{3037\,\log 2}\over{11772894589747200\,\pi^{12}}}$  $-\left({{12049461559623287119}\over{67387701443603967370504175616000\,\pi^{12}}}\right)$ \\} 
} 
\\
\hline
\end{longtable}
\newpage
{\hskip-10em}
\begin{longtable}[l]{m{0.4\textwidth} m{0.16\textwidth} m{0.18\textwidth} m{0.10\textwidth} m{0.10\textwidth} m{0.04\textwidth} }
\caption{$N= 3$ state 5  }
\label{N=3_state_5}
\\\hline\hline
State & & Total width & Channels & &
\endfirsthead
\\
\multirow{2}{0.3\textwidth}{$\, [[2,1,1,1]] $ } 
& & $2.7611e-28$ 
& $16$  
& &
\\
& \multicolumn{5}{c}{
\makecell{${{1787}\over{7002354451176656732160000\,\pi^{12}}}$} 
  } 
\\ 
 \hline
decay & width & ratio  & cl. deg. & q. deg. & s.t. norm.
\\
$\, \begin{pmatrix} \left[ 2 , 1 , 1 , 1 \right] \cr \left[ 2 , 1 \right] \cr \left[ 1 , 1 \right] \cr \ifx\endpmatrix\undefined}\else\end{pmatrix}\fi  $ 
& $ 1.3985e-28 $ 
& [$ 5.0650e-1 $]  
& $ 1 $  
& $ 1 $ 
& $ 1 $  
\\
\multicolumn{6}{ c }{ \makecell{ $+{{3046231\,\log 2}\over{1836423158169600\,\pi^{12}}}$  $-\left({{4147186389977980768213}\over{3606932549154840381531095040000\,\pi^{12}}}\right)$ \\} 
} 
\\
\hline
$\, \begin{pmatrix} \left[ 2 , 1 , 1 , 1 \right] &\left[  \right] \cr \left[ 1 , 2 \right] &\left[  \right] \cr \left[  \right] &\left[ 1 , 1 \right] \cr \ifx\endpmatrix\undefined}\else\end{pmatrix}\fi  $ 
& $ 4.0943e-30 $ 
& [$ 3.4105e-1 $]  
& $ 23 $  
& $ 23 $ 
& $ 1 $  
\\
\multicolumn{6}{ c }{ \makecell{ $+{{3973092770924626961}\over{1048168433087731392923566080000\,\pi^{12}}}$  $-\left({{35149\,\log 2}\over{6427481053593600\,\pi^{12}}}\right)$ \\} 
} 
\\
\hline
$\, \begin{pmatrix} \left[ 2 , 1 , 1 , 1 \right] &\left[  \right] \cr \left[ 1 , 1 \right] &\left[ 1 , 1 \right] \cr \left[ 1 , 1 \right] &\left[  \right] \cr \ifx\endpmatrix\undefined}\else\end{pmatrix}\fi  $ 
& $ 7.4426e-31 $ 
& [$ 6.1997e-2 $]  
& $ 23 $  
& $ 23 $ 
& $ 1 $  
\\
\multicolumn{6}{ c }{ \makecell{ $+{{410325381434534674841}\over{8759693333661755212289802240000\,\pi^{12}}}$  $-\left({{86873\,\log 2}\over{1285496210718720\,\pi^{12}}}\right)$ \\} 
} 
\\
\hline
$\, \begin{pmatrix} \left[ 2 , 1 , 1 , 1 \right] &\left[  \right] \cr \left[  \right] &\left[ 2 , 1 \right] \cr \left[  \right] &\left[ 1 , 1 \right] \cr \ifx\endpmatrix\undefined}\else\end{pmatrix}\fi  $ 
& $ 5.8546e-31 $ 
& [$ 4.8769e-2 $]  
& $ 23 $  
& $ 23 $ 
& $ 1 $  
\\
\multicolumn{6}{ c }{ \makecell{ $+{{208471\,\log 2}\over{5843164594176000\,\pi^{12}}}$  $-\left({{689266628470583785439}\over{27871751516196493857285734400000\,\pi^{12}}}\right)$ \\} 
} 
\\
\hline
$\, \begin{pmatrix} \left[ 2 , 1 , 1 , 1 \right] &\left[  \right] \cr \left[ 2 , 1 \right] &\left[  \right] \cr \left[  \right] &\left[ 1 , 1 \right] \cr \ifx\endpmatrix\undefined}\else\end{pmatrix}\fi  $ 
& $ 1.8620e-31 $ 
& [$ 1.5511e-2 $]  
& $ 23 $  
& $ 23 $ 
& $ 1 $  
\\
\multicolumn{6}{ c }{ \makecell{ $+{{100939\,\log 2}\over{12854962107187200\,\pi^{12}}}$  $-\left({{6810900259429376819}\over{1251384761951679316041400320000\,\pi^{12}}}\right)$ \\} 
} 
\\
\hline
\end{longtable}
\newpage
{\hskip-10em}
\begin{longtable}[l]{m{0.4\textwidth} m{0.16\textwidth} m{0.18\textwidth} m{0.10\textwidth} m{0.10\textwidth} m{0.04\textwidth} }
\caption{$N= 3$ state 6  }
\label{N=3_state_6}
\\\hline\hline
State & & Total width & Channels & &
\endfirsthead
\\
\multirow{2}{0.3\textwidth}{$\, [[1,1],[1,1],[1,1]] $ } 
& & $2.7611e-28$ 
& $22$  
& &
\\
& \multicolumn{5}{c}{
\makecell{${{1787}\over{7002354451176656732160000\,\pi^{12}}}$} 
  } 
\\ 
 \hline
decay & width & ratio  & cl. deg. & q. deg. & s.t. norm.
\\
$\, \begin{pmatrix} \left[ 1 , 1 \right] &\left[ 1 , 1 \right] &\left[ 1 , 1 \right] \cr \left[ 1 , 1 \right] &\left[ 1 , 1 \right] &\left[  \right] \cr \left[  \right] &\left[  \right] &\left[ 1 , 1 \right] \cr \ifx\endpmatrix\undefined}\else\end{pmatrix}\fi  $ 
& $ 8.6976e-29 $ 
& [$ 9.4502e-1 $]  
& $ 3 $  
& $ 3 $ 
& $ 1 $  
\\
\multicolumn{6}{ c }{ \makecell{ $+{{4234103\,\log 2}\over{485631901827072000\,\pi^{12}}}$ $-\left({{6999596824902329210131}\over{1158226118561943189180540518400000\,\pi^{12}}}\right)$ \\} 
} 
\\
\hline
$\, \begin{pmatrix} \left[ 1 , 1 \right] &\left[ 1 , 1 \right] &\left[ 1 , 1 \right] &\left[  \right] \cr \left[ 1 , 1 \right] &\left[ 1 , 1 \right] &\left[  \right] &\left[  \right] \cr \left[  \right] &\left[  \right] &\left[  \right] &\left[ 1 , 1 \right] \cr \ifx\endpmatrix\undefined}\else\end{pmatrix}\fi  $ 
& $ 1.5873e-31 $ 
& [$ 3.6218e-2 $]  
& $ 63 $  
& $ 63 $ 
& $ 1 $  
\\
\multicolumn{6}{ c }{ \makecell{ $+{{3953\,\log 2}\over{112068900421632000\,\pi^{12}}}$  $-\left({{84953597311527883613}\over{3474678355685829567541621555200000\,\pi^{12}}}\right)$ \\} 
} 
\\
\hline
$\, \begin{pmatrix} \left[ 1 , 1 \right] &\left[ 1 , 1 \right] &\left[ 1 , 1 \right] &\left[  \right] \cr \left[ 1 , 1 \right] &\left[  \right] &\left[  \right] &\left[ 1 , 1 \right] \cr \left[  \right] &\left[ 1 , 1 \right] &\left[  \right] &\left[  \right] \cr \ifx\endpmatrix\undefined}\else\end{pmatrix}\fi  $ 
& $ 1.5129e-32 $ 
& [$ 6.9038e-3 $]  
& $ 126 $  
& $ 126 $ 
& $ 1 $  
\\
\multicolumn{6}{ c }{ \makecell{ $+{{7360777\,\log 2}\over{1456895705481216000\,\pi^{12}}}$  $-\left({{10797205109005390487}\over{3083121877272253387348377600000\,\pi^{12}}}\right)$ \\} 
} 
\\
\hline
$\, \begin{pmatrix} \left[ 1 , 1 \right] &\left[ 1 , 1 \right] &\left[ 1 , 1 \right] \cr \left[ 2 , 1 \right] &\left[  \right] &\left[  \right] \cr \left[  \right] &\left[ 1 , 1 \right] &\left[  \right] \cr \ifx\endpmatrix\undefined}\else\end{pmatrix}\fi  $ 
& $ 2.1446e-31 $ 
& [$ 4.6603e-3 $]  
& $ 6 $  
& $ 6 $ 
& $ 1 $  
\\
\multicolumn{6}{ c }{ \makecell{ $+{{130098266081546674199}\over{1419394753139636261250662400000\,\pi^{12}}}$  $-\left({{419719\,\log 2}\over{3174064717824000\,\pi^{12}}}\right)$ \\} 
} 
\\
\hline
$\, \begin{pmatrix} \left[ 1 , 1 \right] &\left[ 1 , 1 \right] &\left[ 1 , 1 \right] \cr \left[ 1 , 1 \right] &\left[ 1 , 1 \right] &\left[  \right] \cr \left[ 1 , 1 \right] &\left[  \right] &\left[  \right] \cr \ifx\endpmatrix\undefined}\else\end{pmatrix}\fi  $ 
& $ 1.7607e-31 $ 
& [$ 3.8262e-3 $]  
& $ 6 $  
& $ 6 $ 
& $ 1 $  
\\
\multicolumn{6}{ c }{ \makecell{ $+{{3196807\,\log 2}\over{485631901827072000\,\pi^{12}}}$  $-\left({{57443407902869900887}\over{12589414332195034665005875200000\,\pi^{12}}}\right)$ \\} 
} 
\\
\hline
\end{longtable}

\newpage

\section{Level $4$}
\label{app:N=4}

{\hskip-10em}
\begin{longtable}[l]{m{0.4\textwidth} m{0.16\textwidth} m{0.18\textwidth} m{0.10\textwidth} m{0.10\textwidth} m{0.04\textwidth} }
\caption{$N= 4$ state 1  }
\label{N=4_state_1}
\\\hline\hline
State & & Total width & Channels & &
\endfirsthead
\\
\multirow{2}{0.3\textwidth}{$\, [[1,1],[1,1],[1,1],[1,1]] $ } 
& & $9.5345e-22$ 
& $69$  
& &
\\
& \multicolumn{5}{c}{
\makecell{${{147733242301247}\over{18718979078800\,6^{{{41}\over{2}}}\,\pi^{12}}}$} 
  } 
\\ 
 \hline
decay & width & ratio  & cl. deg. & q. deg. & s.t. norm.
\\
$\, \begin{pmatrix} \left[ 1 , 1 \right] &\left[ 1 , 1 \right] &\left[ 1 , 1 \right] &\left[ 1 , 1 \right] \cr \left[ 1 , 1 \right] &\left[  \right] &\left[  \right] &\left[  \right] \cr \left[  \right] &\left[ 1 , 1 \right] &\left[  \right] &\left[  \right] \cr \ifx\endpmatrix\undefined}\else\end{pmatrix}\fi  $ 
& $ 1.3144e-22 $ 
& [$ 8.2713e-1 $]  
& $ 12 $  
& $ 6 $ 
& $ 1 $  
\\
\multicolumn{6}{ c }{ \makecell{${{5403753\,\sqrt{6}}\over{108956939595441766400000\,\pi^{12}}}$} 
} 
\\
\hline
$\, \begin{pmatrix} \left[ 1 , 1 \right] &\left[ 1 , 1 \right] &\left[ 1 , 1 \right] &\left[ 1 , 1 \right] &\left[  \right] \cr \left[ 1 , 1 \right] &\left[  \right] &\left[  \right] &\left[  \right] &\left[  \right] \cr \left[  \right] &\left[  \right] &\left[  \right] &\left[  \right] &\left[ 1 , 1 \right] \cr \ifx\endpmatrix\undefined}\else\end{pmatrix}\fi  $ 
& $ 1.8469e-24 $ 
& [$ 1.5497e-1 $]  
& $ 160 $  
& $ 80 $ 
& $ 1 $  
\\
\multicolumn{6}{ c }{ \makecell{${{26044983\,\sqrt{6}}\over{37372230281236525875200000\,\pi^{12}}}$} 
} 
\\
\hline
$\, \begin{pmatrix} \left[ 1 , 1 \right] &\left[ 1 , 1 \right] &\left[ 1 , 1 \right] &\left[ 1 , 1 \right] &\left[  \right] \cr \left[  \right] &\left[  \right] &\left[  \right] &\left[  \right] &\left[ 1 , 1 \right] \cr \left[  \right] &\left[  \right] &\left[  \right] &\left[  \right] &\left[ 1 , 1 \right] \cr \ifx\endpmatrix\undefined}\else\end{pmatrix}\fi  $ 
& $ 7.2555e-25 $ 
& [$ 7.6098e-3 $]  
& $ 20 $  
& $ 20 $ 
& $ {{1}\over{2}} $  
\\
\multicolumn{6}{ c }{ \makecell{${{292329\,\sqrt{6}}\over{1067778008035329310720000\,\pi^{12}}}$} 
} 
\\
\hline
\end{longtable}
\newpage
{\hskip-10em}
\begin{longtable}[l]{m{0.4\textwidth} m{0.16\textwidth} m{0.18\textwidth} m{0.10\textwidth} m{0.10\textwidth} m{0.04\textwidth} }
\caption{$N= 4$ state 2  }
\label{N=4_state_2}
\\\hline\hline
State & & Total width & Channels & &
\endfirsthead
\\
\multirow{2}{0.3\textwidth}{$\, [[1,2],[1,1],[1,1]] $ } 
& & $1.1821e-21$ 
& $136$  
& &
\\
& \multicolumn{5}{c}{
\makecell{${{87274963383734573}\over{6881977602500\,6^{{{49}\over{2}}}\,\pi^{12}}}$} 
  } 
\\ 
 \hline
decay & width & ratio  & cl. deg. & q. deg. & s.t. norm.
\\
$\, \begin{pmatrix} \left[ 1 , 2 \right] &\left[ 1 , 1 \right] &\left[ 1 , 1 \right] \cr \left[  \right] &\left[ 1 , 1 \right] &\left[  \right] \cr \left[ 1 , 1 \right] &\left[  \right] &\left[  \right] \cr \ifx\endpmatrix\undefined}\else\end{pmatrix}\fi  $ 
& $ 3.0309e-22 $ 
& [$ 5.1278e-1 $]  
& $ 4 $  
& $ 2 $ 
& $ 1 $  
\\
\multicolumn{6}{ c }{ \makecell{${{18422721\,\sqrt{6}}\over{161087199488088473600000\,\pi^{12}}}$} 
} 
\\
\hline
$\, \begin{pmatrix} \left[ 1 , 2 \right] &\left[ 1 , 1 \right] &\left[ 1 , 1 \right] \cr \left[  \right] &\left[ 1 , 1 \right] &\left[  \right] \cr \left[  \right] &\left[  \right] &\left[ 1 , 1 \right] \cr \ifx\endpmatrix\undefined}\else\end{pmatrix}\fi  $ 
& $ 1.8822e-22 $ 
& [$ 1.5922e-1 $]  
& $ 2 $  
& $ 1 $ 
& $ 1 $  
\\
\multicolumn{6}{ c }{ \makecell{${{2592027\,\sqrt{6}}\over{36496318634020044800000\,\pi^{12}}}$} 
} 
\\
\hline
$\, \begin{pmatrix} \left[ 1 , 2 \right] &\left[ 1 , 1 \right] &\left[ 1 , 1 \right] &\left[  \right] \cr \left[  \right] &\left[ 1 , 1 \right] &\left[  \right] &\left[  \right] \cr \left[  \right] &\left[  \right] &\left[  \right] &\left[ 1 , 1 \right] \cr \ifx\endpmatrix\undefined}\else\end{pmatrix}\fi  $ 
& $ 3.4704e-24 $ 
& [$ 1.2330e-1 $]  
& $ 84 $  
& $ 42 $ 
& $ 1 $  
\\
\multicolumn{6}{ c }{ \makecell{${{48938499\,\sqrt{6}}\over{37372230281236525875200000\,\pi^{12}}}$} 
} 
\\
\hline
\end{longtable}
\newpage
{\hskip-10em}
\begin{longtable}[l]{m{0.4\textwidth} m{0.16\textwidth} m{0.18\textwidth} m{0.10\textwidth} m{0.10\textwidth} m{0.04\textwidth} }
\caption{$N= 4$ state 3  }
\label{N=4_state_3}
\\\hline\hline
State & & Total width & Channels & &
\endfirsthead
\\
\multirow{2}{0.3\textwidth}{$\, [[1,2],[1,2]] $ } 
& & $1.4468e-21$ 
& $63$  
& &
\\
& \multicolumn{5}{c}{
\makecell{${{31950560275395331}\over{343089792500\,6^{{{51}\over{2}}}\,\pi^{12}}}$} 
  } 
\\ 
 \hline
decay & width & ratio  & cl. deg. & q. deg. & s.t. norm.
\\
$\, \begin{pmatrix} \left[ 1 , 2 \right] &\left[ 1 , 2 \right] \cr \left[ 1 , 1 \right] &\left[  \right] \cr \left[  \right] &\left[ 1 , 1 \right] \cr \ifx\endpmatrix\undefined}\else\end{pmatrix}\fi  $ 
& $ 6.9784e-22 $ 
& [$ 4.8232e-1 $]  
& $ 2 $  
& $ 1 $ 
& $ 1 $  
\\
\multicolumn{6}{ c }{ \makecell{${{7872568443\,\sqrt{6}}\over{29897784224989220700160000\,\pi^{12}}}$} 
} 
\\
\hline
$\, \begin{pmatrix} \left[ 1 , 2 \right] &\left[ 1 , 2 \right] \cr \left[ 1 , 1 \right] &\left[  \right] \cr \left[ 1 , 1 \right] &\left[  \right] \cr \ifx\endpmatrix\undefined}\else\end{pmatrix}\fi  $ 
& $ 3.5505e-22 $ 
& [$ 2.4540e-1 $]  
& $ 2 $  
& $ 2 $ 
& $ {{1}\over{2}} $  
\\
\multicolumn{6}{ c }{ \makecell{${{4005478431\,\sqrt{6}}\over{29897784224989220700160000\,\pi^{12}}}$} 
} 
\\
\hline
$\, \begin{pmatrix} \left[ 1 , 2 \right] &\left[ 1 , 2 \right] &\left[  \right] \cr \left[ 1 , 1 \right] &\left[  \right] &\left[  \right] \cr \left[  \right] &\left[  \right] &\left[ 1 , 1 \right] \cr \ifx\endpmatrix\undefined}\else\end{pmatrix}\fi  $ 
& $ 7.9416e-24 $ 
& [$ 2.4151e-1 $]  
& $ 88 $  
& $ 44 $ 
& $ 1 $  
\\
\multicolumn{6}{ c }{ \makecell{${{89591589\,\sqrt{6}}\over{29897784224989220700160000\,\pi^{12}}}$} 
} 
\\
\hline
\end{longtable}
\newpage
{\hskip-10em}
\begin{longtable}[l]{m{0.4\textwidth} m{0.16\textwidth} m{0.18\textwidth} m{0.10\textwidth} m{0.10\textwidth} m{0.04\textwidth} }
\caption{$N= 4$ state 4  }
\label{N=4_state_4}
\\\hline\hline
State & & Total width & Channels & &
\endfirsthead
\\
\multirow{2}{0.3\textwidth}{$\, [[2,1,1,2]] $ } 
& & $1.4793e-21$ 
& $45$  
& &
\\
& \multicolumn{5}{c}{
\makecell{${{16709643893238636401}\over{29248404810625\,6^{{{53}\over{2}}}\,\pi^{12}}}$} 
  } 
\\ 
 \hline
decay & width & ratio  & cl. deg. & q. deg. & s.t. norm.
\\
$\, \begin{pmatrix} \left[ 2 , 1 , 1 , 2 \right] &\left[  \right] \cr \left[ 1 , 1 \right] &\left[  \right] \cr \left[  \right] &\left[ 1 , 1 \right] \cr \ifx\endpmatrix\undefined}\else\end{pmatrix}\fi  $ 
& $ 3.7909e-23 $ 
& [$ 5.8941e-1 $]  
& $ 46 $  
& $ 23 $ 
& $ 1 $  
\\
\multicolumn{6}{ c }{ \makecell{${{32704239\,\sqrt{6}}\over{2286301146616822759424000\,\pi^{12}}}$} 
} 
\\
\hline
$\, \begin{pmatrix} \left[ 2 , 1 , 1 , 2 \right] \cr \left[ 1 , 1 \right] \cr \left[ 1 , 1 \right] \cr \ifx\endpmatrix\undefined}\else\end{pmatrix}\fi  $ 
& $ 1.0448e-21 $ 
& [$ 3.5314e-1 $]  
& $ 1 $  
& $ 1 $ 
& $ {{1}\over{2}} $  
\\
\multicolumn{6}{ c }{ \makecell{${{4506738363\,\sqrt{6}}\over{11431505733084113797120000\,\pi^{12}}}$} 
} 
\\
\hline
$\, \begin{pmatrix} \left[ 2 , 1 , 1 , 2 \right] &\left[  \right] \cr \left[  \right] &\left[ 1 , 1 \right] \cr \left[  \right] &\left[ 1 , 1 \right] \cr \ifx\endpmatrix\undefined}\else\end{pmatrix}\fi  $ 
& $ 4.2798e-24 $ 
& [$ 3.3271e-2 $]  
& $ 23 $  
& $ 23 $ 
& $ {{1}\over{2}} $  
\\
\multicolumn{6}{ c }{ \makecell{${{13173\,\sqrt{6}}\over{8157204033883340800000\,\pi^{12}}}$} 
} 
\\
\hline
\end{longtable}
\newpage
{\hskip-10em}
\begin{longtable}[l]{m{0.4\textwidth} m{0.16\textwidth} m{0.18\textwidth} m{0.10\textwidth} m{0.10\textwidth} m{0.04\textwidth} }
\caption{$N= 4$ state 5  }
\label{N=4_state_5}
\\\hline\hline
State & & Total width & Channels & &
\endfirsthead
\\
\multirow{2}{0.3\textwidth}{$\, [[1,3],[1,1]] $ } 
& & $1.6396e-21$ 
& $107$  
& &
\\
& \multicolumn{5}{c}{
\makecell{${{685914869096753941}\over{233987238485000\,6^{{{47}\over{2}}}\,\pi^{12}}}$} 
  } 
\\ 
 \hline
decay & width & ratio  & cl. deg. & q. deg. & s.t. norm.
\\
$\, \begin{pmatrix} \left[ 1 , 3 \right] &\left[ 1 , 1 \right] \cr \left[  \right] &\left[ 1 , 1 \right] \cr \left[ 1 , 1 \right] &\left[  \right] \cr \ifx\endpmatrix\undefined}\else\end{pmatrix}\fi  $ 
& $ 6.8228e-22 $ 
& [$ 4.1614e-1 $]  
& $ 2 $  
& $ 1 $ 
& $ 1 $  
\\
\multicolumn{6}{ c }{ \makecell{${{9621361521\,\sqrt{6}}\over{37372230281236525875200000\,\pi^{12}}}$} 
} 
\\
\hline
$\, \begin{pmatrix} \left[ 1 , 3 \right] &\left[ 1 , 1 \right] \cr \left[ 1 , 1 \right] &\left[  \right] \cr \left[ 1 , 1 \right] &\left[  \right] \cr \ifx\endpmatrix\undefined}\else\end{pmatrix}\fi  $ 
& $ 9.6987e-22 $ 
& [$ 2.9577e-1 $]  
& $ 1 $  
& $ 1 $ 
& $ {{1}\over{2}} $  
\\
\multicolumn{6}{ c }{ \makecell{${{210412971\,\sqrt{6}}\over{574957388942100398080000\,\pi^{12}}}$} 
} 
\\
\hline
$\, \begin{pmatrix} \left[ 1 , 3 \right] &\left[ 1 , 1 \right] &\left[  \right] \cr \left[ 1 , 1 \right] &\left[  \right] &\left[  \right] \cr \left[  \right] &\left[  \right] &\left[ 1 , 1 \right] \cr \ifx\endpmatrix\undefined}\else\end{pmatrix}\fi  $ 
& $ 1.2107e-23 $ 
& [$ 1.6245e-1 $]  
& $ 44 $  
& $ 22 $ 
& $ 1 $  
\\
\multicolumn{6}{ c }{ \makecell{${{42680763\,\sqrt{6}}\over{9343057570309131468800000\,\pi^{12}}}$} 
} 
\\
\hline
\end{longtable}
%
%
%

\newpage
\section{Level $5$}
\label{app:N=5}

%
%
{\hskip-10em}
\begin{longtable}[l]{m{0.4\textwidth} m{0.16\textwidth} m{0.18\textwidth} m{0.10\textwidth} m{0.10\textwidth} m{0.04\textwidth} }
\caption{$N= 5$ state 1  }
\label{N=5_state_1}
\\\hline\hline
State & & Total width & Channels & &
\endfirsthead
\\
\multirow{2}{0.3\textwidth}{$\, [[1,1],[1,1],[1,1],[1,1],[1,1]] $ } 
& & $6.9906e-23$ 
& $206$  
& &
\\
& \multicolumn{5}{c}{
\makecell{${{5559422026188769469}\over{422177324717523375\,2^{{{115}\over{2}}}\,\pi^{12}}}$} 
  } 
\\ 
 \hline
decay & width & ratio  & cl. deg. & q. deg. & s.t. norm.
\\
$\, \begin{pmatrix} \left[ 1 , 1 \right] &\left[ 1 , 1 \right] &\left[ 1 , 1 \right] &\left[ 1 , 1 \right] &\left[ 1 , 1 \right] \cr \left[ 1 , 1 \right] &\left[ 1 , 1 \right] &\left[  \right] &\left[  \right] &\left[  \right] \cr \left[  \right] &\left[  \right] &\left[ 1 , 1 \right] &\left[  \right] &\left[  \right] \cr \ifx\endpmatrix\undefined}\else\end{pmatrix}\fi  $ 
&&&&& \\ 
& $ 1.9399e-24 $ 
& [$ 8.3251e-1 $]  
& $ 30 $  
& $ 30 $ 
& $ 1 $  
\\
\multicolumn{6}{ c }{ \makecell{ $+{{4072153\,\log 4}\over{206239658625\,2^{{{53}\over{2}}}\,\pi^{12}}}$  $-\left({{10806114715212488780117005721}\over{11489868616131633853125\,2^{{{123}\over{2}}}\,\pi^{12}}}\right)$ \\} 
} 
\\
\hline
$\, \begin{pmatrix} \left[ 1 , 1 \right] &\left[ 1 , 1 \right] &\left[ 1 , 1 \right] &\left[ 1 , 1 \right] &\left[ 1 , 1 \right] &\left[  \right] \cr \left[ 1 , 1 \right] &\left[ 1 , 1 \right] &\left[  \right] &\left[  \right] &\left[  \right] &\left[  \right] \cr \left[  \right] &\left[  \right] &\left[  \right] &\left[  \right] &\left[  \right] &\left[ 1 , 1 \right] \cr \ifx\endpmatrix\undefined}\else\end{pmatrix}\fi  $ 
&&&&& \\ 
& $ 2.7377e-26 $ 
& [$ 7.4409e-2 $]  
& $ 190 $  
& $ 190 $ 
& $ 1 $  
\\
\multicolumn{6}{ c }{ \makecell{ $+{{2713\,\log 4}\over{41247931725\,2^{{{53}\over{2}}}\,\pi^{12}}}$  $-\left({{35996190889063225175455909}\over{11489868616131633853125\,2^{{{123}\over{2}}}\,\pi^{12}}}\right)$ \\} 
} 
\\
\hline
$\, \begin{pmatrix} \left[ 1 , 1 \right] &\left[ 1 , 1 \right] &\left[ 1 , 1 \right] &\left[ 1 , 1 \right] &\left[ 1 , 1 \right] &\left[  \right] \cr \left[ 1 , 1 \right] &\left[  \right] &\left[  \right] &\left[  \right] &\left[  \right] &\left[ 1 , 1 \right] \cr \left[  \right] &\left[ 1 , 1 \right] &\left[  \right] &\left[  \right] &\left[  \right] &\left[  \right] \cr \ifx\endpmatrix\undefined}\else\end{pmatrix}\fi  $ 
&&&&& \\ 
& $ 8.5424e-27 $ 
& [$ 4.6435e-2 $]  
& $ 380 $  
& $ 380 $ 
& $ 1 $  
\\
\multicolumn{6}{ c }{ \makecell{ $+{{2435263\,\log 4}\over{123743795175\,2^{{{55}\over{2}}}\,\pi^{12}}}$  $-\left({{3231201466955703977420873191}\over{6893921169678980311875\,2^{{{123}\over{2}}}\,\pi^{12}}}\right)$ \\} 
} 
\\
\hline
\end{longtable}
\newpage
{\hskip-10em}
\begin{longtable}[l]{m{0.4\textwidth} m{0.16\textwidth} m{0.18\textwidth} m{0.10\textwidth} m{0.10\textwidth} m{0.04\textwidth} }
\caption{$N= 5$ state 2  }
\label{N=5_state_2}
\\\hline\hline
State & & Total width & Channels & &
\endfirsthead
\\
\multirow{2}{0.3\textwidth}{$\, [[1,2],[1,1],[1,1],[1,1]] $ } 
& & $8.9957e-23$ 
& $447$  
& &
\\
& \multicolumn{5}{c}{
\makecell{${{3716416640375169943}\over{27414111994644375\,2^{{{121}\over{2}}}\,\pi^{12}}}$} 
  } 
\\ 
 \hline
decay & width & ratio  & cl. deg. & q. deg. & s.t. norm.
\\
$\, \begin{pmatrix} \left[ 1 , 2 \right] &\left[ 1 , 1 \right] &\left[ 1 , 1 \right] &\left[ 1 , 1 \right] \cr \left[ 1 , 1 \right] &\left[ 1 , 1 \right] &\left[  \right] &\left[  \right] \cr \left[  \right] &\left[  \right] &\left[ 1 , 1 \right] &\left[  \right] \cr \ifx\endpmatrix\undefined}\else\end{pmatrix}\fi  $ 
& $ 4.3964e-24 $ 
& [$ 2.9323e-1 $]  
& $ 6 $  
& $ 6 $ 
& $ 1 $  
\\
\multicolumn{6}{ c }{ \makecell{ $+{{23213131\,\log 4}\over{41247931725\,2^{{{61}\over{2}}}\,\pi^{12}}}$  $-\left({{1231993650431834471586502331281}\over{11489868616131633853125\,2^{{{135}\over{2}}}\,\pi^{12}}}\right)$ \\} 
} 
\\
\hline
$\, \begin{pmatrix} \left[ 1 , 2 \right] &\left[ 1 , 1 \right] &\left[ 1 , 1 \right] &\left[ 1 , 1 \right] \cr \left[  \right] &\left[ 1 , 1 \right] &\left[ 1 , 1 \right] &\left[  \right] \cr \left[ 1 , 1 \right] &\left[  \right] &\left[  \right] &\left[  \right] \cr \ifx\endpmatrix\undefined}\else\end{pmatrix}\fi  $ 
& $ 4.7600e-24 $ 
& [$ 1.5874e-1 $]  
& $ 3 $  
& $ 3 $ 
& $ 1 $  
\\
\multicolumn{6}{ c }{ \makecell{ $+{{134987603\,\log 4}\over{206239658625\,2^{{{61}\over{2}}}\,\pi^{12}}}$  $-\left({{9881681764056799555076086021}\over{79240473214700923125\,2^{{{135}\over{2}}}\,\pi^{12}}}\right)$ \\} 
} 
\\
\hline
$\, \begin{pmatrix} \left[ 1 , 2 \right] &\left[ 1 , 1 \right] &\left[ 1 , 1 \right] &\left[ 1 , 1 \right] \cr \left[  \right] &\left[ 1 , 1 \right] &\left[ 1 , 1 \right] &\left[  \right] \cr \left[  \right] &\left[  \right] &\left[  \right] &\left[ 1 , 1 \right] \cr \ifx\endpmatrix\undefined}\else\end{pmatrix}\fi  $ 
& $ 4.1089e-24 $ 
& [$ 1.3703e-1 $]  
& $ 3 $  
& $ 3 $ 
& $ 1 $  
\\
\multicolumn{6}{ c }{ \makecell{ $+{{15407059\,\log 4}\over{41247931725\,2^{{{61}\over{2}}}\,\pi^{12}}}$  $-\left({{28196492855452137128593270637}\over{396202366073504615625\,2^{{{135}\over{2}}}\,\pi^{12}}}\right)$ \\} 
} 
\\
\hline
\end{longtable}
\newpage
{\hskip-10em}
\begin{longtable}[l]{m{0.4\textwidth} m{0.16\textwidth} m{0.18\textwidth} m{0.10\textwidth} m{0.10\textwidth} m{0.04\textwidth} }
\caption{$N= 5$ state 3  }
\label{N=5_state_3}
\\\hline\hline
State & & Total width & Channels & &
\endfirsthead
\\
\multirow{2}{0.3\textwidth}{$\, [[1,2],[1,2],[1,1]] $ } 
& & $1.1188e-22$ 
& $415$  
& &
\\
& \multicolumn{5}{c}{
\makecell{${{30616125747596510087}\over{22697705629974375\,2^{{{127}\over{2}}}\,\pi^{12}}}$} 
  } 
\\ 
 \hline
decay & width & ratio  & cl. deg. & q. deg. & s.t. norm.
\\
$\, \begin{pmatrix} \left[ 1 , 2 \right] &\left[ 1 , 2 \right] &\left[ 1 , 1 \right] \cr \left[ 1 , 1 \right] &\left[  \right] &\left[ 1 , 1 \right] \cr \left[  \right] &\left[ 1 , 1 \right] &\left[  \right] \cr \ifx\endpmatrix\undefined}\else\end{pmatrix}\fi  $ 
& $ 1.0778e-23 $ 
& [$ 1.9267e-1 $]  
& $ 2 $  
& $ 2 $ 
& $ 1 $  
\\
\multicolumn{6}{ c }{ \makecell{ $+{{13315633\,\log 4}\over{22915517625\,2^{{{59}\over{2}}}\,\pi^{12}}}$  $-\left({{15596167727288231039066705249}\over{8804497023855658125\,2^{{{141}\over{2}}}\,\pi^{12}}}\right)$ \\} 
} 
\\
\hline
$\, \begin{pmatrix} \left[ 1 , 2 \right] &\left[ 1 , 2 \right] &\left[ 1 , 1 \right] \cr \left[ 1 , 1 \right] &\left[ 1 , 1 \right] &\left[  \right] \cr \left[ 1 , 1 \right] &\left[  \right] &\left[  \right] \cr \ifx\endpmatrix\undefined}\else\end{pmatrix}\fi  $ 
& $ 8.0684e-24 $ 
& [$ 1.4423e-1 $]  
& $ 2 $  
& $ 2 $ 
& $ 1 $  
\\
\multicolumn{6}{ c }{ \makecell{ $+{{278794489\,\log 4}\over{206239658625\,2^{{{61}\over{2}}}\,\pi^{12}}}$  $-\left({{189395524468422809371164429774131}\over{11489868616131633853125\,2^{{{147}\over{2}}}\,\pi^{12}}}\right)$ \\} 
} 
\\
\hline
$\, \begin{pmatrix} \left[ 1 , 2 \right] &\left[ 1 , 2 \right] &\left[ 1 , 1 \right] \cr \left[ 1 , 1 \right] &\left[  \right] &\left[ 1 , 1 \right] \cr \left[ 1 , 1 \right] &\left[  \right] &\left[  \right] \cr \ifx\endpmatrix\undefined}\else\end{pmatrix}\fi  $ 
& $ 7.6111e-24 $ 
& [$ 1.3605e-1 $]  
& $ 2 $  
& $ 2 $ 
& $ 1 $  
\\
\multicolumn{6}{ c }{ \makecell{ $+{{92573471\,\log 4}\over{206239658625\,2^{{{59}\over{2}}}\,\pi^{12}}}$  $-\left({{125776923408167357320765162199627}\over{11489868616131633853125\,2^{{{147}\over{2}}}\,\pi^{12}}}\right)$ \\} 
} 
\\
\hline
\end{longtable}
\newpage
{\hskip-10em}
\begin{longtable}[l]{m{0.4\textwidth} m{0.16\textwidth} m{0.18\textwidth} m{0.10\textwidth} m{0.10\textwidth} m{0.04\textwidth} }
\caption{$N= 5$ state 4  }
\label{N=5_state_4}
\\\hline\hline
State & & Total width & Channels & &
\endfirsthead
\\
\multirow{2}{0.3\textwidth}{$\, [[1,3],[1,1],[1,1]] $ } 
& & $1.3006e-22$ 
& $415$  
& &
\\
& \multicolumn{5}{c}{
\makecell{${{413738481831562699313}\over{2110886623587616875\,2^{{{121}\over{2}}}\,\pi^{12}}}$} 
  } 
\\ 
 \hline
decay & width & ratio  & cl. deg. & q. deg. & s.t. norm.
\\
$\, \begin{pmatrix} \left[ 1 , 3 \right] &\left[ 1 , 1 \right] &\left[ 1 , 1 \right] \cr \left[ 1 , 1 \right] &\left[ 1 , 1 \right] &\left[  \right] \cr \left[ 1 , 1 \right] &\left[  \right] &\left[  \right] \cr \ifx\endpmatrix\undefined}\else\end{pmatrix}\fi  $ 
& $ 1.4799e-23 $ 
& [$ 2.2757e-1 $]  
& $ 2 $  
& $ 2 $ 
& $ 1 $  
\\
\multicolumn{6}{ c }{ \makecell{ $+{{35490541\,\log 4}\over{17677685025\,2^{{{61}\over{2}}}\,\pi^{12}}}$  $-\left({{1318516595997668267086436011141}\over{6893921169678980311875\,2^{{{133}\over{2}}}\,\pi^{12}}}\right)$ \\} 
} 
\\
\hline
$\, \begin{pmatrix} \left[ 1 , 3 \right] &\left[ 1 , 1 \right] &\left[ 1 , 1 \right] \cr \left[ 1 , 1 \right] &\left[ 1 , 1 \right] &\left[  \right] \cr \left[  \right] &\left[  \right] &\left[ 1 , 1 \right] \cr \ifx\endpmatrix\undefined}\else\end{pmatrix}\fi  $ 
& $ 1.3833e-23 $ 
& [$ 2.1271e-1 $]  
& $ 2 $  
& $ 2 $ 
& $ 1 $  
\\
\multicolumn{6}{ c }{ \makecell{ $+{{601797199\,\log 4}\over{618718975875\,2^{{{61}\over{2}}}\,\pi^{12}}}$  $-\left({{6387811422859382791631790383969}\over{34469605848394901559375\,2^{{{135}\over{2}}}\,\pi^{12}}}\right)$ \\} 
} 
\\
\hline
$\, \begin{pmatrix} \left[ 1 , 3 \right] &\left[ 1 , 1 \right] &\left[ 1 , 1 \right] \cr \left[  \right] &\left[ 1 , 1 \right] &\left[ 1 , 1 \right] \cr \left[ 1 , 1 \right] &\left[  \right] &\left[  \right] \cr \ifx\endpmatrix\undefined}\else\end{pmatrix}\fi  $ 
& $ 1.4848e-23 $ 
& [$ 1.1416e-1 $]  
& $ 1 $  
& $ 1 $ 
& $ 1 $  
\\
\multicolumn{6}{ c }{ \makecell{ $+{{245244557\,\log 4}\over{206239658625\,2^{{{61}\over{2}}}\,\pi^{12}}}$  $-\left({{17952877295849081909077417051}\over{79240473214700923125\,2^{{{135}\over{2}}}\,\pi^{12}}}\right)$ \\} 
} 
\\
\hline
\end{longtable}
\newpage
{\hskip-10em}
\begin{longtable}[l]{m{0.4\textwidth} m{0.16\textwidth} m{0.18\textwidth} m{0.10\textwidth} m{0.10\textwidth} m{0.04\textwidth} }
\caption{$N= 5$ state 5  }
\label{N=5_state_5}
\\\hline\hline
State & & Total width & Channels & &
\endfirsthead
\\
\multirow{2}{0.3\textwidth}{$\, [[1,3],[1,2]] $ } 
& & $1.5573e-22$ 
& $309$  
& &
\\
& \multicolumn{5}{c}{
\makecell{${{3963273782637216944497}\over{2110886623587616875\,2^{{{127}\over{2}}}\,\pi^{12}}}$} 
  } 
\\ 
 \hline
decay & width & ratio  & cl. deg. & q. deg. & s.t. norm.
\\
$\, \begin{pmatrix} \left[ 1 , 3 \right] &\left[ 1 , 2 \right] \cr \left[ 1 , 1 \right] &\left[ 1 , 1 \right] \cr \left[ 1 , 1 \right] &\left[  \right] \cr \ifx\endpmatrix\undefined}\else\end{pmatrix}\fi  $ 
& $ 3.3526e-23 $ 
& [$ 2.1528e-1 $]  
& $ 1 $  
& $ 1 $ 
& $ 1 $  
\\
\multicolumn{6}{ c }{ \makecell{ $+{{414524533\,\log 4}\over{123743795175\,2^{{{61}\over{2}}}\,\pi^{12}}}$  $-\left({{56320173711671969725504046121907}\over{1378784233935796062375\,2^{{{147}\over{2}}}\,\pi^{12}}}\right)$ \\} 
} 
\\
\hline
$\, \begin{pmatrix} \left[ 1 , 3 \right] &\left[ 1 , 2 \right] \cr \left[ 1 , 1 \right] &\left[ 1 , 1 \right] \cr \left[  \right] &\left[ 1 , 1 \right] \cr \ifx\endpmatrix\undefined}\else\end{pmatrix}\fi  $ 
& $ 2.5596e-23 $ 
& [$ 1.6435e-1 $]  
& $ 1 $  
& $ 1 $ 
& $ 1 $  
\\
\multicolumn{6}{ c }{ \makecell{ $+{{480491273\,\log 4}\over{206239658625\,2^{{{61}\over{2}}}\,\pi^{12}}}$  $-\left({{326414035937152560491007812528353}\over{11489868616131633853125\,2^{{{147}\over{2}}}\,\pi^{12}}}\right)$ \\} 
} 
\\
\hline
$\, \begin{pmatrix} \left[ 1 , 3 \right] &\left[ 1 , 2 \right] \cr \left[ 1 , 2 \right] &\left[  \right] \cr \left[  \right] &\left[ 1 , 1 \right] \cr \ifx\endpmatrix\undefined}\else\end{pmatrix}\fi  $ 
& $ 1.7443e-23 $ 
& [$ 1.1200e-1 $]  
& $ 1 $  
& $ 1 $ 
& $ 1 $  
\\
\multicolumn{6}{ c }{ \makecell{ $+{{61627513\,\log 4}\over{47593767375\,2^{{{59}\over{2}}}\,\pi^{12}}}$  $-\left({{1046646055186904797642381778207}\over{530301628436844639375\,2^{{{139}\over{2}}}\,\pi^{12}}}\right)$ \\} 
} 
\\
\hline
\end{longtable}
%
%
%

\newpage
\section{Level $6$}
\label{app:N=6}

{\hskip-10em}
\begin{longtable}[l]{m{0.4\textwidth} m{0.16\textwidth} m{0.18\textwidth} m{0.10\textwidth} m{0.10\textwidth} m{0.04\textwidth} }
\caption{$N= 6$ state 1  }
\label{N=6_state_1}
\\\hline\hline
State & & Total width & Channels & &
\endfirsthead
\\
\multirow{2}{0.3\textwidth}{$\, [[1,1],[1,1],[1,1],[1,1],[1,1],[1,1]] $ } 
& & $8.7975e-21$ 
& $602$  
& &
\\
& \multicolumn{5}{c}{
\makecell{ $+{{413678753\,5^{{{33}\over{2}}}}\over{3026167063575207552\,10^{{{45}\over{2}}}\,\pi^{12}}}$ $+{{4228946986401031982162419}\over{16446560128126128\,10^{{{45}\over{2}}}\,\pi^{12}}}$} 
  } 
\\ 
 \hline
decay & width & ratio  & cl. deg. & q. deg. & s.t. norm.
\\
$\, \begin{pmatrix} \left[ 1 , 1 \right] &\left[ 1 , 1 \right] &\left[ 1 , 1 \right] &\left[ 1 , 1 \right] &\left[ 1 , 1 \right] &\left[ 1 , 1 \right] \cr \left[ 1 , 1 \right] &\left[  \right] &\left[  \right] &\left[  \right] &\left[  \right] &\left[  \right] \cr \left[  \right] &\left[ 1 , 1 \right] &\left[  \right] &\left[  \right] &\left[  \right] &\left[  \right] \cr \ifx\endpmatrix\undefined}\else\end{pmatrix}\fi  $ 
&&&&&\\ 
& $ 4.6425e-22 $ 
& [$ 7.9155e-1 $]  
& $ 30 $  
& $ 15 $ 
& $ 1 $  
\\
\multicolumn{6}{ c }{ \makecell{${{20026455078125\,\sqrt{10}}\over{147590304194636078025145319424\,\pi^{12}}}$} 
} 
\\
\hline
$\, \begin{pmatrix} \left[ 1 , 1 \right] &\left[ 1 , 1 \right] &\left[ 1 , 1 \right] &\left[ 1 , 1 \right] &\left[ 1 , 1 \right] &\left[ 1 , 1 \right] &\left[  \right] \cr \left[ 1 , 1 \right] &\left[  \right] &\left[  \right] &\left[  \right] &\left[  \right] &\left[  \right] &\left[  \right] \cr \left[  \right] &\left[  \right] &\left[  \right] &\left[  \right] &\left[  \right] &\left[  \right] &\left[ 1 , 1 \right] \cr \ifx\endpmatrix\undefined}\else\end{pmatrix}\fi  $ 
&&&&&\\ 
& $ 1.5138e-23 $ 
& [$ 1.8584e-1 $]  
& $ 216 $  
& $ 108 $ 
& $ 1 $  
\\
\multicolumn{6}{ c }{ \makecell{${{356201171875\,\sqrt{10}}\over{80503802287983315286442901504\,\pi^{12}}}$} 
} 
\\
\hline
$\, \begin{pmatrix} \left[ 1 , 1 \right] &\left[ 1 , 1 \right] &\left[ 1 , 1 \right] &\left[ 1 , 1 \right] &\left[ 1 , 1 \right] &\left[ 1 , 1 \right] \cr \left[ 1 , 1 \right] &\left[  \right] &\left[  \right] &\left[  \right] &\left[  \right] &\left[  \right] \cr \left[ 1 , 1 \right] &\left[  \right] &\left[  \right] &\left[  \right] &\left[  \right] &\left[  \right] \cr \ifx\endpmatrix\undefined}\else\end{pmatrix}\fi  $ 
&&&&&\\ 
& $ 2.4790e-23 $ 
& [$ 8.4536e-3 $]  
& $ 6 $  
& $ 6 $ 
& $ {{1}\over{2}} $  
\\
\multicolumn{6}{ c }{ \makecell{${{8837890625\,\sqrt{10}}\over{1219754580120959322521862144\,\pi^{12}}}$} 
} 
\\
\hline
\end{longtable}
\newpage
{\hskip-10em}
\begin{longtable}[l]{m{0.4\textwidth} m{0.16\textwidth} m{0.18\textwidth} m{0.10\textwidth} m{0.10\textwidth} m{0.04\textwidth} }
\caption{$N= 6$ state 2  }
\label{N=6_state_2}
\\\hline\hline
State & & Total width & Channels & &
\endfirsthead
\\
\multirow{2}{0.3\textwidth}{$\, [[1,2],[1,1],[1,1],[1,1],[1,1]] $ } 
& & $9.5349e-21$ 
& $1380$  
& &
\\
& \multicolumn{5}{c}{
\makecell{ $+{{1514902789\,5^{{{33}\over{2}}}}\over{48809146186696896\,10^{{{49}\over{2}}}\,\pi^{12}}}$ $+{{9566097453968103187998941}\over{343258514470872\,10^{{{49}\over{2}}}\,\pi^{12}}}$} 
  } 
\\ 
 \hline
decay & width & ratio  & cl. deg. & q. deg. & s.t. norm.
\\
$\, \begin{pmatrix} \left[ 1 , 2 \right] &\left[ 1 , 1 \right] &\left[ 1 , 1 \right] &\left[ 1 , 1 \right] &\left[ 1 , 1 \right] \cr \left[  \right] &\left[ 1 , 1 \right] &\left[  \right] &\left[  \right] &\left[  \right] \cr \left[ 1 , 1 \right] &\left[  \right] &\left[  \right] &\left[  \right] &\left[  \right] \cr \ifx\endpmatrix\undefined}\else\end{pmatrix}\fi  $ 
&&&&&\\ 
& $ 1.0085e-21 $ 
& [$ 4.2307e-1 $]  
& $ 8 $  
& $ 4 $ 
& $ 1 $  
\\
\multicolumn{6}{ c }{ \makecell{${{87006318359375\,\sqrt{10}}\over{295180608389272156050290638848\,\pi^{12}}}$} 
} 
\\
\hline
$\, \begin{pmatrix} \left[ 1 , 2 \right] &\left[ 1 , 1 \right] &\left[ 1 , 1 \right] &\left[ 1 , 1 \right] &\left[ 1 , 1 \right] \cr \left[  \right] &\left[ 1 , 1 \right] &\left[  \right] &\left[  \right] &\left[  \right] \cr \left[  \right] &\left[  \right] &\left[ 1 , 1 \right] &\left[  \right] &\left[  \right] \cr \ifx\endpmatrix\undefined}\else\end{pmatrix}\fi  $ 
&&&&&\\ 
& $ 4.7807e-22 $ 
& [$ 3.0083e-1 $]  
& $ 12 $  
& $ 6 $ 
& $ 1 $  
\\
\multicolumn{6}{ c }{ \makecell{${{4582837890625\,\sqrt{10}}\over{32797845376585795116698959872\,\pi^{12}}}$} 
} 
\\
\hline
$\, \begin{pmatrix} \left[ 1 , 2 \right] &\left[ 1 , 1 \right] &\left[ 1 , 1 \right] &\left[ 1 , 1 \right] &\left[ 1 , 1 \right] &\left[  \right] \cr \left[  \right] &\left[ 1 , 1 \right] &\left[  \right] &\left[  \right] &\left[  \right] &\left[  \right] \cr \left[  \right] &\left[  \right] &\left[  \right] &\left[  \right] &\left[  \right] &\left[ 1 , 1 \right] \cr \ifx\endpmatrix\undefined}\else\end{pmatrix}\fi  $ 
&&&&&\\ 
& $ 1.7354e-23 $ 
& [$ 1.3832e-1 $]  
& $ 152 $  
& $ 76 $ 
& $ 1 $  
\\
\multicolumn{6}{ c }{ \makecell{${{15595703125\,\sqrt{10}}\over{3074798004054918292190527488\,\pi^{12}}}$} 
} 
\\
\hline
\end{longtable}
\newpage
{\hskip-10em}
\begin{longtable}[l]{m{0.4\textwidth} m{0.16\textwidth} m{0.18\textwidth} m{0.10\textwidth} m{0.10\textwidth} m{0.04\textwidth} }
\caption{$N= 6$ state 3  }
\label{N=6_state_3}
\\\hline\hline
State & & Total width & Channels & &
\endfirsthead
\\
\multirow{2}{0.3\textwidth}{$\, [[1,2],[1,2],[1,1],[1,1]] $ } 
& & $1.0266e-20$ 
& $1638$  
& &
\\
& \multicolumn{5}{c}{
\makecell{ $+{{244673638481\,5^{{{33}\over{2}}}}\over{504361177262534592\,10^{{{51}\over{2}}}\,\pi^{12}}}$ $+{{113502198488307450267000153437}\over{378270882946900944\,10^{{{51}\over{2}}}\,\pi^{12}}}$} 
  } 
\\ 
 \hline
decay & width & ratio  & cl. deg. & q. deg. & s.t. norm.
\\
$\, \begin{pmatrix} \left[ 1 , 2 \right] &\left[ 1 , 2 \right] &\left[ 1 , 1 \right] &\left[ 1 , 1 \right] \cr \left[  \right] &\left[  \right] &\left[ 1 , 1 \right] &\left[  \right] \cr \left[ 1 , 1 \right] &\left[  \right] &\left[  \right] &\left[  \right] \cr \ifx\endpmatrix\undefined}\else\end{pmatrix}\fi  $ 
&&&&&\\ 
& $ 1.0286e-21 $ 
& [$ 4.0078e-1 $]  
& $ 8 $  
& $ 4 $ 
& $ 1 $  
\\
\multicolumn{6}{ c }{ \makecell{${{118325083984375\,\sqrt{10}}\over{393574144519029541400387518464\,\pi^{12}}}$} 
} 
\\
\hline
$\, \begin{pmatrix} \left[ 1 , 2 \right] &\left[ 1 , 2 \right] &\left[ 1 , 1 \right] &\left[ 1 , 1 \right] \cr \left[ 1 , 1 \right] &\left[  \right] &\left[  \right] &\left[  \right] \cr \left[  \right] &\left[ 1 , 1 \right] &\left[  \right] &\left[  \right] \cr \ifx\endpmatrix\undefined}\else\end{pmatrix}\fi  $ 
&&&&&\\ 
& $ 2.1853e-21 $ 
& [$ 2.1286e-1 $]  
& $ 2 $  
& $ 1 $ 
& $ 1 $  
\\
\multicolumn{6}{ c }{ \makecell{${{2327580078125\,\sqrt{10}}\over{3644205041842866124077662208\,\pi^{12}}}$} 
} 
\\
\hline
$\, \begin{pmatrix} \left[ 1 , 2 \right] &\left[ 1 , 2 \right] &\left[ 1 , 1 \right] &\left[ 1 , 1 \right] &\left[  \right] \cr \left[ 1 , 1 \right] &\left[  \right] &\left[  \right] &\left[  \right] &\left[  \right] \cr \left[  \right] &\left[  \right] &\left[  \right] &\left[  \right] &\left[ 1 , 1 \right] \cr \ifx\endpmatrix\undefined}\else\end{pmatrix}\fi  $ 
&&&&&\\ 
& $ 3.8031e-23 $ 
& [$ 1.4818e-1 $]  
& $ 80 $  
& $ 40 $ 
& $ 1 $  
\\
\multicolumn{6}{ c }{ \makecell{${{1640552734375\,\sqrt{10}}\over{147590304194636078025145319424\,\pi^{12}}}$} 
} 
\\
\hline
\end{longtable}
\newpage
{\hskip-10em}
\begin{longtable}[l]{m{0.4\textwidth} m{0.16\textwidth} m{0.18\textwidth} m{0.10\textwidth} m{0.10\textwidth} m{0.04\textwidth} }
\caption{$N= 6$ state 4  }
\label{N=6_state_4}
\\\hline\hline
State & & Total width & Channels & &
\endfirsthead
\\
\multirow{2}{0.3\textwidth}{$\, [[1,2],[1,2],[1,2]] $ } 
& & $1.0976e-20$ 
& $574$  
& &
\\
& \multicolumn{5}{c}{
\makecell{ $+{{1749104285857\,5^{{{33}\over{2}}}}\over{26545325119080768\,10^{{{55}\over{2}}}\,\pi^{12}}}$ $+{{1011242734226719357487293669603}\over{31522573578908412\,10^{{{55}\over{2}}}\,\pi^{12}}}$} 
  } 
\\ 
 \hline
decay & width & ratio  & cl. deg. & q. deg. & s.t. norm.
\\
$\, \begin{pmatrix} \left[ 1 , 2 \right] &\left[ 1 , 2 \right] &\left[ 1 , 2 \right] \cr \left[ 1 , 1 \right] &\left[  \right] &\left[  \right] \cr \left[  \right] &\left[ 1 , 1 \right] &\left[  \right] \cr \ifx\endpmatrix\undefined}\else\end{pmatrix}\fi  $ 
&&&&&\\ 
& $ 2.2077e-21 $ 
& [$ 6.0341e-1 $]  
& $ 6 $  
& $ 3 $ 
& $ 1 $  
\\
\multicolumn{6}{ c }{ \makecell{${{253954155078125\,\sqrt{10}}\over{393574144519029541400387518464\,\pi^{12}}}$} 
} 
\\
\hline
$\, \begin{pmatrix} \left[ 1 , 2 \right] &\left[ 1 , 2 \right] &\left[ 1 , 2 \right] &\left[  \right] \cr \left[ 1 , 1 \right] &\left[  \right] &\left[  \right] &\left[  \right] \cr \left[  \right] &\left[  \right] &\left[  \right] &\left[ 1 , 1 \right] \cr \ifx\endpmatrix\undefined}\else\end{pmatrix}\fi  $ 
&&&&&\\ 
& $ 4.1831e-23 $ 
& [$ 2.4010e-1 $]  
& $ 126 $  
& $ 63 $ 
& $ 1 $  
\\
\multicolumn{6}{ c }{ \makecell{${{209214453125\,\sqrt{10}}\over{17111919326914327886973370368\,\pi^{12}}}$} 
} 
\\
\hline
$\, \begin{pmatrix} \left[ 1 , 2 \right] &\left[ 1 , 2 \right] &\left[ 1 , 2 \right] \cr \left[ 1 , 1 \right] &\left[  \right] &\left[  \right] \cr \left[ 1 , 1 \right] &\left[  \right] &\left[  \right] \cr \ifx\endpmatrix\undefined}\else\end{pmatrix}\fi  $ 
&&&&&\\ 
& $ 9.6901e-22 $ 
& [$ 1.3243e-1 $]  
& $ 3 $  
& $ 3 $ 
& $ {{1}\over{2}} $  
\\
\multicolumn{6}{ c }{ \makecell{${{111468300265625\,\sqrt{10}}\over{393574144519029541400387518464\,\pi^{12}}}$} 
} 
\\
\hline
\end{longtable}
\newpage
{\hskip-10em}
\begin{longtable}[l]{m{0.4\textwidth} m{0.16\textwidth} m{0.18\textwidth} m{0.10\textwidth} m{0.10\textwidth} m{0.04\textwidth} }
\caption{$N= 6$ state 5  }
\label{N=6_state_5}
\\\hline\hline
State & & Total width & Channels & &
\endfirsthead
\\
\multirow{2}{0.3\textwidth}{$\, [[1,3],[1,1],[1,1],[1,1]] $ } 
& & $1.1010e-20$ 
& $1347$  
& &
\\
& \multicolumn{5}{c}{
\makecell{ $+{{99518084077\,5^{{{33}\over{2}}}}\over{1513083531787603776\,10^{{{49}\over{2}}}\,\pi^{12}}}$ $+{{6086181022686901010788685773}\over{189135441473450472\,10^{{{49}\over{2}}}\,\pi^{12}}}$} 
  } 
\\ 
 \hline
decay & width & ratio  & cl. deg. & q. deg. & s.t. norm.
\\
$\, \begin{pmatrix} \left[ 1 , 3 \right] &\left[ 1 , 1 \right] &\left[ 1 , 1 \right] &\left[ 1 , 1 \right] \cr \left[  \right] &\left[ 1 , 1 \right] &\left[  \right] &\left[  \right] \cr \left[ 1 , 1 \right] &\left[  \right] &\left[  \right] &\left[  \right] \cr \ifx\endpmatrix\undefined}\else\end{pmatrix}\fi  $ 
&&&&&\\ 
& $ 1.6826e-21 $ 
& [$ 4.5850e-1 $]  
& $ 6 $  
& $ 3 $ 
& $ 1 $  
\\
\multicolumn{6}{ c }{ \makecell{${{4949005859375\,\sqrt{10}}\over{10062975285997914410805362688\,\pi^{12}}}$} 
} 
\\
\hline
$\, \begin{pmatrix} \left[ 1 , 3 \right] &\left[ 1 , 1 \right] &\left[ 1 , 1 \right] &\left[ 1 , 1 \right] \cr \left[ 1 , 1 \right] &\left[  \right] &\left[  \right] &\left[  \right] \cr \left[ 1 , 1 \right] &\left[  \right] &\left[  \right] &\left[  \right] \cr \ifx\endpmatrix\undefined}\else\end{pmatrix}\fi  $ 
&&&&&\\ 
& $ 3.2001e-21 $ 
& [$ 1.4533e-1 $]  
& $ 1 $  
& $ 1 $ 
& $ {{1}\over{2}} $  
\\
\multicolumn{6}{ c }{ \makecell{${{9101810546875\,\sqrt{10}}\over{9731228847997983166493097984\,\pi^{12}}}$} 
} 
\\
\hline
$\, \begin{pmatrix} \left[ 1 , 3 \right] &\left[ 1 , 1 \right] &\left[ 1 , 1 \right] &\left[ 1 , 1 \right] \cr \left[  \right] &\left[ 1 , 1 \right] &\left[  \right] &\left[  \right] \cr \left[  \right] &\left[  \right] &\left[ 1 , 1 \right] &\left[  \right] \cr \ifx\endpmatrix\undefined}\else\end{pmatrix}\fi  $ 
&&&&&\\ 
& $ 5.0572e-22 $ 
& [$ 1.3780e-1 $]  
& $ 6 $  
& $ 3 $ 
& $ 1 $  
\\
\multicolumn{6}{ c }{ \makecell{${{43630802734375\,\sqrt{10}}\over{295180608389272156050290638848\,\pi^{12}}}$} 
} 
\\
\hline
\end{longtable}
%
%
%

\newpage
\section{Level $7$}
\label{app:N=7}

{\hskip-10em}
\begin{longtable}[l]{m{0.4\textwidth} m{0.16\textwidth} m{0.18\textwidth} m{0.10\textwidth} m{0.10\textwidth} m{0.04\textwidth} }
\caption{$N= 7$ state 1  }
\label{N=7_state_1}
\\\hline\hline
State & & Total width & Channels & &
\endfirsthead
\\
\multirow{2}{0.3\textwidth}{$\, [[1,1],[1,1],[1,1],[1,1],[1,1],[1,1],[1,1]] $ } 
& & $2.7415e-21$ 
& $1698$  
& &
\\
& \multicolumn{5}{c}{
\makecell{${{1127823499734097220348667859}\over{101098902522355436371312640000\,3^{{{53}\over{2}}}\,\pi^{12}}}$} 
  } 
\\ 
 \hline
decay & width & ratio  & cl. deg. & q. deg. & s.t. norm.
\\
$\, \begin{pmatrix} \left[ 1 , 1 \right] &\left[ 1 , 1 \right] &\left[ 1 , 1 \right] &\left[ 1 , 1 \right] &\left[ 1 , 1 \right] &\left[ 1 , 1 \right] &\left[ 1 , 1 \right] \cr \left[ 1 , 1 \right] &\left[ 1 , 1 \right] &\left[  \right] &\left[  \right] &\left[  \right] &\left[  \right] &\left[  \right] \cr \left[  \right] &\left[  \right] &\left[ 1 , 1 \right] &\left[  \right] &\left[  \right] &\left[  \right] &\left[  \right] \cr \ifx\endpmatrix\undefined}\else\end{pmatrix}\fi  $ 
&&&&&\\ 
& $ 1.9988e-23 $ 
& [$ 7.6556e-1 $]  
& $ 105 $  
& $ 105 $ 
& $ 1 $  
\\
\multicolumn{6}{ c }{ \makecell{ $+{{278426273\,\log 6}\over{10631941225185280000\,3^{{{17}\over{2}}}\,\pi^{12}}}$ $-\left({{6654102634062839364274285236629731}\over{1103057935530857048276451589095424000\,3^{{{51}\over{2}}}\,\pi^{12}}}\right)$} 
} 
\\
\hline
$\, \begin{pmatrix} \left[ 1 , 1 \right] &\left[ 1 , 1 \right] &\left[ 1 , 1 \right] &\left[ 1 , 1 \right] &\left[ 1 , 1 \right] &\left[ 1 , 1 \right] &\left[ 1 , 1 \right] &\left[  \right] \cr \left[ 1 , 1 \right] &\left[ 1 , 1 \right] &\left[  \right] &\left[  \right] &\left[  \right] &\left[  \right] &\left[  \right] &\left[  \right] \cr \left[  \right] &\left[  \right] &\left[  \right] &\left[  \right] &\left[  \right] &\left[  \right] &\left[  \right] &\left[ 1 , 1 \right] \cr \ifx\endpmatrix\undefined}\else\end{pmatrix}\fi  $ 
&&&&&\\ 
& $ 6.6513e-25 $ 
& [$ 8.6615e-2 $]  
& $ 357 $  
& $ 357 $ 
& $ 1 $  
\\
\multicolumn{6}{ c }{ \makecell{ $+{{764213\,\log 6}\over{10631941225185280000\,3^{{{17}\over{2}}}\,\pi^{12}}}$ $-\left({{3061910247496032238138411226749}\over{64885760913579826369203034652672000\,3^{{{53}\over{2}}}\,\pi^{12}}}\right)$} 
} 
\\
\hline
$\, \begin{pmatrix} \left[ 1 , 1 \right] &\left[ 1 , 1 \right] &\left[ 1 , 1 \right] &\left[ 1 , 1 \right] &\left[ 1 , 1 \right] &\left[ 1 , 1 \right] &\left[ 1 , 1 \right] &\left[  \right] \cr \left[ 1 , 1 \right] &\left[  \right] &\left[  \right] &\left[  \right] &\left[  \right] &\left[  \right] &\left[  \right] &\left[ 1 , 1 \right] \cr \left[  \right] &\left[ 1 , 1 \right] &\left[  \right] &\left[  \right] &\left[  \right] &\left[  \right] &\left[  \right] &\left[  \right] \cr \ifx\endpmatrix\undefined}\else\end{pmatrix}\fi  $ 
&&&&&\\ 
& $ 2.9950e-25 $ 
& [$ 7.8003e-2 $]  
& $ 714 $  
& $ 714 $ 
& $ 1 $  
\\
\multicolumn{6}{ c }{ \makecell{ $+{{7328401\,\log 6}\over{625408307363840000\,3^{{{17}\over{2}}}\,\pi^{12}}}$ $-\left({{175902067667267432653812919261147}\over{64885760913579826369203034652672000\,3^{{{51}\over{2}}}\,\pi^{12}}}\right)$} 
} 
\\
\hline
\end{longtable}
\newpage
{\hskip-10em}
\begin{longtable}[l]{m{0.4\textwidth} m{0.16\textwidth} m{0.18\textwidth} m{0.10\textwidth} m{0.10\textwidth} m{0.04\textwidth} }
\caption{$N= 7$ state 2  }
\label{N=7_state_2}
\\\hline\hline
State & & Total width & Channels & &
\endfirsthead
\\
\multirow{2}{0.3\textwidth}{$\, [[1,2],[1,1],[1,1],[1,1],[1,1],[1,1]] $ } 
& & $3.1250e-21$ 
& $4092$  
& &
\\
& \multicolumn{5}{c}{
\makecell{${{178957685532417566936506623767}\over{4690989077037292247628906496000\,3^{{{55}\over{2}}}\,\pi^{12}}}$} 
  } 
\\ 
 \hline
decay & width & ratio  & cl. deg. & q. deg. & s.t. norm.
\\
$\, \begin{pmatrix} \left[ 1 , 2 \right] &\left[ 1 , 1 \right] &\left[ 1 , 1 \right] &\left[ 1 , 1 \right] &\left[ 1 , 1 \right] &\left[ 1 , 1 \right] \cr \left[ 1 , 1 \right] &\left[ 1 , 1 \right] &\left[  \right] &\left[  \right] &\left[  \right] &\left[  \right] \cr \left[  \right] &\left[  \right] &\left[ 1 , 1 \right] &\left[  \right] &\left[  \right] &\left[  \right] \cr \ifx\endpmatrix\undefined}\else\end{pmatrix}\fi  $ 
&&&&&\\ 
& $ 4.3936e-23 $ 
& [$ 2.8119e-1 $]  
& $ 20 $  
& $ 20 $ 
& $ 1 $  
\\
\multicolumn{6}{ c }{ \makecell{ $+{{376467451\,\log 6}\over{21263882450370560000\,3^{{{15}\over{2}}}\,\pi^{12}}}$ $-\left({{121416279242946637582345589416166147}\over{1103057935530857048276451589095424000\,3^{{{55}\over{2}}}\,\pi^{12}}}\right)$} 
} 
\\
\hline
$\, \begin{pmatrix} \left[ 1 , 2 \right] &\left[ 1 , 1 \right] &\left[ 1 , 1 \right] &\left[ 1 , 1 \right] &\left[ 1 , 1 \right] &\left[ 1 , 1 \right] \cr \left[  \right] &\left[ 1 , 1 \right] &\left[ 1 , 1 \right] &\left[  \right] &\left[  \right] &\left[  \right] \cr \left[  \right] &\left[  \right] &\left[  \right] &\left[ 1 , 1 \right] &\left[  \right] &\left[  \right] \cr \ifx\endpmatrix\undefined}\else\end{pmatrix}\fi  $ 
&&&&&\\ 
& $ 2.2855e-23 $ 
& [$ 2.1940e-1 $]  
& $ 30 $  
& $ 30 $ 
& $ 1 $  
\\
\multicolumn{6}{ c }{ \makecell{ $+{{15721061\,\log 6}\over{4252776490074112000\,3^{{{13}\over{2}}}\,\pi^{12}}}$ $-\left({{5074462227205403259585415824315317}\over{220611587106171409655290317819084800\,3^{{{53}\over{2}}}\,\pi^{12}}}\right)$} 
} 
\\
\hline
$\, \begin{pmatrix} \left[ 1 , 2 \right] &\left[ 1 , 1 \right] &\left[ 1 , 1 \right] &\left[ 1 , 1 \right] &\left[ 1 , 1 \right] &\left[ 1 , 1 \right] \cr \left[  \right] &\left[ 1 , 1 \right] &\left[ 1 , 1 \right] &\left[  \right] &\left[  \right] &\left[  \right] \cr \left[ 1 , 1 \right] &\left[  \right] &\left[  \right] &\left[  \right] &\left[  \right] &\left[  \right] \cr \ifx\endpmatrix\undefined}\else\end{pmatrix}\fi  $ 
&&&&&\\ 
& $ 4.5316e-23 $ 
& [$ 1.4501e-1 $]  
& $ 10 $  
& $ 10 $ 
& $ 1 $  
\\
\multicolumn{6}{ c }{ \makecell{ $+{{77140709\,\log 6}\over{4252776490074112000\,3^{{{15}\over{2}}}\,\pi^{12}}}$ $-\left({{548783761216259287095154143145579}\over{1622144022839495659230075866316800\,3^{{{57}\over{2}}}\,\pi^{12}}}\right)$} 
} 
\\
\hline
\end{longtable}
\newpage
{\hskip-10em}
\begin{longtable}[l]{m{0.4\textwidth} m{0.16\textwidth} m{0.18\textwidth} m{0.10\textwidth} m{0.10\textwidth} m{0.04\textwidth} }
\caption{$N= 7$ state 3  }
\label{N=7_state_3}
\\\hline\hline
State & & Total width & Channels & &
\endfirsthead
\\
\multirow{2}{0.3\textwidth}{$\, [[1,2],[1,2],[1,1],[1,1],[1,1]] $ } 
& & $3.5383e-21$ 
& $5361$  
& &
\\
& \multicolumn{5}{c}{
\makecell{${{389037859377003123487165704121}\over{3002233009303867038482500157440\,3^{{{57}\over{2}}}\,\pi^{12}}}$} 
  } 
\\ 
 \hline
decay & width & ratio  & cl. deg. & q. deg. & s.t. norm.
\\
$\, \begin{pmatrix} \left[ 1 , 2 \right] &\left[ 1 , 2 \right] &\left[ 1 , 1 \right] &\left[ 1 , 1 \right] &\left[ 1 , 1 \right] \cr \left[ 1 , 1 \right] &\left[  \right] &\left[ 1 , 1 \right] &\left[  \right] &\left[  \right] \cr \left[  \right] &\left[  \right] &\left[  \right] &\left[ 1 , 1 \right] &\left[  \right] \cr \ifx\endpmatrix\undefined}\else\end{pmatrix}\fi  $ 
&&&&&\\ 
& $ 4.9794e-23 $ 
& [$ 1.6887e-1 $]  
& $ 12 $  
& $ 12 $ 
& $ 1 $  
\\
\multicolumn{6}{ c }{ \makecell{ $+{{11648333\,\log 6}\over{42527764900741120000\,3^{{{7}\over{2}}}\,\pi^{12}}}$ $-\left({{4871026790838146753876369132147729921}\over{35297853936987425544846450851053568000\,3^{{{55}\over{2}}}\,\pi^{12}}}\right)$} 
} 
\\
\hline
$\, \begin{pmatrix} \left[ 1 , 2 \right] &\left[ 1 , 2 \right] &\left[ 1 , 1 \right] &\left[ 1 , 1 \right] &\left[ 1 , 1 \right] \cr \left[ 1 , 1 \right] &\left[  \right] &\left[ 1 , 1 \right] &\left[  \right] &\left[  \right] \cr \left[  \right] &\left[ 1 , 1 \right] &\left[  \right] &\left[  \right] &\left[  \right] \cr \ifx\endpmatrix\undefined}\else\end{pmatrix}\fi  $ 
&&&&&\\ 
& $ 9.9453e-23 $ 
& [$ 1.6865e-1 $]  
& $ 6 $  
& $ 6 $ 
& $ 1 $  
\\
\multicolumn{6}{ c }{ \makecell{ $+{{312535261\,\log 6}\over{8505552980148224000\,3^{{{15}\over{2}}}\,\pi^{12}}}$ $-\left({{14508387175709250956085156051338424373}\over{7059570787397485108969290170210713600\,3^{{{59}\over{2}}}\,\pi^{12}}}\right)$} 
} 
\\
\hline
$\, \begin{pmatrix} \left[ 1 , 2 \right] &\left[ 1 , 2 \right] &\left[ 1 , 1 \right] &\left[ 1 , 1 \right] &\left[ 1 , 1 \right] \cr \left[  \right] &\left[  \right] &\left[ 1 , 1 \right] &\left[ 1 , 1 \right] &\left[  \right] \cr \left[ 1 , 1 \right] &\left[  \right] &\left[  \right] &\left[  \right] &\left[  \right] \cr \ifx\endpmatrix\undefined}\else\end{pmatrix}\fi  $ 
&&&&&\\ 
& $ 5.1044e-23 $ 
& [$ 8.6557e-2 $]  
& $ 6 $  
& $ 6 $ 
& $ 1 $  
\\
\multicolumn{6}{ c }{ \makecell{ $+{{195858109\,\log 6}\over{8505552980148224000\,3^{{{15}\over{2}}}\,\pi^{12}}}$ $-\left({{3033598876294889203474754414296221179}\over{7059570787397485108969290170210713600\,3^{{{57}\over{2}}}\,\pi^{12}}}\right)$} 
} 
\\
\hline
\end{longtable}
\newpage
{\hskip-10em}
\begin{longtable}[l]{m{0.4\textwidth} m{0.16\textwidth} m{0.18\textwidth} m{0.10\textwidth} m{0.10\textwidth} m{0.04\textwidth} }
\caption{$N= 7$ state 4  }
\label{N=7_state_4}
\\\hline\hline
State & & Total width & Channels & &
\endfirsthead
\\
\multirow{2}{0.3\textwidth}{$\, [[1,3],[1,1],[1,1],[1,1],[1,1]] $ } 
& & $3.8921e-21$ 
& $4059$  
& &
\\
& \multicolumn{5}{c}{
\makecell{${{371478323785466724440751232259}\over{23454945385186461238144532480000\,3^{{{53}\over{2}}}\,\pi^{12}}}$} 
  } 
\\ 
 \hline
decay & width & ratio  & cl. deg. & q. deg. & s.t. norm.
\\
$\, \begin{pmatrix} \left[ 1 , 3 \right] &\left[ 1 , 1 \right] &\left[ 1 , 1 \right] &\left[ 1 , 1 \right] &\left[ 1 , 1 \right] \cr \left[ 1 , 1 \right] &\left[ 1 , 1 \right] &\left[  \right] &\left[  \right] &\left[  \right] \cr \left[  \right] &\left[  \right] &\left[ 1 , 1 \right] &\left[  \right] &\left[  \right] \cr \ifx\endpmatrix\undefined}\else\end{pmatrix}\fi  $ 
&&&&&\\ 
& $ 8.0411e-23 $ 
& [$ 2.4792e-1 $]  
& $ 12 $  
& $ 12 $ 
& $ 1 $  
\\
\multicolumn{6}{ c }{ \makecell{ $+{{62963171\,\log 6}\over{607539498582016000\,3^{{{17}\over{2}}}\,\pi^{12}}}$ $-\left({{125386941692390284535940602016101}\over{15757970507583672118235022701363200\,3^{{{49}\over{2}}}\,\pi^{12}}}\right)$} 
} 
\\
\hline
$\, \begin{pmatrix} \left[ 1 , 3 \right] &\left[ 1 , 1 \right] &\left[ 1 , 1 \right] &\left[ 1 , 1 \right] &\left[ 1 , 1 \right] \cr \left[ 1 , 1 \right] &\left[ 1 , 1 \right] &\left[  \right] &\left[  \right] &\left[  \right] \cr \left[ 1 , 1 \right] &\left[  \right] &\left[  \right] &\left[  \right] &\left[  \right] \cr \ifx\endpmatrix\undefined}\else\end{pmatrix}\fi  $ 
&&&&&\\ 
& $ 1.4203e-22 $ 
& [$ 1.4597e-1 $]  
& $ 4 $  
& $ 4 $ 
& $ 1 $  
\\
\multicolumn{6}{ c }{ \makecell{ $+{{37584737\,\log 6}\over{607539498582016000\,3^{{{15}\over{2}}}\,\pi^{12}}}$ $-\left({{39685256797436548496921611283309639}\over{11460342187333579722352743782809600\,3^{{{59}\over{2}}}\,\pi^{12}}}\right)$} 
} 
\\
\hline
$\, \begin{pmatrix} \left[ 1 , 3 \right] &\left[ 1 , 1 \right] &\left[ 1 , 1 \right] &\left[ 1 , 1 \right] &\left[ 1 , 1 \right] \cr \left[  \right] &\left[ 1 , 1 \right] &\left[ 1 , 1 \right] &\left[  \right] &\left[  \right] \cr \left[ 1 , 1 \right] &\left[  \right] &\left[  \right] &\left[  \right] &\left[  \right] \cr \ifx\endpmatrix\undefined}\else\end{pmatrix}\fi  $ 
&&&&&\\ 
& $ 8.4614e-23 $ 
& [$ 1.3044e-1 $]  
& $ 6 $  
& $ 6 $ 
& $ 1 $  
\\
\multicolumn{6}{ c }{ \makecell{ $+{{1839073\,\log 6}\over{17358271388057600\,3^{{{17}\over{2}}}\,\pi^{12}}}$ $-\left({{76899910270360386638918323483423}\over{3151594101516734423647004540272640\,3^{{{51}\over{2}}}\,\pi^{12}}}\right)$} 
} 
\\
\hline
\end{longtable}
\newpage
{\hskip-10em}
\begin{longtable}[l]{m{0.4\textwidth} m{0.16\textwidth} m{0.18\textwidth} m{0.10\textwidth} m{0.10\textwidth} m{0.04\textwidth} }
\caption{$N= 7$ state 5  }
\label{N=7_state_5}
\\\hline\hline
State & & Total width & Channels & &
\endfirsthead
\\
\multirow{2}{0.3\textwidth}{$\, [[1,2],[1,2],[1,2],[1,1]] $ } 
& & $3.9824e-21$ 
& $3925$  
& &
\\
& \multicolumn{5}{c}{
\makecell{${{13683379731543053357751584673829}\over{93819781540745844952578129920000\,3^{{{57}\over{2}}}\,\pi^{12}}}$} 
  } 
\\ 
 \hline
decay & width & ratio  & cl. deg. & q. deg. & s.t. norm.
\\
$\, \begin{pmatrix} \left[ 1 , 2 \right] &\left[ 1 , 2 \right] &\left[ 1 , 2 \right] &\left[ 1 , 1 \right] \cr \left[ 1 , 1 \right] &\left[  \right] &\left[  \right] &\left[ 1 , 1 \right] \cr \left[  \right] &\left[ 1 , 1 \right] &\left[  \right] &\left[  \right] \cr \ifx\endpmatrix\undefined}\else\end{pmatrix}\fi  $ 
&&&&&\\ 
& $ 1.1102e-22 $ 
& [$ 1.6726e-1 $]  
& $ 6 $  
& $ 6 $ 
& $ 1 $  
\\
\multicolumn{6}{ c }{ \makecell{ $+{{260861149\,\log 6}\over{17011105960296448000\,3^{{{13}\over{2}}}\,\pi^{12}}}$ $-\left({{145398846512922704552916713612288938709}\over{56476566299179880871754321361685708800\,3^{{{59}\over{2}}}\,\pi^{12}}}\right)$} 
} 
\\
\hline
$\, \begin{pmatrix} \left[ 1 , 2 \right] &\left[ 1 , 2 \right] &\left[ 1 , 2 \right] &\left[ 1 , 1 \right] \cr \left[ 1 , 1 \right] &\left[ 1 , 1 \right] &\left[  \right] &\left[  \right] \cr \left[  \right] &\left[  \right] &\left[ 1 , 1 \right] &\left[  \right] \cr \ifx\endpmatrix\undefined}\else\end{pmatrix}\fi  $ 
&&&&&\\ 
& $ 2.1815e-22 $ 
& [$ 1.6433e-1 $]  
& $ 3 $  
& $ 3 $ 
& $ 1 $  
\\
\multicolumn{6}{ c }{ \makecell{ $+{{82598741\,\log 6}\over{3402221192059289600\,3^{{{13}\over{2}}}\,\pi^{12}}}$ $-\left({{137958920678797696078419155638582315843}\over{11295313259835976174350864272337141760\,3^{{{61}\over{2}}}\,\pi^{12}}}\right)$} 
} 
\\
\hline
$\, \begin{pmatrix} \left[ 1 , 2 \right] &\left[ 1 , 2 \right] &\left[ 1 , 2 \right] &\left[ 1 , 1 \right] \cr \left[ 1 , 1 \right] &\left[ 1 , 1 \right] &\left[  \right] &\left[  \right] \cr \left[ 1 , 1 \right] &\left[  \right] &\left[  \right] &\left[  \right] \cr \ifx\endpmatrix\undefined}\else\end{pmatrix}\fi  $ 
&&&&&\\ 
& $ 9.6283e-23 $ 
& [$ 1.4506e-1 $]  
& $ 6 $  
& $ 6 $ 
& $ 1 $  
\\
\multicolumn{6}{ c }{ \makecell{ $+{{18235909\,\log 6}\over{347165427761152000\,3^{{{15}\over{2}}}\,\pi^{12}}}$ $-\left({{1282165135158462972746366315492586799619}\over{16136161799765680249072663246195916800\,3^{{{65}\over{2}}}\,\pi^{12}}}\right)$} 
} 
\\
\hline
\end{longtable}
%
%
%

\newpage
\section{Level $8$}
\label{app:N=8}

\begin{longtable}[l]{m{0.4\textwidth} m{0.16\textwidth} m{0.18\textwidth} m{0.10\textwidth} m{0.10\textwidth} m{0.04\textwidth} }
\caption{$N= 8$ state 1  }
\label{N=8_state_1}
\\\hline\hline
State & & Total width & Channels & &
\endfirsthead
\\
\multirow{2}{0.3\textwidth}{$\, [[2,1],[2,1],[2,1],[1,1],[1,1]] $ } 
& & $2.0447e-20$ 
& $4794$  
& &
\\
& \multicolumn{5}{c}{
\makecell{ $+{{281377201789\,3^{{{15}\over{2}}}\,7^{{{35}\over{2}}}}\over{54628857311600000\,14^{{{49}\over{2}}}\,\pi^{12}}}$ $+{{97326858944034849422061519899786173}\over{452097869808820950000\,14^{{{49}\over{2}}}\,\pi^{12}}}$} 
  } 
\\ 
 \hline
decay & width & ratio  & cl. deg. & q. deg. & s.t. norm.
\\
$\, \begin{pmatrix} \left[ 2 , 1 \right] &\left[ 2 , 1 \right] &\left[ 2 , 1 \right] &\left[ 1 , 1 \right] &\left[ 1 , 1 \right] \cr \left[ 1 , 1 \right] &\left[  \right] &\left[  \right] &\left[  \right] &\left[  \right] \cr \left[  \right] &\left[ 1 , 1 \right] &\left[  \right] &\left[  \right] &\left[  \right] \cr \ifx\endpmatrix\undefined}\else\end{pmatrix}\fi  $ 
&&&&&\\ 
& $ 3.1912e-21 $ 
& [$ 4.6820e-1 $]  
& $ 6 $  
& $ 3 $ 
& $ 1 $  
\\
\multicolumn{6}{ c }{ \makecell{${{616878770449417\,\sqrt{14}}\over{782556941236138745108889600000\,\pi^{12}}}$} 
} 
\\
\hline
$\, \begin{pmatrix} \left[ 2 , 1 \right] &\left[ 2 , 1 \right] &\left[ 2 , 1 \right] &\left[ 1 , 1 \right] &\left[ 1 , 1 \right] \cr \left[ 1 , 1 \right] &\left[ 1 , 1 \right] &\left[  \right] &\left[ 1 , 1 \right] &\left[  \right] \cr \left[  \right] &\left[  \right] &\left[ 1 , 1 \right] &\left[  \right] &\left[  \right] \cr \ifx\endpmatrix\undefined}\else\end{pmatrix}\fi  $ 
&&&&&\\ 
& $ 6.7145e-22 $ 
& [$ 1.9703e-1 $]  
& $ 6 $  
& $ 6 $ 
& $ 1 $  
\\
\multicolumn{6}{ c }{ \makecell{ $+{{240700692559484490361\,\log \left({{2}\over{7}}\right)}\over{21044863125\,14^{{{31}\over{2}}}\,\pi^{12}}}$ $+{{3205695921502472827816383209175731197921231}\over{2122393575464617568256000\,14^{{{45}\over{2}}}\,\pi^{12}}}$} 
} 
\\
\hline
$\, \begin{pmatrix} \left[ 2 , 1 \right] &\left[ 2 , 1 \right] &\left[ 2 , 1 \right] &\left[ 1 , 1 \right] &\left[ 1 , 1 \right] \cr \left[  \right] &\left[  \right] &\left[  \right] &\left[ 1 , 1 \right] &\left[  \right] \cr \left[ 1 , 1 \right] &\left[  \right] &\left[  \right] &\left[  \right] &\left[  \right] \cr \ifx\endpmatrix\undefined}\else\end{pmatrix}\fi  $ 
&&&&&\\ 
& $ 2.6551e-22 $ 
& [$ 7.7910e-2 $]  
& $ 12 $  
& $ 6 $ 
& $ 1 $  
\\
\multicolumn{6}{ c }{ \makecell{${{180416659061051\,\sqrt{14}}\over{2750806217678548316140339200000\,\pi^{12}}}$} 
} 
\\
\hline
\end{longtable}
\newpage
{\hskip-10em}
\begin{longtable}[l]{m{0.4\textwidth} m{0.16\textwidth} m{0.18\textwidth} m{0.10\textwidth} m{0.10\textwidth} m{0.04\textwidth} }
\caption{$N= 8$ state 2  }
\label{N=8_state_2}
\\\hline\hline
State & & Total width & Channels & &
\endfirsthead
\\
\multirow{2}{0.3\textwidth}{$\, [[2,2],[2,1],[1,1],[1,1]] $ } 
& & $2.1143e-20$ 
& $12381$  
& &
\\
& \multicolumn{5}{c}{
\makecell{ $+{{35547339990013\,3^{{{11}\over{2}}}\,7^{{{35}\over{2}}}}\over{54628857311600000\,14^{{{51}\over{2}}}\,\pi^{12}}}$ $+{{15524779503782813084984375438371012589}\over{4973076567897030450000\,14^{{{51}\over{2}}}\,\pi^{12}}}$} 
  } 
\\ 
 \hline
decay & width & ratio  & cl. deg. & q. deg. & s.t. norm.
\\
$\, \begin{pmatrix} \left[ 2 , 2 \right] &\left[ 2 , 1 \right] &\left[ 1 , 1 \right] &\left[ 1 , 1 \right] \cr \left[  \right] &\left[ 1 , 1 \right] &\left[  \right] &\left[  \right] \cr \left[ 1 , 1 \right] &\left[  \right] &\left[  \right] &\left[  \right] \cr \ifx\endpmatrix\undefined}\else\end{pmatrix}\fi  $ 
&&&&&\\ 
& $ 6.4110e-21 $ 
& [$ 3.0321e-1 $]  
& $ 2 $  
& $ 1 $ 
& $ 1 $  
\\
\multicolumn{6}{ c }{ \makecell{${{71879063991928291\,\sqrt{14}}\over{45388302591696047216315596800000\,\pi^{12}}}$} 
} 
\\
\hline
$\, \begin{pmatrix} \left[ 2 , 2 \right] &\left[ 2 , 1 \right] &\left[ 1 , 1 \right] &\left[ 1 , 1 \right] \cr \left[ 1 , 1 \right] &\left[  \right] &\left[  \right] &\left[  \right] \cr \left[ 1 , 1 \right] &\left[  \right] &\left[  \right] &\left[  \right] \cr \ifx\endpmatrix\undefined}\else\end{pmatrix}\fi  $ 
&&&&&\\ 
& $ 6.3817e-21 $ 
& [$ 1.5092e-1 $]  
& $ 1 $  
& $ 1 $ 
& $ {{1}\over{2}} $  
\\
\multicolumn{6}{ c }{ \makecell{${{214654133478175531\,\sqrt{14}}\over{136164907775088141648946790400000\,\pi^{12}}}$} 
} 
\\
\hline
$\, \begin{pmatrix} \left[ 2 , 2 \right] &\left[ 2 , 1 \right] &\left[ 1 , 1 \right] &\left[ 1 , 1 \right] \cr \left[ 1 , 1 \right] &\left[ 1 , 1 \right] &\left[ 1 , 1 \right] &\left[  \right] \cr \left[ 1 , 1 \right] &\left[  \right] &\left[  \right] &\left[  \right] \cr \ifx\endpmatrix\undefined}\else\end{pmatrix}\fi  $ 
&&&&&\\ 
& $ 1.3555e-21 $ 
& [$ 1.2822e-1 $]  
& $ 2 $  
& $ 2 $ 
& $ 1 $  
\\
\multicolumn{6}{ c }{ \makecell{ $+{{8892671599121294546009\,\log \left({{2}\over{7}}\right)}\over{12626917875\,14^{{{33}\over{2}}}\,\pi^{12}}}$ $+{{16354690203627064202705597921277735967391836931}\over{897193647810042881126400\,14^{{{51}\over{2}}}\,\pi^{12}}}$} 
} 
\\
\hline
\end{longtable}
\newpage
{\hskip-10em}
\begin{longtable}[l]{m{0.4\textwidth} m{0.16\textwidth} m{0.18\textwidth} m{0.10\textwidth} m{0.10\textwidth} m{0.04\textwidth} }
\caption{$N= 8$ state 3  }
\label{N=8_state_3}
\\\hline\hline
State & & Total width & Channels & &
\endfirsthead
\\
\multirow{2}{0.3\textwidth}{$\, [[2,1],[2,1],[2,1],[1,2]] $ } 
& & $2.1235e-20$ 
& $18452$  
& &
\\
& \multicolumn{5}{c}{
\makecell{ $+{{14538429406537\,3^{{{13}\over{2}}}\,7^{{{35}\over{2}}}}\over{54628857311600000\,14^{{{51}\over{2}}}\,\pi^{12}}}$ $+{{30811124927425162173102770068129233101}\over{9946153135794060900000\,14^{{{51}\over{2}}}\,\pi^{12}}}$} 
  } 
\\ 
 \hline
decay & width & ratio  & cl. deg. & q. deg. & s.t. norm.
\\
$\, \begin{pmatrix} \left[ 2 , 1 \right] &\left[ 2 , 1 \right] &\left[ 2 , 1 \right] &\left[ 1 , 2 \right] \cr \left[ 1 , 1 \right] &\left[  \right] &\left[  \right] &\left[  \right] \cr \left[  \right] &\left[ 1 , 1 \right] &\left[  \right] &\left[  \right] \cr \ifx\endpmatrix\undefined}\else\end{pmatrix}\fi  $ 
&&&&&\\ 
& $ 3.0773e-21 $ 
& [$ 4.3475e-1 $]  
& $ 6 $  
& $ 3 $ 
& $ 1 $  
\\
\multicolumn{6}{ c }{ \makecell{${{117958476047447\,\sqrt{14}}\over{155173684074174520397660160000\,\pi^{12}}}$} 
} 
\\
\hline
$\, \begin{pmatrix} \left[ 2 , 1 \right] &\left[ 2 , 1 \right] &\left[ 2 , 1 \right] &\left[ 1 , 2 \right] \cr \left[ 1 , 1 \right] &\left[ 1 , 1 \right] &\left[  \right] &\left[ 1 , 1 \right] \cr \left[  \right] &\left[  \right] &\left[ 1 , 1 \right] &\left[  \right] \cr \ifx\endpmatrix\undefined}\else\end{pmatrix}\fi  $ 
&&&&&\\ 
& $ 1.4615e-21 $ 
& [$ 2.0647e-1 $]  
& $ 3 $  
& $ 3 $ 
& $ 1 $  
\\
\multicolumn{6}{ c }{ \makecell{ $+{{6993506060264984939\,\log \left({{2}\over{7}}\right)}\over{4208972625\,14^{{{29}\over{2}}}\,\pi^{12}}}$ $+{{47160289269075390221381178263833240494156187}\over{1096570013990052410265600\,14^{{{47}\over{2}}}\,\pi^{12}}}$} 
} 
\\
\hline
$\, \begin{pmatrix} \left[ 2 , 1 \right] &\left[ 2 , 1 \right] &\left[ 2 , 1 \right] &\left[ 1 , 2 \right] \cr \left[  \right] &\left[  \right] &\left[  \right] &\left[ 1 , 1 \right] \cr \left[ 1 , 1 \right] &\left[  \right] &\left[  \right] &\left[  \right] \cr \ifx\endpmatrix\undefined}\else\end{pmatrix}\fi  $ 
&&&&&\\ 
& $ 5.2273e-22 $ 
& [$ 7.3848e-2 $]  
& $ 6 $  
& $ 3 $ 
& $ 1 $  
\\
\multicolumn{6}{ c }{ \makecell{${{23442980505501977\,\sqrt{14}}\over{181553210366784188865262387200000\,\pi^{12}}}$} 
} 
\\
\hline
\end{longtable}
\newpage
{\hskip-10em}
\begin{longtable}[l]{m{0.4\textwidth} m{0.16\textwidth} m{0.18\textwidth} m{0.10\textwidth} m{0.10\textwidth} m{0.04\textwidth} }
\caption{$N= 8$ state 4  }
\label{N=8_state_4}
\\\hline\hline
State & & Total width & Channels & &
\endfirsthead
\\
\multirow{2}{0.3\textwidth}{$\, [[1,1],[1,1],[1,1],[1,1],[1,1],[1,1],[1,1],[1,1]] $ } 
& & $2.1431e-20$ 
& $8672$  
& &
\\
& \multicolumn{5}{c}{
\makecell{ $+{{7099921\,3^{{{17}\over{2}}}\,7^{{{37}\over{2}}}}\over{218515429246400000\,14^{{{45}\over{2}}}\,\pi^{12}}}$ $+{{6585265421998849133555222258746039}\over{5425174437705851400000\,14^{{{45}\over{2}}}\,\pi^{12}}}$} 
  } 
\\ 
 \hline
decay & width & ratio  & cl. deg. & q. deg. & s.t. norm.
\\
$\, \begin{pmatrix} \left[ 1 , 1 \right] &\left[ 1 , 1 \right] &\left[ 1 , 1 \right] &\left[ 1 , 1 \right] &\left[ 1 , 1 \right] &\left[ 1 , 1 \right] &\left[ 1 , 1 \right] &\left[ 1 , 1 \right] \cr \left[ 1 , 1 \right] &\left[  \right] &\left[  \right] &\left[  \right] &\left[  \right] &\left[  \right] &\left[  \right] &\left[  \right] \cr \left[  \right] &\left[ 1 , 1 \right] &\left[  \right] &\left[  \right] &\left[  \right] &\left[  \right] &\left[  \right] &\left[  \right] \cr \ifx\endpmatrix\undefined}\else\end{pmatrix}\fi  $ 
&&&&&\\ 
& $ 5.8573e-22 $ 
& [$ 7.6526e-1 $]  
& $ 56 $  
& $ 28 $ 
& $ 1 $  
\\
\multicolumn{6}{ c }{ \makecell{${{17383729358213533\,\sqrt{14}}\over{120145506860371889690247168000000\,\pi^{12}}}$} 
} 
\\
\hline
$\, \begin{pmatrix} \left[ 1 , 1 \right] &\left[ 1 , 1 \right] &\left[ 1 , 1 \right] &\left[ 1 , 1 \right] &\left[ 1 , 1 \right] &\left[ 1 , 1 \right] &\left[ 1 , 1 \right] &\left[ 1 , 1 \right] &\left[  \right] \cr \left[ 1 , 1 \right] &\left[  \right] &\left[  \right] &\left[  \right] &\left[  \right] &\left[  \right] &\left[  \right] &\left[  \right] &\left[  \right] \cr \left[  \right] &\left[  \right] &\left[  \right] &\left[  \right] &\left[  \right] &\left[  \right] &\left[  \right] &\left[  \right] &\left[ 1 , 1 \right] \cr \ifx\endpmatrix\undefined}\else\end{pmatrix}\fi  $ 
&&&&&\\ 
& $ 3.1160e-23 $ 
& [$ 1.8610e-1 $]  
& $ 256 $  
& $ 128 $ 
& $ 1 $  
\\
\multicolumn{6}{ c }{ \makecell{${{17570725148299043\,\sqrt{14}}\over{2282764630347065904114696192000000\,\pi^{12}}}$} 
} 
\\
\hline
$\, \begin{pmatrix} \left[ 1 , 1 \right] &\left[ 1 , 1 \right] &\left[ 1 , 1 \right] &\left[ 1 , 1 \right] &\left[ 1 , 1 \right] &\left[ 1 , 1 \right] &\left[ 1 , 1 \right] &\left[ 1 , 1 \right] \cr \left[ 1 , 1 \right] &\left[ 1 , 1 \right] &\left[ 1 , 1 \right] &\left[  \right] &\left[  \right] &\left[  \right] &\left[  \right] &\left[  \right] \cr \left[  \right] &\left[  \right] &\left[  \right] &\left[ 1 , 1 \right] &\left[  \right] &\left[  \right] &\left[  \right] &\left[  \right] \cr \ifx\endpmatrix\undefined}\else\end{pmatrix}\fi  $ 
&&&&&\\ 
& $ 1.4519e-24 $ 
& [$ 1.8970e-2 $]  
& $ 280 $  
& $ 280 $ 
& $ 1 $  
\\
\multicolumn{6}{ c }{ \makecell{ $-\left({{1238299675956119987\,\log \left({{2}\over{7}}\right)}\over{136521804375\,14^{{{27}\over{2}}}\,\pi^{12}}}\right)$ $-\left({{8545160977278587313054474801088508546811097}\over{99875596874213973526990848000\,14^{{{39}\over{2}}}\,\pi^{12}}}\right)$} 
} 
\\
\hline
\end{longtable}
\newpage
{\hskip-10em}
\begin{longtable}[l]{m{0.4\textwidth} m{0.16\textwidth} m{0.18\textwidth} m{0.10\textwidth} m{0.10\textwidth} m{0.04\textwidth} }
\caption{$N= 8$ state 5  }
\label{N=8_state_5}
\\\hline\hline
State & & Total width & Channels & &
\endfirsthead
\\
\multirow{2}{0.3\textwidth}{$\, [[1,2],[1,1],[1,1],[1,1],[1,1],[1,1],[1,1]] $ } 
& & $2.1863e-20$ 
& $3636$  
& &
\\
& \multicolumn{5}{c}{
\makecell{ $+{{6282893\,3^{{{23}\over{2}}}\,7^{{{37}\over{2}}}}\over{19865039022400000\,14^{{{49}\over{2}}}\,\pi^{12}}}$ $+{{579116804097938830107205925778418697}\over{2387076752590574616000\,14^{{{49}\over{2}}}\,\pi^{12}}}$} 
  } 
\\ 
 \hline
decay & width & ratio  & cl. deg. & q. deg. & s.t. norm.
\\
$\, \begin{pmatrix} \left[ 1 , 2 \right] &\left[ 1 , 1 \right] &\left[ 1 , 1 \right] &\left[ 1 , 1 \right] &\left[ 1 , 1 \right] &\left[ 1 , 1 \right] &\left[ 1 , 1 \right] \cr \left[  \right] &\left[ 1 , 1 \right] &\left[  \right] &\left[  \right] &\left[  \right] &\left[  \right] &\left[  \right] \cr \left[  \right] &\left[  \right] &\left[ 1 , 1 \right] &\left[  \right] &\left[  \right] &\left[  \right] &\left[  \right] \cr \ifx\endpmatrix\undefined}\else\end{pmatrix}\fi  $ 
&&&&&\\ 
& $ 5.7072e-22 $ 
& [$ 3.9156e-1 $]  
& $ 30 $  
& $ 15 $ 
& $ 1 $  
\\
\multicolumn{6}{ c }{ \makecell{${{71516762367461887\,\sqrt{14}}\over{507281028966014645358821376000000\,\pi^{12}}}$} 
} 
\\
\hline
$\, \begin{pmatrix} \left[ 1 , 2 \right] &\left[ 1 , 1 \right] &\left[ 1 , 1 \right] &\left[ 1 , 1 \right] &\left[ 1 , 1 \right] &\left[ 1 , 1 \right] &\left[ 1 , 1 \right] \cr \left[  \right] &\left[ 1 , 1 \right] &\left[  \right] &\left[  \right] &\left[  \right] &\left[  \right] &\left[  \right] \cr \left[ 1 , 1 \right] &\left[  \right] &\left[  \right] &\left[  \right] &\left[  \right] &\left[  \right] &\left[  \right] \cr \ifx\endpmatrix\undefined}\else\end{pmatrix}\fi  $ 
&&&&&\\ 
& $ 1.2064e-21 $ 
& [$ 3.3109e-1 $]  
& $ 12 $  
& $ 6 $ 
& $ 1 $  
\\
\multicolumn{6}{ c }{ \makecell{${{272120827515923281\,\sqrt{14}}\over{913105852138826361645878476800000\,\pi^{12}}}$} 
} 
\\
\hline
$\, \begin{pmatrix} \left[ 1 , 2 \right] &\left[ 1 , 1 \right] &\left[ 1 , 1 \right] &\left[ 1 , 1 \right] &\left[ 1 , 1 \right] &\left[ 1 , 1 \right] &\left[ 1 , 1 \right] &\left[  \right] \cr \left[  \right] &\left[ 1 , 1 \right] &\left[  \right] &\left[  \right] &\left[  \right] &\left[  \right] &\left[  \right] &\left[  \right] \cr \left[  \right] &\left[  \right] &\left[  \right] &\left[  \right] &\left[  \right] &\left[  \right] &\left[  \right] &\left[ 1 , 1 \right] \cr \ifx\endpmatrix\undefined}\else\end{pmatrix}\fi  $ 
&&&&&\\ 
& $ 3.1473e-23 $ 
& [$ 1.4684e-1 $]  
& $ 204 $  
& $ 102 $ 
& $ 1 $  
\\
\multicolumn{6}{ c }{ \makecell{${{35495192073594043\,\sqrt{14}}\over{4565529260694131808229392384000000\,\pi^{12}}}$} 
} 
\\
\hline
\end{longtable}

\end{document}

%% file: Tables_from_level2_to_8state1_num_width_channels.V1.0.tex

%
\begin{longtable}[l]{ m{0.5\textwidth} m{0.25\textwidth} m{0.25\textwidth} }
\caption{Summary level $ 2 $ }
\\
\label{summaryL2}
State & decay width & channels
\\
\endfirsthead
\\
State & decay width & channels
\\
\endhead 
 $ [[1,2]] $ & $5.3406e-25$ & $ 4 $ 
\\
 $ [[2,1]] $ & $5.3406e-25$ & $ 4 $ 
\\
 $ [[1,1],[1,1]] $ & $5.3406e-25$ & $ 5 $ 
\\
\end{longtable}

\newpage
%
\begin{longtable}[l]{ m{0.5\textwidth} m{0.25\textwidth} m{0.25\textwidth} }
\caption{Summary level $ 3 $ }
\\
\label{summaryL3}
State & decay width & channels
\\
\endfirsthead
\\
State & decay width & channels
\\
\endhead 
 $ [[2,1],[1,1]] $ & $1.9357e-28$ & $ 35 $ 
\\
 $ [[3,1]] $ & $2.4516e-28$ & $ 16 $ 
\\
 $ [[1,3]] $ & $2.6063e-28$ & $ 16 $ 
\\
 $ [[1,2],[1,1]] $ & $2.7095e-28$ & $ 35 $ 
\\
 $ [[2,1,1,1]] $ & $2.7611e-28$ & $ 16 $ 
\\
 $ [[1,1],[1,1],[1,1]] $ & $2.7611e-28$ & $ 22 $ 
\\
\end{longtable}

\newpage
%
\begin{longtable}[l]{ m{0.5\textwidth} m{0.25\textwidth} m{0.25\textwidth} }
\caption{Summary level $ 4 $ }
\\
\label{summaryL4}
State & decay width & channels
\\
\endfirsthead
\\
State & decay width & channels
\\
\endhead 
 $ [[1,1],[1,1],[1,1],[1,1]] $ & $9.5345e-22$ & $ 69 $ 
\\
 $ [[1,2],[1,1],[1,1]] $ & $1.1821e-21$ & $ 136 $ 
\\
 $ [[1,2],[1,2]] $ & $1.4468e-21$ & $ 63 $ 
\\
 $ [[2,1,1,2]] $ & $1.4793e-21$ & $ 45 $ 
\\
 $ [[1,3],[1,1]] $ & $1.6396e-21$ & $ 107 $ 
\\
 $ [[4,1]] $ & $2.1882e-21$ & $ 45 $ 
\\
 $ [[2,1,1,1],[1,1]] $ & $2.3380e-21$ & $ 107 $ 
\\
 $ [[1,4]] $ & $2.3796e-21$ & $ 45 $ 
\\
 $ [[2,1],[1,1],[1,1]] $ & $3.4640e-21$ & $ 136 $ 
\\
 $ [[2,1],[1,2]] $ & $4.1971e-21$ & $ 107 $ 
\\
 $ [[3,1],[1,1]] $ & $5.8146e-21$ & $ 107 $ 
\\
 $ [[3,1,1,1]] $ & $7.1202e-21$ & $ 45 $ 
\\
 $ [[2,1],[2,1]] $ & $1.3951e-20$ & $ 63 $ 
\\
 $ [[2,2]] $ & $1.4077e-20$ & $ 45 $ 
\\
\end{longtable}

\newpage
%
\begin{longtable}[l]{ m{0.5\textwidth} m{0.25\textwidth} m{0.25\textwidth} }
\caption{Summary level $ 5 $ }
\\
\label{summaryL5}
State & decay width & channels
\\
\endfirsthead
\\
State & decay width & channels
\\
\endhead 
 $ [[1,1],[1,1],[1,1],[1,1],[1,1]] $ & $6.9906e-23$ & $ 206 $ 
\\
 $ [[1,2],[1,1],[1,1],[1,1]] $ & $8.9957e-23$ & $ 447 $ 
\\
 $ [[1,2],[1,2],[1,1]] $ & $1.1188e-22$ & $ 415 $ 
\\
 $ [[1,3],[1,1],[1,1]] $ & $1.3006e-22$ & $ 415 $ 
\\
 $ [[1,3],[1,2]] $ & $1.5573e-22$ & $ 309 $ 
\\
 $ [[5,1]] $ & $1.8530e-22$ & $ 121 $ 
\\
 $ [[2,1,1,3]] $ & $1.8701e-22$ & $ 121 $ 
\\
 $ [[1,4],[1,1]] $ & $1.9303e-22$ & $ 309 $ 
\\
 $ [[2,1,1,2],[1,1]] $ & $2.0071e-22$ & $ 309 $ 
\\
 $ [[2,1,1,1],[1,1],[1,1]] $ & $2.2311e-22$ & $ 415 $ 
\\
 $ [[2,1,1,1],[1,2]] $ & $2.6625e-22$ & $ 309 $ 
\\
 $ [[2,1],[1,1],[1,1],[1,1]] $ & $2.7313e-22$ & $ 447 $ 
\\
 $ [[1,5]] $ & $2.8448e-22$ & $ 121 $ 
\\
 $ [[2,1],[1,2],[1,1]] $ & $3.2257e-22$ & $ 687 $ 
\\
 $ [[2,1],[1,3]] $ & $4.2147e-22$ & $ 309 $ 
\\
 $ [[4,1],[1,1]] $ & $5.1420e-22$ & $ 309 $ 
\\
 $ [[3,1],[1,1],[1,1]] $ & $5.7756e-22$ & $ 415 $ 
\\
 $ [[3,1,1,1],[1,1]] $ & $5.9861e-22$ & $ 309 $ 
\\
 $ [[4,1,1,1]] $ & $6.1637e-22$ & $ 121 $ 
\\
 $ [[3,1],[1,2]] $ & $6.3940e-22$ & $ 309 $ 
\\
 $ [[3,1,1,2]] $ & $6.5036e-22$ & $ 121 $ 
\\
 $ [[3,1,2,1]] $ & $7.1357e-22$ & $ 121 $ 
\\
 $ [[2,2,1,1]] $ & $7.6615e-22$ & $ 121 $ 
\\
 $ [[2,1,1,1],[2,1]] $ & $8.5195e-22$ & $ 309 $ 
\\
 $ [[2,1],[2,1],[1,1]] $ & $9.2177e-22$ & $ 415 $ 
\\
 $ [[2,2],[1,1]] $ & $9.3379e-22$ & $ 309 $ 
\\
 $ [[3,1],[2,1]] $ & $1.4223e-21$ & $ 309 $ 
\\
\end{longtable}

\newpage
%
\begin{longtable}[l]{ m{0.5\textwidth} m{0.25\textwidth} m{0.25\textwidth} }
\caption{Summary level $ 6 $ }
\\
\label{summaryL6}
State & decay width & channels
\\
\endfirsthead
\\
State & decay width & channels
\\
\endhead 
 $ [[1,1],[1,1],[1,1],[1,1],[1,1],[1,1]] $ & $8.7975e-21$ & $ 602 $ 
\\
 $ [[1,2],[1,1],[1,1],[1,1],[1,1]] $ & $9.5349e-21$ & $ 1380 $ 
\\
 $ [[1,2],[1,2],[1,1],[1,1]] $ & $1.0266e-20$ & $ 1638 $ 
\\
 $ [[1,2],[1,2],[1,2]] $ & $1.0976e-20$ & $ 574 $ 
\\
 $ [[1,3],[1,1],[1,1],[1,1]] $ & $1.1010e-20$ & $ 1347 $ 
\\
 $ [[1,3],[1,2],[1,1]] $ & $1.1728e-20$ & $ 2051 $ 
\\
 $ [[1,3],[1,3]] $ & $1.3166e-20$ & $ 500 $ 
\\
 $ [[1,4],[1,1],[1,1]] $ & $1.3212e-20$ & $ 1221 $ 
\\
 $ [[1,4],[1,2]] $ & $1.3890e-20$ & $ 868 $ 
\\
 $ [[3,1,2,1,1,1]] $ & $1.4130e-20$ & $ 322 $ 
\\
 $ [[6,1]] $ & $1.4538e-20$ & $ 322 $ 
\\
 $ [[2,1],[2,1],[2,1]] $ & $1.5307e-20$ & $ 574 $ 
\\
 $ [[2,1,1,3],[1,1]] $ & $1.5503e-20$ & $ 868 $ 
\\
 $ [[2,1,1,2],[1,1],[1,1]] $ & $1.5709e-20$ & $ 1221 $ 
\\
 $ [[2,2],[2,1]] $ & $1.5930e-20$ & $ 868 $ 
\\
 $ [[1,5],[1,1]] $ & $1.6125e-20$ & $ 868 $ 
\\
 $ [[2,1,1,4]] $ & $1.6336e-20$ & $ 322 $ 
\\
 $ [[2,1,1,2],[1,2]] $ & $1.6719e-20$ & $ 868 $ 
\\
 $ [[2,1,1,1],[1,1],[1,1],[1,1]] $ & $1.6794e-20$ & $ 1347 $ 
\\
 $ [[2,3]] $ & $1.7175e-20$ & $ 322 $ 
\\
 $ [[2,1,1,1],[1,2],[1,1]] $ & $1.7825e-20$ & $ 2051 $ 
\\
 $ [[2,1],[1,1],[1,1],[1,1],[1,1]] $ & $1.8643e-20$ & $ 1380 $ 
\\
 $ [[2,1],[1,2],[1,1],[1,1]] $ & $1.9636e-20$ & $ 2685 $ 
\\
 $ [[1,6]] $ & $1.9680e-20$ & $ 322 $ 
\\
 $ [[2,1,1,1],[1,3]] $ & $1.9889e-20$ & $ 868 $ 
\\
 $ [[2,1],[1,2],[1,2]] $ & $2.0479e-20$ & $ 1221 $ 
\\
 $ [[2,1],[1,3],[1,1]] $ & $2.1624e-20$ & $ 2051 $ 
\\
 $ [[2,1],[1,4]] $ & $2.4378e-20$ & $ 868 $ 
\\
 $ [[5,1],[1,1]] $ & $2.5138e-20$ & $ 868 $ 
\\
 $ [[3,1,2,1],[1,1]] $ & $2.5598e-20$ & $ 868 $ 
\\
 $ [[4,1,2,1]] $ & $2.5711e-20$ & $ 322 $ 
\\
 $ [[4,1,1,2]] $ & $2.5971e-20$ & $ 322 $ 
\\
 $ [[5,1,1,1]] $ & $2.5977e-20$ & $ 322 $ 
\\
 $ [[3,2]] $ & $2.6316e-20$ & $ 322 $ 
\\
 $ [[4,1,1,1],[1,1]] $ & $2.6412e-20$ & $ 868 $ 
\\
 $ [[2,2,1,2]] $ & $2.6834e-20$ & $ 322 $ 
\\
 $ [[3,1],[3,1]] $ & $2.6912e-20$ & $ 500 $ 
\\
 $ [[2,1,1,1],[2,1,1,1]] $ & $2.6936e-20$ & $ 500 $ 
\\
 $ [[4,1],[1,1],[1,1]] $ & $2.7207e-20$ & $ 1221 $ 
\\
 $ [[3,1,1,3]] $ & $2.8015e-20$ & $ 322 $ 
\\
 $ [[4,1],[1,2]] $ & $2.8244e-20$ & $ 868 $ 
\\
 $ [[3,1,1,2],[1,1]] $ & $2.8594e-20$ & $ 868 $ 
\\
 $ [[3,1,1,1],[1,1],[1,1]] $ & $2.8809e-20$ & $ 1221 $ 
\\
 $ [[3,1,1,1],[1,2]] $ & $2.9630e-20$ & $ 868 $ 
\\
 $ [[3,1],[1,1],[1,1],[1,1]] $ & $2.9686e-20$ & $ 1347 $ 
\\
 $ [[3,1],[1,2],[1,1]] $ & $3.0850e-20$ & $ 2051 $ 
\\
 $ [[2,1],[2,1],[1,2]] $ & $3.1455e-20$ & $ 1221 $ 
\\
 $ [[2,1],[2,1],[1,1],[1,1]] $ & $3.1724e-20$ & $ 1638 $ 
\\
 $ [[2,1,1,1],[2,1],[1,1]] $ & $3.2026e-20$ & $ 2051 $ 
\\
 $ [[2,2],[1,2]] $ & $3.2083e-20$ & $ 868 $ 
\\
 $ [[2,2],[1,1],[1,1]] $ & $3.2341e-20$ & $ 1221 $ 
\\
 $ [[2,2,1,1],[1,1]] $ & $3.2448e-20$ & $ 868 $ 
\\
 $ [[2,1,1,2],[2,1]] $ & $3.2857e-20$ & $ 868 $ 
\\
 $ [[3,1],[2,1,1,1]] $ & $3.3029e-20$ & $ 868 $ 
\\
 $ [[3,1],[1,3]] $ & $3.3177e-20$ & $ 868 $ 
\\
 $ [[3,1,1,1],[2,1]] $ & $3.7890e-20$ & $ 868 $ 
\\
 $ [[4,1],[2,1]] $ & $4.4150e-20$ & $ 868 $ 
\\
 $ [[3,1],[2,1],[1,1]] $ & $4.7721e-20$ & $ 2051 $ 
\\
\end{longtable}

\newpage
%
\begin{longtable}[l]{ m{0.5\textwidth} m{0.25\textwidth} m{0.25\textwidth} }
\caption{Summary level $ 7 $ }
\\
\label{summaryL7}
State & decay width & channels
\\
\endfirsthead
\\
State & decay width & channels
\\
\endhead 
 $ [[1,1],[1,1],[1,1],[1,1],[1,1],[1,1],[1,1]] $ & $2.7415e-21$ & $ 828 $ 
\\
 $ [[1,2],[1,1],[1,1],[1,1],[1,1],[1,1]] $ & $3.1250e-21$ & $ 1698 $ 
\\
 $ [[1,2],[1,2],[1,1],[1,1],[1,1]] $ & $3.5383e-21$ & $ 4092 $ 
\\
 $ [[1,3],[1,1],[1,1],[1,1],[1,1]] $ & $3.8921e-21$ & $ 5361 $ 
\\
 $ [[1,2],[1,2],[1,2],[1,1]] $ & $3.9824e-21$ & $ 4059 $ 
\\
 $ [[1,3],[1,2],[1,1],[1,1]] $ & $4.3648e-21$ & $ 3925 $ 
\\
 $ [[1,3],[1,2],[1,2]] $ & $4.8706e-21$ & $ 8065 $ 
\\
 $ [[1,4],[1,1],[1,1],[1,1]] $ & $5.0874e-21$ & $ 3472 $ 
\\
 $ [[1,3],[1,3],[1,1]] $ & $5.3102e-21$ & $ 3925 $ 
\\
 $ [[7,1]] $ & $5.5342e-21$ & $ 3472 $ 
\\
 $ [[1,4],[1,2],[1,1]] $ & $5.6509e-21$ & $ 828 $ 
\\
 $ [[2,1,1,1],[1,1],[1,1],[1,1],[1,1]] $ & $6.3245e-21$ & $ 5919 $ 
\\
 $ [[2,1,1,2],[1,1],[1,1],[1,1]] $ & $6.5221e-21$ & $ 4059 $ 
\\
 $ [[1,4],[1,3]] $ & $6.7779e-21$ & $ 3925 $ 
\\
 $ [[1,5],[1,1],[1,1]] $ & $6.7999e-21$ & $ 2364 $ 
\\
 $ [[2,1],[1,1],[1,1],[1,1],[1,1],[1,1]] $ & $6.8344e-21$ & $ 3472 $ 
\\
 $ [[2,1,1,1],[1,2],[1,1],[1,1]] $ & $7.0513e-21$ & $ 4092 $ 
\\
 $ [[2,1,1,3],[1,1],[1,1]] $ & $7.1997e-21$ & $ 8065 $ 
\\
 $ [[2,1,1,2],[1,2],[1,1]] $ & $7.2156e-21$ & $ 3472 $ 
\\
 $ [[1,5],[1,2]] $ & $7.4890e-21$ & $ 5919 $ 
\\
 $ [[2,1],[1,2],[1,1],[1,1],[1,1]] $ & $7.6371e-21$ & $ 2364 $ 
\\
 $ [[2,1,1,1],[1,2],[1,2]] $ & $7.8281e-21$ & $ 8835 $ 
\\
 $ [[2,1,1,3],[1,2]] $ & $7.8981e-21$ & $ 3472 $ 
\\
 $ [[2,1,1,4],[1,1]] $ & $8.1802e-21$ & $ 2364 $ 
\\
 $ [[2,1],[1,2],[1,2],[1,1]] $ & $8.4912e-21$ & $ 2364 $ 
\\
 $ [[2,1,1,1],[1,3],[1,1]] $ & $8.5048e-21$ & $ 8065 $ 
\\
 $ [[2,1,1,2],[1,3]] $ & $8.6025e-21$ & $ 5919 $ 
\\
 $ [[1,6],[1,1]] $ & $9.1661e-21$ & $ 2364 $ 
\\
 $ [[2,1],[1,3],[1,1],[1,1]] $ & $9.2423e-21$ & $ 2364 $ 
\\
 $ [[2,1,1,5]] $ & $9.3257e-21$ & $ 8065 $ 
\\
 $ [[2,1],[1,3],[1,2]] $ & $1.0200e-20$ & $ 828 $ 
\\
 $ [[6,1],[1,1]] $ & $1.0441e-20$ & $ 5919 $ 
\\
 $ [[2,1,1,1],[1,4]] $ & $1.0760e-20$ & $ 2364 $ 
\\
 $ [[6,1,1,1]] $ & $1.1401e-20$ & $ 2364 $ 
\\
 $ [[3,1],[1,1],[1,1],[1,1],[1,1]] $ & $1.1725e-20$ & $ 828 $ 
\\
 $ [[2,1],[1,4],[1,1]] $ & $1.1727e-20$ & $ 4059 $ 
\\
 $ [[3,1,1,1],[1,1],[1,1],[1,1]] $ & $1.2249e-20$ & $ 5919 $ 
\\
 $ [[1,7]] $ & $1.2376e-20$ & $ 3925 $ 
\\
 $ [[3,1],[1,2],[1,1],[1,1]] $ & $1.2800e-20$ & $ 828 $ 
\\
 $ [[2,2,1,2],[1,1]] $ & $1.3238e-20$ & $ 8065 $ 
\\
 $ [[2,2,1,1],[1,1],[1,1]] $ & $1.3363e-20$ & $ 2364 $ 
\\
 $ [[3,1,1,1],[1,2],[1,1]] $ & $1.3367e-20$ & $ 3472 $ 
\\
 $ [[3,1,1,2],[1,1],[1,1]] $ & $1.3478e-20$ & $ 5919 $ 
\\
 $ [[3,1],[1,2],[1,2]] $ & $1.3938e-20$ & $ 3472 $ 
\\
 $ [[2,1,1,1],[2,1],[1,1],[1,1]] $ & $1.4444e-20$ & $ 3472 $ 
\\
 $ [[2,2,1,1],[1,2]] $ & $1.4654e-20$ & $ 8065 $ 
\\
 $ [[2,1,1,2],[2,1],[1,1]] $ & $1.4658e-20$ & $ 2364 $ 
\\
 $ [[3,1,1,2],[1,2]] $ & $1.4683e-20$ & $ 5919 $ 
\\
 $ [[2,1,1,1],[2,1,1,1],[1,1]] $ & $1.4699e-20$ & $ 2364 $ 
\\
 $ [[3,1],[1,3],[1,1]] $ & $1.4949e-20$ & $ 3472 $ 
\\
 $ [[2,2,1,3]] $ & $1.5117e-20$ & $ 5919 $ 
\\
 $ [[2,1],[1,5]] $ & $1.5247e-20$ & $ 828 $ 
\\
 $ [[2,1],[2,1],[1,1],[1,1],[1,1]] $ & $1.5495e-20$ & $ 2364 $ 
\\
 $ [[3,1,1,3],[1,1]] $ & $1.5542e-20$ & $ 5361 $ 
\\
 $ [[3,1,1,1],[1,3]] $ & $1.5603e-20$ & $ 2364 $ 
\\
 $ [[2,1,1,3],[2,1]] $ & $1.5745e-20$ & $ 2364 $ 
\\
 $ [[2,2],[1,1],[1,1],[1,1]] $ & $1.5749e-20$ & $ 2364 $ 
\\
 $ [[2,1,1,1],[2,1],[1,2]] $ & $1.5873e-20$ & $ 3925 $ 
\\
 $ [[2,1,1,2],[2,1,1,1]] $ & $1.6281e-20$ & $ 5919 $ 
\\
 $ [[5,1],[1,1],[1,1]] $ & $1.6420e-20$ & $ 2364 $ 
\\
 $ [[4,1],[1,1],[1,1],[1,1]] $ & $1.6658e-20$ & $ 3472 $ 
\\
 $ [[2,1],[2,1],[1,2],[1,1]] $ & $1.7055e-20$ & $ 3925 $ 
\\
 $ [[5,1,1,1],[1,1]] $ & $1.7177e-20$ & $ 8065 $ 
\\
 $ [[4,1,1,1],[1,1],[1,1]] $ & $1.7183e-20$ & $ 2364 $ 
\\
 $ [[2,2],[1,2],[1,1]] $ & $1.7317e-20$ & $ 3472 $ 
\\
 $ [[5,1],[1,2]] $ & $1.7627e-20$ & $ 5919 $ 
\\
 $ [[4,1],[1,2],[1,1]] $ & $1.7980e-20$ & $ 2364 $ 
\\
 $ [[3,1],[1,4]] $ & $1.8270e-20$ & $ 5919 $ 
\\
 $ [[4,1,1,1],[1,2]] $ & $1.8510e-20$ & $ 2364 $ 
\\
 $ [[3,1,1,4]] $ & $1.8605e-20$ & $ 2364 $ 
\\
 $ [[4,1,1,2],[1,1]] $ & $1.8692e-20$ & $ 828 $ 
\\
 $ [[3,1,2,1],[1,1],[1,1]] $ & $1.8767e-20$ & $ 2364 $ 
\\
 $ [[5,1,1,2]] $ & $1.8815e-20$ & $ 3472 $ 
\\
 $ [[3,1,2,1,1,1],[1,1]] $ & $1.8972e-20$ & $ 828 $ 
\\
 $ [[2,1],[2,1],[1,3]] $ & $2.0174e-20$ & $ 2364 $ 
\\
 $ [[3,1,2,1],[1,2]] $ & $2.0362e-20$ & $ 3472 $ 
\\
 $ [[3,1,2,1,1,2]] $ & $2.0379e-20$ & $ 2364 $ 
\\
 $ [[2,2],[1,3]] $ & $2.0453e-20$ & $ 828 $ 
\\
 $ [[4,1],[1,3]] $ & $2.0625e-20$ & $ 2364 $ 
\\
 $ [[4,1,1,3]] $ & $2.1201e-20$ & $ 2364 $ 
\\
 $ [[3,1],[2,1],[1,1],[1,1]] $ & $2.3666e-20$ & $ 828 $ 
\\
 $ [[3,1],[2,1,1,1],[1,1]] $ & $2.4731e-20$ & $ 8065 $ 
\\
 $ [[3,1,1,1],[2,1],[1,1]] $ & $2.4835e-20$ & $ 5919 $ 
\\
 $ [[3,1],[2,1],[1,2]] $ & $2.5639e-20$ & $ 5919 $ 
\\
 $ [[3,1,1,1],[2,1,1,1]] $ & $2.6028e-20$ & $ 5919 $ 
\\
 $ [[2,3,1,1]] $ & $2.6650e-20$ & $ 2364 $ 
\\
 $ [[3,1,1,2],[2,1]] $ & $2.7268e-20$ & $ 828 $ 
\\
 $ [[3,1],[2,1,1,2]] $ & $2.7526e-20$ & $ 2364 $ 
\\
 $ [[3,1,2,2]] $ & $2.7873e-20$ & $ 2364 $ 
\\
 $ [[2,2,1,1],[2,1]] $ & $2.8883e-20$ & $ 828 $ 
\\
 $ [[4,1,2,1],[1,1]] $ & $3.0057e-20$ & $ 2364 $ 
\\
 $ [[5,1],[2,1]] $ & $3.0749e-20$ & $ 2364 $ 
\\
 $ [[5,1,2,1]] $ & $3.0763e-20$ & $ 2364 $ 
\\
 $ [[4,1,3,1]] $ & $3.0943e-20$ & $ 828 $ 
\\
 $ [[2,1,1,1],[2,1],[2,1]] $ & $3.1265e-20$ & $ 828 $ 
\\
 $ [[4,1],[2,1],[1,1]] $ & $3.1372e-20$ & $ 3472 $ 
\\
 $ [[2,2],[2,1,1,1]] $ & $3.1676e-20$ & $ 5919 $ 
\\
 $ [[4,1,1,1],[2,1]] $ & $3.2320e-20$ & $ 2364 $ 
\\
 $ [[2,1],[2,1],[2,1],[1,1]] $ & $3.3527e-20$ & $ 2364 $ 
\\
 $ [[2,2],[2,1],[1,1]] $ & $3.3847e-20$ & $ 3925 $ 
\\
 $ [[2,3],[1,1]] $ & $3.4489e-20$ & $ 5919 $ 
\\
 $ [[4,1,2,1,1,1]] $ & $3.5932e-20$ & $ 2364 $ 
\\
 $ [[3,1,2,1],[2,1]] $ & $3.7184e-20$ & $ 828 $ 
\\
 $ [[4,1],[2,1,1,1]] $ & $3.7762e-20$ & $ 2364 $ 
\\
 $ [[3,1],[3,1],[1,1]] $ & $3.8443e-20$ & $ 2364 $ 
\\
 $ [[3,2],[1,1]] $ & $3.8540e-20$ & $ 3472 $ 
\\
 $ [[3,1,1,1],[3,1]] $ & $4.0146e-20$ & $ 2364 $ 
\\
 $ [[3,2,1,1]] $ & $4.1686e-20$ & $ 2364 $ 
\\
 $ [[3,1],[2,1],[2,1]] $ & $4.6710e-20$ & $ 828 $ 
\\
 $ [[3,1],[2,2]] $ & $4.7224e-20$ & $ 3472 $ 
\\
 $ [[4,1],[3,1]] $ & $5.6626e-20$ & $ 2364 $ 
\\
\end{longtable}

\newpage
%
\begin{longtable}[l]{ m{0.5\textwidth} m{0.25\textwidth} m{0.25\textwidth} }
\caption{Summary level $ 8 $ }
\\
\label{summaryL8}
State & decay width & channels
\\
\endfirsthead
\\
State & decay width & channels
\\
\endhead 
 $ [[2,1],[2,1],[2,1],[1,1],[1,1]] $ & $2.0447e-20$ & $ 4794 $ 
\\
 $ [[2,2],[2,1],[1,1],[1,1]] $ & $2.1143e-20$ & $ 12381 $ 
\\
 $ [[2,1],[2,1],[2,1],[1,2]] $ & $2.1235e-20$ & $ 18452 $ 
\\
$ [[1,1],[1,1],[1,1],[1,1],[1,1],[1,1],[1,1],[1,1]] $ && \\
& $2.1431e-20$ & $ 8672 $ 
\\
$ [[1,2],[1,1],[1,1],[1,1],[1,1],[1,1],[1,1]] $ && \\
& $2.1863e-20$ & $ 3636 $ 
\\
 $ [[2,2],[2,1],[1,2]] $ & $2.1912e-20$ & $ 9257 $ 
\\
 $ [[1,2],[1,2],[1,1],[1,1],[1,1],[1,1]] $ & $2.2261e-20$ & $ 12880 $ 
\\
 $ [[1,2],[1,2],[1,2],[1,1],[1,1]] $ & $2.2630e-20$ & $ 12874 $ 
\\
 $ [[1,3],[1,1],[1,1],[1,1],[1,1],[1,1]] $ & $2.2727e-20$ & $ 12381 $ 
\\
 $ [[1,2],[1,2],[1,2],[1,2]] $ & $2.2976e-20$ & $ 9225 $ 
\\
 $ [[1,3],[1,2],[1,1],[1,1],[1,1]] $ & $2.3056e-20$ & $ 3535 $ 
\\
 $ [[1,3],[1,2],[1,2],[1,1]] $ & $2.3368e-20$ & $ 20766 $ 
\\
 $ [[2,1,1,1],[2,1],[2,1],[1,1]] $ & $2.3626e-20$ & $ 18452 $ 
\\
 $ [[1,3],[1,3],[1,1],[1,1]] $ & $2.3715e-20$ & $ 18452 $ 
\\
 $ [[1,4],[1,1],[1,1],[1,1],[1,1]] $ & $2.3971e-20$ & $ 10878 $ 
\\
 $ [[1,3],[1,3],[1,2]] $ & $2.3992e-20$ & $ 9100 $ 
\\
 $ [[2,2],[2,1,1,1],[1,1]] $ & $2.4089e-20$ & $ 7430 $ 
\\
 $ [[1,4],[1,2],[1,1],[1,1]] $ & $2.4206e-20$ & $ 12880 $ 
\\
 $ [[1,4],[1,2],[1,2]] $ & $2.4441e-20$ & $ 18452 $ 
\\
 $ [[1,4],[1,3],[1,1]] $ & $2.4677e-20$ & $ 7430 $ 
\\
 $ [[2,1,1,1],[2,1,1,1],[2,1]] $ & $2.5372e-20$ & $ 12880 $ 
\\
 $ [[1,4],[1,4]] $ & $2.5382e-20$ & $ 7430 $ 
\\
 $ [[1,5],[1,1],[1,1],[1,1]] $ & $2.5493e-20$ & $ 2688 $ 
\\
 $ [[1,5],[1,2],[1,1]] $ & $2.5626e-20$ & $ 8672 $ 
\\
 $ [[2,1,1,2],[2,1],[2,1]] $ & $2.6871e-20$ & $ 12880 $ 
\\
 $ [[2,2],[2,1,1,2]] $ & $2.7139e-20$ & $ 7430 $ 
\\
 $ [[2,2,1,1],[2,1],[1,1]] $ & $2.7686e-20$ & $ 4794 $ 
\\
 $ [[3,1,2,1],[1,2],[1,1]] $ & $2.8398e-20$ & $ 12880 $ 
\\
 $ [[3,1,2,1],[1,1],[1,1],[1,1]] $ & $2.8398e-20$ & $ 12880 $ 
\\
 $ [[2,1,1,3],[1,1],[1,1],[1,1]] $ & $2.8458e-20$ & $ 8672 $ 
\\
 $ [[2,1,1,3],[1,2],[1,1]] $ & $2.8736e-20$ & $ 8672 $ 
\\
 $ [[2,1,1,2],[1,1],[1,1],[1,1],[1,1]] $ & $2.9362e-20$ & $ 12880 $ 
\\
 $ [[2,1,1,6]] $ & $2.9426e-20$ & $ 9100 $ 
\\
 $ [[2,1,1,2],[1,2],[1,1],[1,1]] $ & $2.9600e-20$ & $ 1569 $ 
\\
 $ [[3,1,2,1],[2,1],[1,1]] $ & $2.9776e-20$ & $ 18452 $ 
\\
 $ [[2,1,1,2],[1,2],[1,2]] $ & $2.9804e-20$ & $ 12880 $ 
\\
 $ [[2,1,1,2],[1,3],[1,1]] $ & $3.0076e-20$ & $ 7430 $ 
\\
 $ [[1,8]] $ & $3.0331e-20$ & $ 12880 $ 
\\
 $ [[2,1,1,2],[2,1,1,2]] $ & $3.0389e-20$ & $ 1569 $ 
\\
 $ [[2,1,1,1],[1,1],[1,1],[1,1],[1,1],[1,1]] $ & $3.0510e-20$ & $ 2688 $ 
\\
 $ [[2,1,1,1],[1,2],[1,1],[1,1],[1,1]] $ & $3.0654e-20$ & $ 9225 $ 
\\
 $ [[2,1,1,2],[1,4]] $ & $3.0740e-20$ & $ 20766 $ 
\\
 $ [[2,1,1,1],[1,2],[1,2],[1,1]] $ & $3.0771e-20$ & $ 4794 $ 
\\
 $ [[2,1,1,1],[1,3],[1,1],[1,1]] $ & $3.0940e-20$ & $ 18452 $ 
\\
 $ [[2,1,1,1],[1,3],[1,2]] $ & $3.1005e-20$ & $ 18452 $ 
\\
 $ [[2,1],[2,1],[2,1],[2,1]] $ & $3.1174e-20$ & $ 12880 $ 
\\
 $ [[2,1,1,1],[1,4],[1,1]] $ & $3.1330e-20$ & $ 3535 $ 
\\
 $ [[4,1],[3,1,1,1]] $ & $3.1336e-20$ & $ 12880 $ 
\\
 $ [[4,1],[2,1,1,2]] $ & $3.1455e-20$ & $ 4794 $ 
\\
 $ [[2,2,1,4]] $ & $3.1477e-20$ & $ 4794 $ 
\\
 $ [[2,1],[1,1],[1,1],[1,1],[1,1],[1,1],[1,1]] $ & $3.1614e-20$ & $ 1569 $ 
\\
 $ [[2,1],[1,2],[1,1],[1,1],[1,1],[1,1]] $ & $3.1660e-20$ & $ 9257 $ 
\\
 $ [[2,1],[1,2],[1,2],[1,1],[1,1]] $ & $3.1698e-20$ & $ 21494 $ 
\\
 $ [[2,1],[1,2],[1,2],[1,2]] $ & $3.1742e-20$ & $ 25370 $ 
\\
 $ [[2,1],[1,3],[1,1],[1,1],[1,1]] $ & $3.1752e-20$ & $ 8672 $ 
\\
 $ [[2,1],[1,3],[1,3]] $ & $3.1817e-20$ & $ 20766 $ 
\\
 $ [[2,1],[1,4],[1,1],[1,1]] $ & $3.1878e-20$ & $ 7430 $ 
\\
 $ [[2,1],[1,4],[1,2]] $ & $3.1897e-20$ & $ 18452 $ 
\\
 $ [[2,2],[2,1],[2,1]] $ & $3.1898e-20$ & $ 12880 $ 
\\
 $ [[2,1],[1,5],[1,1]] $ & $3.2014e-20$ & $ 7430 $ 
\\
 $ [[3,1,1,1],[2,1,1,2]] $ & $3.2025e-20$ & $ 12880 $ 
\\
 $ [[4,1],[3,1],[1,1]] $ & $3.2081e-20$ & $ 4794 $ 
\\
 $ [[3,1,1,1],[3,1,1,1]] $ & $3.2395e-20$ & $ 12880 $ 
\\
 $ [[2,1],[2,1],[1,4]] $ & $3.2532e-20$ & $ 2688 $ 
\\
 $ [[2,2],[2,2]] $ & $3.2579e-20$ & $ 7430 $ 
\\
 $ [[3,1],[2,1],[2,1],[1,1]] $ & $3.2696e-20$ & $ 2688 $ 
\\
 $ [[3,1],[2,2],[1,1]] $ & $3.2792e-20$ & $ 18452 $ 
\\
 $ [[3,1,1,1],[3,1],[1,1]] $ & $3.2986e-20$ & $ 12880 $ 
\\
 $ [[2,2],[1,4]] $ & $3.3167e-20$ & $ 12880 $ 
\\
 $ [[2,1],[2,1],[1,3],[1,1]] $ & $3.3720e-20$ & $ 4794 $ 
\\
 $ [[5,1],[1,1],[1,1],[1,1]] $ & $3.3912e-20$ & $ 18452 $ 
\\
 $ [[2,3,1,2]] $ & $3.3946e-20$ & $ 8672 $ 
\\
 $ [[5,1],[1,2],[1,1]] $ & $3.3986e-20$ & $ 1569 $ 
\\
 $ [[2,1,1,2],[2,1,1,1],[1,1]] $ & $3.4014e-20$ & $ 12880 $ 
\\
 $ [[2,1],[2,1],[1,2],[1,2]] $ & $3.4341e-20$ & $ 12880 $ 
\\
 $ [[2,2],[1,3],[1,1]] $ & $3.4387e-20$ & $ 10878 $ 
\\
 $ [[2,1],[2,1],[1,2],[1,1],[1,1]] $ & $3.4662e-20$ & $ 12880 $ 
\\
 $ [[4,1,1,1],[1,1],[1,1],[1,1]] $ & $3.4815e-20$ & $ 25370 $ 
\\
 $ [[4,1,1,1],[1,2],[1,1]] $ & $3.4923e-20$ & $ 8672 $ 
\\
 $ [[2,2],[1,2],[1,2]] $ & $3.5019e-20$ & $ 12880 $ 
\\
 $ [[2,1,1,1],[2,1,1,1],[1,2]] $ & $3.5094e-20$ & $ 7430 $ 
\\
 $ [[2,1],[2,1],[1,1],[1,1],[1,1],[1,1]] $ & $3.5133e-20$ & $ 7430 $ 
\\
 $ [[2,1,1,1],[2,1],[1,3]] $ & $3.5335e-20$ & $ 12874 $ 
\\
 $ [[2,2],[1,2],[1,1],[1,1]] $ & $3.5350e-20$ & $ 12880 $ 
\\
 $ [[2,1,1,1],[2,1,1,1],[1,1],[1,1]] $ & $3.5395e-20$ & $ 18452 $ 
\\
 $ [[3,1],[2,1,1,1],[2,1]] $ & $3.5823e-20$ & $ 10878 $ 
\\
 $ [[2,2],[1,1],[1,1],[1,1],[1,1]] $ & $3.5832e-20$ & $ 12880 $ 
\\
 $ [[2,1,1,1],[2,1],[1,2],[1,1]] $ & $3.6254e-20$ & $ 9100 $ 
\\
 $ [[3,1],[3,1],[1,1],[1,1]] $ & $3.6707e-20$ & $ 31042 $ 
\\
 $ [[2,1,1,1],[2,1],[1,1],[1,1],[1,1]] $ & $3.6713e-20$ & $ 10878 $ 
\\
 $ [[4,1],[2,1,1,1],[1,1]] $ & $3.7071e-20$ & $ 20766 $ 
\\
 $ [[4,1],[1,1],[1,1],[1,1],[1,1]] $ & $3.7105e-20$ & $ 12880 $ 
\\
 $ [[4,1],[1,2],[1,1],[1,1]] $ & $3.7124e-20$ & $ 9100 $ 
\\
 $ [[4,1],[1,3],[1,1]] $ & $3.7160e-20$ & $ 18452 $ 
\\
 $ [[4,1],[1,2],[1,2]] $ & $3.7186e-20$ & $ 12880 $ 
\\
 $ [[3,1],[3,1],[1,2]] $ & $3.7207e-20$ & $ 7430 $ 
\\
 $ [[4,1],[1,4]] $ & $3.7281e-20$ & $ 7430 $ 
\\
 $ [[3,1],[2,1,1,2],[1,1]] $ & $3.7429e-20$ & $ 4794 $ 
\\
 $ [[3,1,1,1],[2,1,1,1],[1,1]] $ & $3.7565e-20$ & $ 12880 $ 
\\
 $ [[2,1,1,2],[2,1],[1,2]] $ & $3.7578e-20$ & $ 12880 $ 
\\
 $ [[2,1,1,2],[2,1],[1,1],[1,1]] $ & $3.7946e-20$ & $ 12880 $ 
\\
 $ [[2,2,1,1],[1,2],[1,1]] $ & $3.8466e-20$ & $ 18452 $ 
\\
 $ [[2,1,1,3],[2,1],[1,1]] $ & $3.8866e-20$ & $ 12880 $ 
\\
 $ [[2,2,1,1],[1,1],[1,1],[1,1]] $ & $3.8914e-20$ & $ 12880 $ 
\\
 $ [[3,1],[3,1],[2,1]] $ & $3.9309e-20$ & $ 8672 $ 
\\
 $ [[3,1,1,2],[1,2],[1,1]] $ & $3.9392e-20$ & $ 7430 $ 
\\
 $ [[3,1,1,2],[1,1],[1,1],[1,1]] $ & $3.9436e-20$ & $ 12880 $ 
\\
 $ [[3,1,1,2],[2,1],[1,1]] $ & $3.9447e-20$ & $ 8672 $ 
\\
 $ [[3,1,1,1],[2,1],[2,1]] $ & $4.0404e-20$ & $ 12880 $ 
\\
 $ [[3,1,1,1],[2,2]] $ & $4.0467e-20$ & $ 7430 $ 
\\
 $ [[4,1,1,1],[2,1],[1,1]] $ & $4.1419e-20$ & $ 4794 $ 
\\
 $ [[3,1,1,1],[1,4]] $ & $4.1786e-20$ & $ 12880 $ 
\\
 $ [[3,1,1,1],[1,3],[1,1]] $ & $4.1999e-20$ & $ 4794 $ 
\\
 $ [[3,1,1,1],[1,2],[1,2]] $ & $4.2103e-20$ & $ 12880 $ 
\\
 $ [[3,1,1,1],[1,2],[1,1],[1,1]] $ & $4.2163e-20$ & $ 7430 $ 
\\
 $ [[3,1,1,1],[1,1],[1,1],[1,1],[1,1]] $ & $4.2245e-20$ & $ 18452 $ 
\\
 $ [[3,1],[2,1,1,1],[1,2]] $ & $4.2512e-20$ & $ 9100 $ 
\\
 $ [[3,1],[2,1,1,1],[1,1],[1,1]] $ & $4.2659e-20$ & $ 12880 $ 
\\
 $ [[3,1,1,1],[2,1],[1,2]] $ & $4.3287e-20$ & $ 18452 $ 
\\
 $ [[3,1,1,1],[2,1],[1,1],[1,1]] $ & $4.3505e-20$ & $ 12880 $ 
\\
 $ [[5,1],[2,1],[1,1]] $ & $4.4329e-20$ & $ 18452 $ 
\\
 $ [[3,1],[1,4],[1,1]] $ & $4.4435e-20$ & $ 12880 $ 
\\
 $ [[4,1],[2,1],[1,1],[1,1]] $ & $4.4535e-20$ & $ 12880 $ 
\\
 $ [[4,1],[2,1],[1,2]] $ & $4.4641e-20$ & $ 18452 $ 
\\
 $ [[3,1],[1,3],[1,2]] $ & $4.4680e-20$ & $ 12880 $ 
\\
 $ [[3,1],[1,3],[1,1],[1,1]] $ & $4.4756e-20$ & $ 12880 $ 
\\
 $ [[3,1],[1,2],[1,2],[1,1]] $ & $4.4881e-20$ & $ 18452 $ 
\\
 $ [[3,1],[1,2],[1,1],[1,1],[1,1]] $ & $4.4982e-20$ & $ 18452 $ 
\\
 $ [[3,1],[1,1],[1,1],[1,1],[1,1],[1,1]] $ & $4.5095e-20$ & $ 20766 $ 
\\
 $ [[4,1],[4,1]] $ & $4.5562e-20$ & $ 9225 $ 
\\
 $ [[3,1],[2,1],[1,3]] $ & $4.6443e-20$ & $ 2688 $ 
\\
 $ [[4,1],[2,2]] $ & $4.8012e-20$ & $ 12880 $ 
\\
 $ [[3,1],[2,1],[1,1],[1,1],[1,1]] $ & $4.8018e-20$ & $ 4794 $ 
\\
 $ [[4,1],[2,1],[2,1]] $ & $4.8298e-20$ & $ 20766 $ 
\\
\end{longtable}